\documentclass[12pt]{iopart}
\usepackage[square,numbers]{natbib}

\usepackage{amsmath}
\usepackage{amssymb}
\usepackage{xcolor}
\usepackage{bm}
\usepackage{graphicx}
\usepackage{subcaption}
\usepackage{hyperref}
\newcommand{\partder}[2]{\frac{\partial  #1}{\partial  #2}\displaystyle} 
\begin{document}

\title[Energetic particle loss mechanisms in equilibria close to quasisymmetry]{Energetic particle loss mechanisms in reactor-scale equilibria close to quasisymmetry}

\author{E. J. Paul\textsuperscript{1}, A. Bhattacharjee\textsuperscript{1}, M. Landreman\textsuperscript{2}, D. Alex\textsuperscript{3}, J. L. Velasco\textsuperscript{4}, and R. Nies\textsuperscript{1}}
\address{\textsuperscript{1}Department of Astrophysical Sciences, Princeton University, 4 Ivy Lane, Princeton NJ 08544, USA\\ 
\textsuperscript{2}Institute for Research in Electronics and Applied Physics, University of Maryland, College Park MD 20742,
USA \\
\textsuperscript{3}Department of Aerospace and Ocean Engineering, Virginia Tech, 460 Old Turner Street, Blacksburg VA 24060, USA \\
\textsuperscript{4}Laboratorio Nacional de Fusion, CIEMAT, Madrid, Spain}
\ead{epaul@princeton.edu}
\vspace{10pt}

\begin{abstract}
Collisionless physics primarily determines the transport of fusion-born alpha particles in 3D equilibria. Several transport mechanisms have been implicated in stellarator configurations, including stochastic diffusion due to class transitions, ripple trapping, and banana drift-convective orbits. Given the guiding center dynamics in a set of six quasihelical and quasiaxisymmetric equilibria, we perform a classification of trapping states and transport mechanisms. In addition to banana drift convection and ripple transport, we observe substantial non-conservation of the parallel adiabatic invariant which can cause losses through diffusive banana tip motion. Furthermore, many lost trajectories undergo transitions between trapping classes on longer time scales, either with periodic or irregular behavior. We discuss possible optimization strategies for each of the relevant transport mechanisms. We perform a comparison between fast ion losses and metrics for the prevalence of mechanisms such as banana-drift convection \cite{2021Velasco}, transitioning orbits, and wide orbit widths. Quasihelical configurations are found to have natural protection against ripple-trapping and diffusive banana tip motion leading to a reduction in prompt losses.
\end{abstract}

%
%
%
%
%

\section{Introduction}

Magnetic confinement reactors must confine both the thermal and energetic particle populations as excessive losses of fusion-born alpha particles can damage material structures and reduce the heating of the bulk plasma. When three-dimensional (3D) magnetic fields are introduced, such as in a stellarator or rippled tokamak, collisionless particle orbits are no longer automatically confined and weakly 
collisional transport is generally enhanced. 

Mitigating alpha particle losses is one of the critical issues in stellarator reactor design. The alpha energy confinement impacts the power balance: an alpha heating efficiency of at least 90\% \cite{2010Sagara,2022Alonso} or 95\% \cite{2008Sagara} is often assumed in stellarator reactor studies. Unconfined alpha orbits can also lead to localized heating of divertor plates or other plasma-facing surfaces, leading to blistering, erosion, and degradation of mechanical properties \cite{2003Ueda}. Although the alpha confinement was a target in the ARIES-CS design, the alpha losses contributed 5 MW/m$^2$ to the maximum divertor heat flux \cite{2008Mau}. In a four-field-period HSR4/18 Helias reactor study \citep{2004Andreeva}, the alpha particle energy loss was reduced to 2.5\% of the heating power \citep{2001Beidlerb}.
While small enough not to impact power balance significantly, these losses may still be a concern due to wall loading. At fixed power, operating a stellarator reactor at higher field strength requires a reduction in density and an increase in temperature \cite{2022Alonso}. While the decrease in beta may alleviate stability and equilibrium limits, the increase in the slowing-down time necessitates stricter requirements on the alpha confinement.  

The confinement of fusion-born alpha particles is largely determined by collisionless physics. 
Comparisons of collisionless with collisional alpha particle losses have found that pitch-angle scattering effects are insignificant for losses before the slowing-down time of $\approx 0.05$ seconds in a W7-X reactor configuration \citep{2021Lazerson}. Generally, the difference between collisionless and collisional calculations is configuration dependent \cite{2021Bader}. Radial electric fields are not likely to have a large impact on the fusion-born alpha particle dynamics long before their slowing down, given that potential differences on the order of MV are required to change the particle energy by order unity. While some calculations have accounted for the ambipolar electric field \cite{2019Henneberg}, its calculation introduces additional complications into the configuration comparisons through the choice of profiles. 
Alpha particles can also be transported due to their interaction with Alfv\'{e}nic instabilities \citep{2002Kolesnichenko, 2006Spong}. We will not consider this mechanism in this work, as this requires knowledge of the thermal and fast ion profiles and more detailed modeling of the mode structure of the instabilities. In this work we focus on the collisionless dynamics without electric fields or perturbations in order to cleanly distinguish the orbital dynamics between configurations without knowledge of the thermal and fast ion profiles.

Many potential collisionless loss mechanisms have been identified in 3D magnetic fields. 
If 3D perturbations are large enough that local wells form along magnetic field lines, local ripple trapping can occur, generally leading to unconfined orbits. Particles trapped in the main toroidal field, analogous to banana orbits in a tokamak, can exhibit banana diffusion and convection \citep{1981Goldston,1996White}. As trajectories transition between trapping in the ripple field and the main toroidal field, collisionless trapping and detrapping leads to substantial transport in both rippled tokamaks \cite{1984Mynick} and stellarators \cite{1983Mynick}. In strongly 3D configurations where multiple trapping can occur, stochastic transitioning behavior is present, leading to diffusive transport \citep{2001Beidler,1992Lotz,2021Tykhyy}. In-surface precessional drift motion can also alter the confinement properties of fast ions. If a particle precesses sufficiently on a surface, the perturbation to the radial drift is phase-mixed, reducing transport. On the other hand, if a particle remains localized near a particular perturbation, a directed radial drift may result, termed banana drift convection transport \citep{2005Nemov}. Similarly, if the bounce or passing motion is in resonance with a perturbation in the equilibrium, a directed radial drift can occur \cite{2009Park,2022White}.

Previous studies have discussed some aspects of fast ion loss mechanisms in 3D configurations, such as the distribution of losses on the first wall \cite{1992Lotz,2008Ku,2016Faustin} or in velocity space \cite{2014Nemov,2014Drevlak,2021Bader}. 
Comparisons of guiding center losses with banana drift convection metrics \cite{2021Bader,2021Velasco} and theoretical predictions from diffusion coefficients \citep{2001Beidler} have been made. However, a more quantitative analysis of the dominant loss channels has not been performed. 

In this work, we perform a detailed analysis of guiding-center confinement in 3D magnetic fields to quantify the relevant loss mechanisms on prompt and non-prompt time scales. We focus our attention on configurations close to quasisymmetry as it eases the classification of orbit topology. 
(For example, the mirror points of trapped orbits in quasi-isodynamic configurations may differ from those in a quasisymmetric configuration.)
By highlighting the most pertinent loss mechanisms in each configuration, we can identify avenues for improving collisionless confinement. Furthermore, a deeper understanding of fast ion orbits in 3D magnetic fields may indicate possible resonant interactions with external perturbations \citep{2021White,2022White}, such as MHD modes or error fields. Finally, more effective optimization strategies can be elucidated by isolating the most critical loss channels. Recent work has indicated that magnetic fields with precise quasisymmetry exist \cite{2022Landreman,2022Wechsung}, yielding excellent guiding center confinement. However, relaxing the requirement of precise symmetry may ease the accommodation of other constraints, such as engineering requirements or microstability. 

Section \ref{sec:loss_mechanisms} provides an overview of fast ion loss mechanisms in 3D fields and classification strategies. In Section \ref{sec:guiding_center} we describe the guiding center calculations and the configurations under consideration. Section \ref{sec:summary_data} describes the classification results for the configurations close to quasiaxisymmety and quasihelical symmetry, including relevant example trajectories. In light of these observations, in Section \ref{sec:optimization_techniques} we discuss equilibrium properties that reduce the identified loss mechanisms. In Section \ref{sec:summary} we summarize the results, and in Section \ref{sec:conclusions} we conclude. 

\section{Loss mechanisms and classification}
\label{sec:loss_mechanisms}

We analyze the lost particles from guiding-center calculations to elucidate the loss mechanisms. The positions of all turning points (where $v_{\|}=0$) along the trajectory are identified. We define a \textit{half-bounce segment} as the segment of the entire trajectory between two consecutive turning points and a \textit{bounce segment} as two consecutive half-bounce segments such that the particle returns to approximately the same position to form a closed orbit. Given the position of each turning point, we categorize the orbit type for each half-bounce segment as banana trapped (Section \ref{sec:banana_trapping}), ripple trapped (Section \ref{sec:ripple_trapping_mechanism}), or barely trapped (Section \ref{sec:barely_trapped}). If a particle never reflects, it is categorized as passing (Section \ref{sec:passing}). Below we discuss the classification of each orbit type and their corresponding loss mechanisms.

\begin{figure}
    \centering 
    \begin{subfigure}[b]{0.32\textwidth}   
    \centering 
    \includegraphics[width=1.0\textwidth]{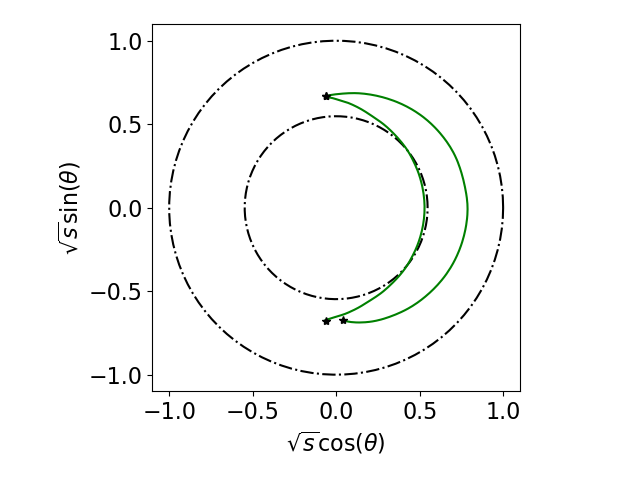}
    \caption{}
    \label{fig:banana_example}
    \end{subfigure}
    \begin{subfigure}[b]{0.32\textwidth}   
    \centering 
    \includegraphics[width=1.0\textwidth]{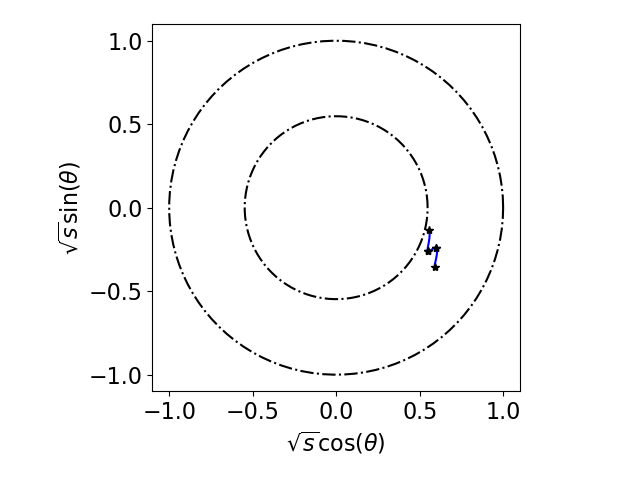}
    \caption{}
    \label{fig:ripple_example}
    \end{subfigure}
    \begin{subfigure}[b]{0.32\textwidth}   
    \centering 
    \includegraphics[width=1.0\textwidth]{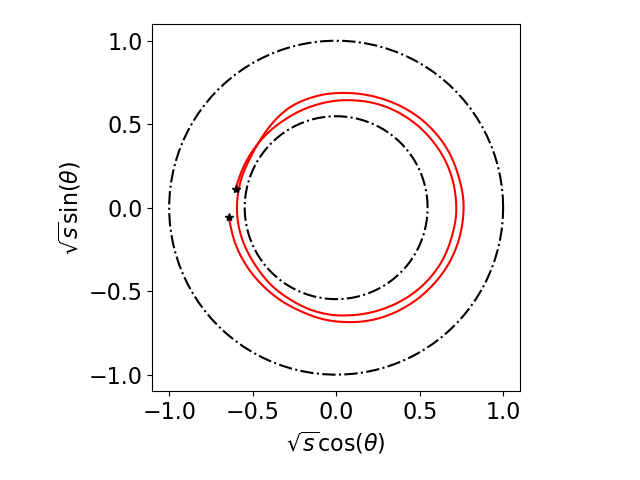}
    \caption{}
    \label{fig:passing_example}
    \end{subfigure}
    \caption{Example ARIES-CS bounce segments in the (a) banana, (b) ripple, and (c) barely-trapped classes are shown in the poloidal cross-section. Black stars indicate bounce points. Black dashed lines indicate the initial magnetic surface and plasma boundary. 
    }
\end{figure}

\subsection{Banana trapping}
\label{sec:banana_trapping}

We use the term banana to refer to a particle trapped in the primary well along a field line. This orbit type is analogous to the trapped particles in a tokamak whose trajectories are banana-shaped in the $(\sqrt{s} \cos\theta, \sqrt{s} \sin\theta)$ plane. Here $s = \psi/\psi_0$ is the normalized toroidal flux and $\theta$ and $\zeta$ are the Boozer poloidal and toroidal angles, respectively. Similarly, for a quasisymmetry magnetic field  $B(s,\chi)$ where $\chi = M \theta - N \zeta$, analogous banana motion appears in the $(\sqrt{s} \cos\chi, \sqrt{s} \sin\chi)$ plane. We assume throughout that the minimum of the field strength occurs at $\chi = 0$. In perfect quasisymmetry,
the banana motion closes in the $(\sqrt{s} \cos\chi, \sqrt{s} \sin\chi)$ plane without any net radial drift. Deviation from quasisymmetry or omnigeneity introduces perturbations that allow for transport similar to that in a rippled tokamak.

A typical banana orbit is centered about the midplane (has an average value of $\chi$ close to zero), and the trajectory does not make a full transit through the field strength variation (the change in $\chi$ between bounces, $\Delta \chi$, is less than $2\pi$). We therefore compare the mean value of $\chi$ along the half-bounce segment, $\mathrm{mean}(\chi) \in [-\pi,\pi)$, to the position of the field minimum, $\chi = 0$. We classify banana half-bounce segments as those not satisfying the conditions for ripple trapping or barely trapping as described in Sections \ref{sec:ripple_trapping_mechanism} and \ref{sec:barely_trapped}.
See Figure \ref{fig:banana_example} for an example of a banana orbit segment from ARIES-CS. Throughout the text, banana segments are indicated by green. 

\subsubsection{Wide bananas}

If the width of banana orbits is sufficiently large, particle orbits may intersect the wall even without symmetry breaking. The timescale of this transport mechanism, $\tau_{\mathrm{loss}}$, scales with the bounce time, $\tau_{\mathrm{b}} \sim 10^{-5}$ seconds. With symmetry breaking, wide orbits can lead to rapid diffusion, given that the orbit width sets the typical step size for banana diffusion. Even in a precisely quasisymmetric field, a small number of fast ions may be unconfined due to these effects \cite{2022Landreman}. 

\subsubsection{Banana-drift convection}
\label{sec:superbanana}

We will use the term banana-drift convection to refer to transport on trajectories that do not precess rapidly compared to their radial drift. Since these particles remain confined in toroidal angle, they may experience a directed radial drift and be quickly lost, see Figure \ref{fig:aten_sb}. (The term superbanana has been used in the literature to denote this drift convective motion \cite{2021Velasco}, although we avoid this term since it can also be used to refer to particles that are ripple-trapped during some segment of their orbit \citep{2006Mynick}.) The timescale for this transport is associated with the bounce-averaged radial drift frequency, $\overline{\textbf{v}_{\mathrm{m}} \cdot \nabla s}$, as $\tau_{\mathrm{loss}} \sim (1 - s_0)/\overline{\textbf{v}_{\mathrm{m}} \cdot \nabla s}$, where $1-s_0$ is the distance in normalized flux over which the particle is transported from initial surface $s_0$. This type of transport can also occur for ripple-trapped and barely-trapped particles. In the case of a strong radial electric field, the electric and magnetic precession drifts can balance each other, leading to the banana-drift convection condition \citep{2006Kolesnichenko}. However, the electric precession drift is typically not significant near the fusion-born alpha birth energy. 

A parameter that has been used in the literature for classifying drift convection is \citep{2008Nemov},
\begin{equation}
    \gamma_{\mathrm{c}} = \frac{2}{\pi} \mathrm{arctan} \left(\frac{\overline{\textbf{v}_{\mathrm{m}} \cdot \nabla s}}{|\overline{\textbf{v}_{\mathrm{m}} \cdot \nabla \alpha}|} \right).
    \label{eq:gamma_c}
\end{equation}
Here $\alpha = \theta - \iota(s) \zeta$ is a field-line label, $\textbf{v}_{\mathrm{m}}$ is the magnetic drift,
\begin{equation}
    \textbf{v}_{\mathrm{m}} =  \frac{v_{\perp}^2}{2 \Omega} \hat{\textbf{b}} \times \nabla \ln(B) + \frac{v_{\|}^2}{\Omega} \hat{\textbf{b}} \times \bm{\kappa},
\end{equation}
$v_{\|}$ is the parallel velocity, $v_{\perp}$ is the perpendicular velocity, $\Omega = q B/m$ is the gyrofrequency, $q$ is the charge, $m$ is the mass, $\hat{\textbf{b}} = \textbf{B}/B$ is the unit vector in the direction of the magnetic field, and $\bm{\kappa} = \hat{\textbf{b}} \cdot \nabla \hat{\textbf{b}}$. In \eqref{eq:gamma_c}, overlines indicate averages of the lowest-order bounce motion along a field line,
\begin{equation}
    \overline{A} = \frac{\oint \frac{dl}{v_{\|}} \, A }{\oint \frac{dl}{v_{\|}}},
\end{equation}
where $l$ measures length along a field line and integration is performed along the closed orbit between bounce points. The quantity $\gamma_{\mathrm{c}}$ takes values in $[-1,1]$, with large positive values indicating outward convective transport. Small values of $|\gamma_{\mathrm{c}}|$ promote closure of the contours of the adiabatic invariant,
\begin{equation}
    J_{\|} = \oint dl \, v_{\|}.
    \label{eq:jpar}
\end{equation}
Since $\overline{\textbf{v}_{\mathrm{m}} \cdot \nabla s}/\overline{\textbf{v}_{\mathrm{m}} \cdot \nabla \alpha} = \left(\partial J_{\|}/\partial \alpha\right)/\left(\partial J_{\|}/\partial s \right)$, small values of $|\gamma_{\mathrm{c}}|$ indicate small angles between $J_{\|}$ contours and flux surfaces. 

To identify banana-drift convection among lost trajectories, we estimate the $\gamma_{\mathrm{c}}$ parameter \eqref{eq:gamma_c} as, 
\begin{equation}
    \gamma_{\mathrm{c}} \approx \frac{2}{\pi} \tan^{-1} \left( \frac{\Delta s}{|\Delta \theta - \mathrm{mean}(\iota) \Delta \zeta |} \right),
    \label{eq:gammac_traj}
\end{equation}
where $\Delta$ represents the difference in the coordinate along the bounce segment and $\mathrm{mean}(\iota)$ is the mean value of $\iota$ along the bounce segment. Each bounce segment is classified as drift convective (DC) if it satisfies $|\gamma_{\mathrm{c}}| > 0.2$ \cite{2021Velasco}. The choice of this critical value is somewhat arbitrary, as small positive values of $\gamma_{\mathrm{c}}$ may allow for convection along $J_{\|}$ contours, although on a longer timescale. A DC segment can be in the ripple, banana, or barely-trapped class. However, we will focus on the distinction between DC and non-DC banana bounce segments due to the prevalence of DC transport among ripple-trapped trajectories and the small number of barely-trapped trajectories. A bounce segment classified as a non-DC banana indicates that diffusive banana tip motion is dominant, resonant banana motion is present, or $J_{\|}$ contours align well with flux surfaces and convective radial transport is relatively slow. 

\subsubsection{Resonant banana-drift convection}

On the drift timescale, convective loss of bananas can be driven by the resonance between the bounce motion and toroidal precession \citep{1996White,2013Park}, see Figure \ref{fig:iter_passing_resonance}. Here the net drift in the toroidal angle, $\Delta \zeta$, between successive bounces (an almost-closed orbit) satisfies the condition,
\begin{equation}
l\Delta \zeta = k \frac{2\pi}{N_P},
\label{eq:bounce_transit_resonance}
\end{equation}
where $N_P$ is the number of field periods and $l$ and $k$ are integers such that after $l$ full bounce orbits, the bounce tip returns to the same location in toroidal angle. Since the same perturbation phase is applied upon successive bounce orbits, a directed radial drift can occur if a perturbation with the same mode numbers is present in the equilibrium. Similar to banana-drift convection, the timescale for this transport is associated with the bounce-averaged radial drift frequency, $\tau_{\mathrm{loss}} \sim (1 - s_0)/\overline{\textbf{v}_{\mathrm{m}} \cdot \nabla s}$. This transport mechanism has been suggested to be present in stellarator equilibria \citep{2004Kolesnichenko}, although there has been little numerical or experimental evidence. A special case of this resonance condition with $l = 1$, $k = 0$ corresponds with banana-drift convective orbits, discussed in Section \ref{sec:superbanana}. If the resonance condition is satisfied for $l \ne 1$, then the shape of the magnetic field strength well may change substantially on successive bounce orbits, leading to possible non-conservation of $J_{\|}$.

\subsubsection{Banana-drift diffusion}

When several resonances in the banana-drift motion begin to overlap, diffusive transport of banana tips begins to occur \citep{1981Goldston,1996White}, see Figure \ref{fig:ariescs_diffusive_banana}. Here the criterion for resonant overlap occurs when the radial displacement of the banana tips, $\Delta s$, changes the net toroidal drift, $\Delta \zeta$, during a full bounce orbit,
\begin{equation}
    \left(\Delta \zeta\right)'(s) \Delta s \gtrsim \frac{2\pi}{N_P}.
\end{equation}
This resonant overlap condition is more likely to occur for wide banana orbits. If strong overlap is present such that diffusive motion arises, the timescale for loss scales as $\tau_{\mathrm{loss}} \sim \left(1 - s_0\right)^2/D$. Using the radial magnetic drift $\overline{ \textbf{v}_{\mathrm{m}} \cdot \nabla s}$ and the bounce time $\tau_{\mathrm{b}}$ as characteristic scales, we can postulate the diffusion coefficient to scale as $D \sim \left(\overline{ \textbf{v}_{\mathrm{m}} \cdot \nabla s}\right)^2 \tau_{\mathrm{b}}$.

While banana-drift convection arises due to trajectories that follow $J_{\|}$ contours that are misaligned with surfaces, banana-drift diffusive yields stochastic motion due to non-conservation of $J_{\|}$. This feature will be discussed further in Section \ref{sec:diffusive_transport}. To distinguish this mechanism, we evaluate the parallel adiabatic invariant \eqref{eq:jpar} if the particle remains within the same trapping class for the full bounce orbit. Integration is performed along the GC trajectory rather than along a field line between successive bounces.

\subsection{Ripple trapping}
\label{sec:ripple_trapping_mechanism}

If a given perturbation to quasisymmetry is large enough, local ripple wells in the field strength can form along a field line. Since ripple-trapped particles tend to be localized in poloidal and toroidal angles, the averaged radial drift typically does not vanish over a bounce period. For a configuration close to QS, the dominant contribution to the magnetic drift comes from the unperturbed quasisymmetric field, leading to $\textbf{v}_{\mathrm{m}} \cdot \nabla s \propto \partial B(s,\chi)/\partial \chi$. If the ripple well is localized away from the field maxima or minima on a surface, this can lead to rapid radial convection. Due to their localization, ripple-trapped trajectories often have features in common with banana-drift convection trajectories, including large values of $|\gamma_{\mathrm{c}}|$, see Figure \ref{fig:ariescs_ripple}. Similar to banana-drift convection, the timescale for this transport is associated with the bounce-averaged radial drift frequency, $\tau_{\mathrm{loss}} \sim (1 - s_0)/\overline{\textbf{v}_{\mathrm{m}} \cdot \nabla s}$.

A ripple-trapped segment is characterized by localization in $\chi$ or small values of $\Delta \chi$ along a half-bounce segment. To distinguish a ripple-trapped segment from a deeply-trapped banana segment, we also compare the mean value of $\chi$ along the half-bounce segment, $\mathrm{mean}(\chi) \in [-\pi,\pi)$, to the location of the field minimum, $\chi = 0$. Specifically we classify a ripple-trapped half-bounce segment as satisfying the inequalities $\mathrm{mean}(\chi)/(2\pi) > 0.04$ and $\Delta \chi/(2\pi) < 1/8$. (In ITER, we instead use the condition $\Delta \chi/(2\pi) < 1/30$ or ($\mathrm{mean}(\chi)/(2\pi) > 0.04$ and $\Delta \chi/(2\pi) < 1/8$) due to the prevalence of deeply-trapped banana trajectories.) See Figure \ref{fig:ripple_example} for an example of a ripple-trapped bounce segment in ARIES-CS. Throughout the text, ripple-trapped segments are indicated by blue. 

\subsection{Barely trapping}
\label{sec:barely_trapped}

We use the term barely trapped to refer to a trapped-particle trajectory that can pass through a full period of the field strength variation, $\Delta \chi = 2\pi$, without mirroring. This behavior occurs because of the variation of the field strength maxima with respect to the field line label or magnetic surface label. 
We classify barely-trapped half-bounce segments as satisfying the inequality $\Delta \chi/(2\pi)> 1.25$ to prevent the misclassification of banana-trapped segments as barely-trapped segments. See Figure \ref{fig:passing_example} for an example of a barely-trapped half-bounce segment in ARIES-CS. Throughout the text, barely-trapped segments are indicated by red. 

Many barely-trapped trajectories nearly close in the $(\sqrt{s}\cos\chi,\sqrt{s}\sin\chi)$ plane during a bounce segment, see Figure \ref{fig:aten_sb_passing}. If a barely-trapped trajectory makes few toroidal transits between bounces, the radial drift may not vanish under an orbit average. This mechanism is similar to banana-drift convection. An example can be seen in Figure \ref{fig:aten_sb_passing}.

\subsection{Passing trajectories}
\label{sec:passing}

Passing trajectories do not mirror. If the radial width of such a passing orbit is sufficiently large, it may intersect the plasma boundary \cite{1992Grieger}, similar to the case of wide bananas. Even if the orbit width is small, passing orbits can be transported due to resonant phenomena discussed in Section \ref{sec:passing_resonance}.

\subsubsection{Resonant passing trajectories} 
\label{sec:passing_resonance}

While passing particles on irrational field lines at low energy stay confined since the net radial drift vanishes on a time average, passing particles on rational trajectories may experience a net drift due to perturbations that resonate with the orbit periodicity \citep{1980Boozer,1993Mynicka,1993Mynickb,2021White}. The condition for such a resonance to occur is,
\begin{equation}
    l = k \frac{\Delta \theta}{2\pi},
    \label{eq:passing_resonance_condition}
\end{equation}
where $\Delta \theta$ is the change in the poloidal angle after a transit through one toroidal field period, $2\pi/N_P$. After a transit through $k$ field periods, $l$ full poloidal turns occur. At low energy, resonance will occur for passing particles near a rational surface with $\iota = N_Pl/k$. At higher energy, such resonances can persist in the neighborhood of a rational surface due to the magnetic drifts, see Figure \ref{fig:iter_passing_resonance}. In analogy with banana-drift diffusion, stochastic transport can occur if these resonances begin to overlap \citep{1993Mynicka, 1993Mynickb, 2022White}.

\subsection{Transitions between classes}

We use the term class to refer to the collection of particles trapped within a given well of the magnetic field strength. Since the local maximum of the field strength may vary between field lines when symmetry is broken, a particle may become collisionlessly detrapped or transition between trapping classes through its cross-field drifts. For example, banana-trapped particles can become entrapped in ripple wells or locally detrapped and make a transit through a full period of $\chi$. Since these entrapping and detrapping processes are associated with a radial displacement, it has been suggested that this leads to diffusive transport \citep{1992Lotz,2001Beidler}. This irregular transitioning behavior can be seen in Figure \ref{fig:ariescs_transitions_driven}. The characteristic timescale for this diffusive loss scales as $\tau_{\mathrm{loss}} \sim \left(1 - s_0\right)^2/D$. Using the radial step size associated with the transition, $\delta s$, and the timescale associated with the separatrix crossing, $\tau_{\mathrm{cross}}$ (which scales with the precession frequency $\overline{\textbf{v}_{\mathrm{m}} \cdot \nabla \alpha}$), as characteristic scales, the characteristic diffusion coefficient scales as $D \sim (\delta s)^2 P/\tau_{\mathrm{cross}}$. Here $P$ is the transition probability.

A trajectory is classified as ripple-trapped, banana-trapped, or barely-trapped between two bounce points. Transitioning trajectories have segments that pass through more than one of these three classes or between barely trapped segments that pass through a different number of field periods. The number of transitions is evaluated along each trajectory. Transitioning behavior can also be distinguished by jumps in the parallel adiabatic invariant \eqref{eq:jpar} associated with crossing a separatrix. 

\section{Guiding center calculations}
\label{sec:guiding_center}

\subsection{Numerical considerations}
\label{sec:numerics}

We perform integration of the guiding center (GC) trajectories of fusion-born alpha particles (3.52 MeV) for 0.2 seconds, which corresponds with about three slowing-down times based on the density and temperature profiles of ARIES-CS \citep{2008Ku}. The GC approximation is likely appropriate for the parameters under consideration, given that the value of $\rho_* = v_{\mathrm{th}}/(aq B/m) \approx 0.03$ for ARIES-CS fusion-born alpha parameters, where $v_{\mathrm{th}} = \sqrt{2 E/m}$ is the thermal velocity, $E=3.52$ MeV is the energy, and $a$ is the effective minor radius.

The GC equations obtained from the Littlejohn Lagrangian \citep{1983Littlejohn} have been implemented in the SIMSOPT code \cite{2021Landreman}. The equations of motion are integrated in Boozer coordinates, $(s, \theta, \zeta)$ \citep{1981Boozer}. A benchmark between the guiding center integration routines in SIMSOPT, BEAMS3D \citep{2014Mcmillan}, and ANTS \citep{2014Drevlak} has been performed. Despite slight differences in the GC equations adopted in the three codes (BEAMS3D and ANTS integrate in cylindrical coordinates and neither uses the Langrangian formalism), the loss fractions computed for the three codes typically agree within 1\%. Integration is performed with the Runge-Kutta Dormand-Prince 5 method provided by the BOOST library. 

Given an initial energy $E_0$ and coordinates ($v_{\|,0},s_0,\theta_0,\zeta_0$), the initial magnetic moment $\mu_0 = (E_0/m -  v_{\|,0}^2/2)/B(s_0,\theta_0,\zeta_0)$ is computed and passed to the integration routine as a parameter held fixed throughout.
The error in energy conservation at any point in time can be evaluated by comparing $E = \frac{1}{2} mv_{\|}^2 + m\mu_0 B(s,\theta,\zeta)$ with $E_0$. Alternatively, the error in magnetic moment conservation at any point can be evaluated by comparing $\mu = (E_0/m -  v_{\|}^2/2)/B(s,\theta,\zeta)$ with $\mu_0$. Energy and magnetic moment are conserved to within a fractional error of $\approx 10^{-3}$ throughout the timescales of interest. 

Integration in Boozer coordinates as opposed to cylindrical coordinates eases the analysis of the dynamics with respect to magnetic surfaces. One complication of integration in flux coordinates is the presence of the coordinate singularity at $s = 0$. (While a similar complication also exists in cylindrical coordinates, the singularity typically lies outside the domain of interest.) Therefore, we do not analyze the orbits that pass through the $s = 0.01$ surface. The number of particles that are discarded for this reason are tabulated in Tables \ref{tab:qa_losses} and \ref{tab:qh_losses}.

\subsection{Configurations}
\label{sec:configurations}

\begin{figure}
    \centering
    \begin{subfigure}[b]{0.32\textwidth}
    \includegraphics[width=1.0\textwidth]{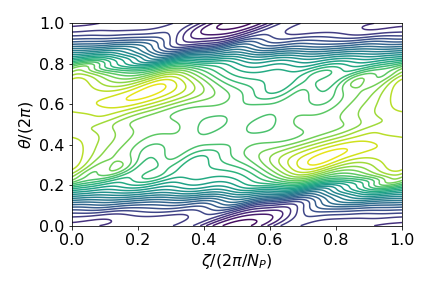}
    \caption{ARIES-CS}
    \end{subfigure}
    \begin{subfigure}[b]{0.32\textwidth}
    \includegraphics[width=1.0\textwidth]{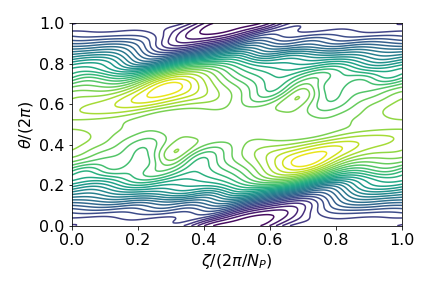}
    \caption{NCSX}
    \end{subfigure}
    \begin{subfigure}[b]{0.32\textwidth}
    \includegraphics[width=1.0\textwidth]{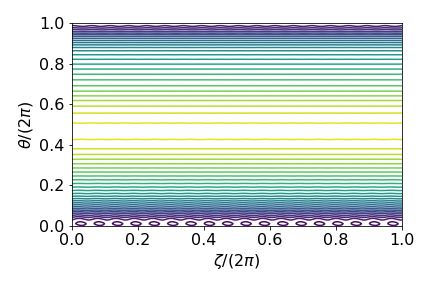}
    \caption{ITER}
    \end{subfigure}
    \begin{subfigure}[b]{0.32\textwidth}
    \includegraphics[width=1.0\textwidth]{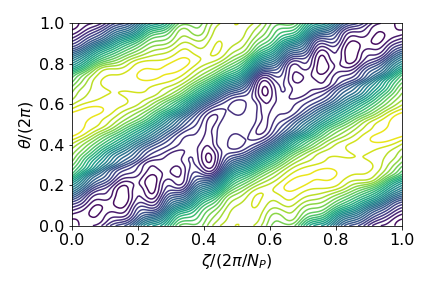}
    \caption{HSX}
    \end{subfigure}
    \begin{subfigure}[b]{0.32\textwidth}
    \includegraphics[width=1.0\textwidth]{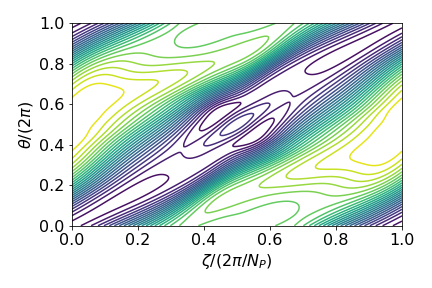}
    \caption{Wistell-A}
    \end{subfigure}
    \begin{subfigure}[b]{0.32\textwidth}
    \includegraphics[width=1.0\textwidth]{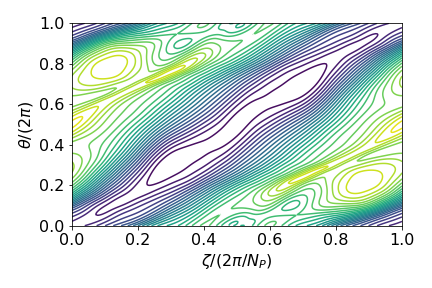}
    \caption{Ku5}
    \end{subfigure}
    \caption{Contours of the magnetic field strength are plotted as a function of the Boozer angles on the plasma boundary.}
    \label{fig:modB}
\end{figure}

\begin{figure}
    \centering
    \begin{subfigure}[b]{1.0\textwidth}
    \centering 
    \includegraphics[width=0.32\textwidth]{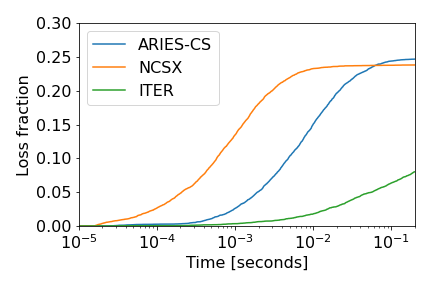}
    \includegraphics[width=0.32\textwidth]{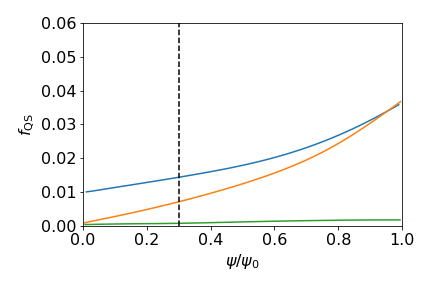}
    \includegraphics[width=0.32\textwidth]{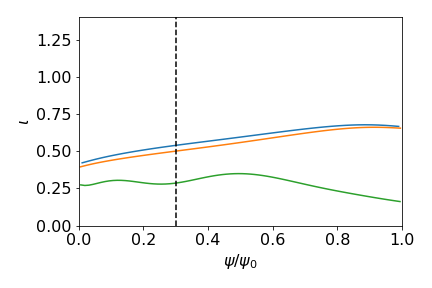}
    \caption{QA configurations}
    \end{subfigure}
    \begin{subfigure}[b]{1.0\textwidth}
    \centering 
    \includegraphics[width=0.32\textwidth]{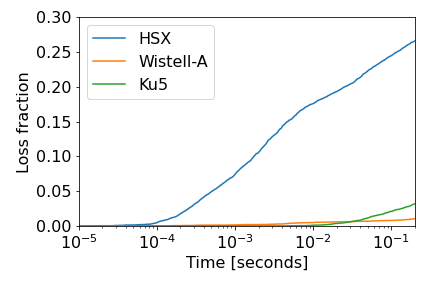}
    \includegraphics[width=0.32\textwidth]{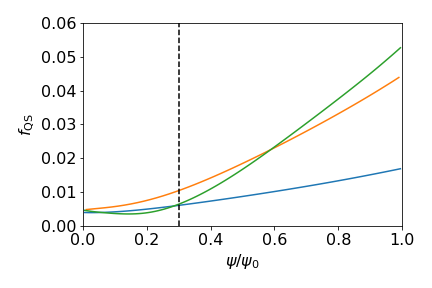}
    \includegraphics[width=0.32\textwidth]{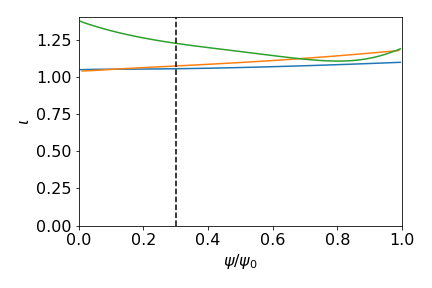}
    \caption{QH configurations}
    \end{subfigure}
    \caption{Collisionless loss fractions (left), quasisymmetry error (center) defined in \eqref{eq:qs_error}, and rotational transform profile (right) for the QA and QH configurations under investigation.}
    \label{fig:loss_times}
\end{figure}

We consider a set of equilibria (Table \ref{tab:configurations}) which are scaled to the same volume-averaged field strength (5.86 T) and volume (444 m$^3$) as ARIES-CS. We use the same scaling protocol described in \citep{2021Bader}. Throughout the text, we use the terms QA and QH to refer to configurations close to quasiaxisymmetry and quasihelical symmetry, respectively, but may have finite deviations from exact quasisymmetry.
All of the equilibria are fixed-boundary except for the ITER equilibrium. In the case of ITER, the free-boundary equilibrium is computed without ferritic inserts or test blanket modules. The ITER profiles correspond to a steady-state scenario \cite{2014Poli}, which features a smaller rotational transform than high-performance scenarios \cite{2015Fafiq}. The HSX fixed-boundary equilibrium includes the effects of coil ripple since a fixed-boundary equilibrium predating the free-boundary equilibrium was not available. The ARIES-CS configuration is the \texttt{n3are} equilibrium, and the NCSX configuration is the \texttt{li383} equilibrium. The collisionless and collisional confinement of the ARIES-CS, Wistell-A, Ku5, and NCSX equilibria were discussed in \citep{2021Bader}. In Figure \ref{fig:loss_times} we show the rotational transform and quasisymmetry error in each configuration, defined by
\begin{equation}
    f_{QS}(s) = \frac{\sqrt{\sum_{Mn\ne Nm} \left(B_{m,n}^c(s)\right)^2 + \left(B_{m,n}^s(s)\right)^2}}{\sqrt{\left(B_{00}^c(s)\right)^2 + \left(B_{00}^s(s)\right)^2}},
    \label{eq:qs_error}
\end{equation}
where the field strength in Boozer coordinates is $B(s,\theta,\zeta) = \sum_{m,n}B_{m,n}^c(s) \cos(m \theta - n \zeta) + B_{m,n}^s(s)\sin(m \theta - n \zeta)$. All configurations except for ITER are stellarator symmetric, so their $B_{m,n}^s$ coefficients vanish. 
The contours of the magnetic field strength on the plasma boundary are shown in Figure \ref{fig:modB}. 

\begin{table}
\centering
\begin{tabular}{|c||c|c|c|c|}
    \hline 
   \textbf{Configuration} & $N_{P}$ & Symmetry & Aspect ratio & Avg. $\iota$ \\ \hline 
   ARIES-CS \citep{2006Mynickb,2008Ku} & 3 & QA & 4.55 & 0.57 \\ \hline 
   NCSX \citep{2003Koniges,2002Mynick} & 3 & QA & 4.37 & 0.55 \\ \hline
   ITER \citep{2017Paul} & 1 & QA & 2.31 & 0.28 \\ \hline 
   HSX \citep{1995Anderson} & 4 & QH & 9.97 & 1.06 \\ \hline 
   Ku5 \citep{2010Ku} & 5 & QH & 10.01 & 1.18 \\ \hline 
   Wistell-A \citep{2020Bader} & 4 & QH & 6.72 & 1.08 \\ \hline 
\end{tabular}
\caption{Quasiaxisymmetric and quasihelically symmetric equilibria under consideration. The number of field periods of the nearby quasisymmetric field is $N_P$. The aspect ratio is computed with the same definition used in the VMEC code, given in \cite{2019Landreman}. The average of $\iota$ is computed with respect to the toroidal flux. 
}
\label{tab:configurations}
\end{table}

In each configuration, $10^4$ particles are initialized with a uniform distribution in $v_{\|}/v_{\mathrm{th}}$ and position on the $s = 0.3$ surface as described in \citep{2019Bader}, which corresponds to an isotropic distribution in 3D velocity space. Particles are considered lost if they pass through the $s = 1$ surface. The loss fractions as a function of time are presented in Figure \ref{fig:loss_times}. While the quasisymmetry errors in the ARIES-CS and NCSX configurations are of similar magnitudes, ARIES-CS has significantly fewer prompt losses than NCSX due to its further optimization. While the Wistell-A and Ku5 configurations have quasisymmetry errors comparable to ARIES-CS and NCSX, the loss fractions in these two QH configurations are significantly reduced. HSX has reduced $f_{QS}$ compared with Wistell-A and Ku5, but the loss fraction is elevated. The rippled ITER configuration has the smallest quasisymmetry error but a substantial number of losses compared to the Wistell-A and Ku5 configurations. (With ferritic inserts and test blanket modules in addition to the coil ripple, there are no collisionless losses from ITER \cite{2021Bader}.) We will analyze the loss mechanisms and configuration characteristics responsible for these differences in confinement. 

\section{Summary of data}
\label{sec:summary_data}

\subsection{Quasiaxisymmetric configurations}

\begin{table}
\centering
\begin{tabular}{|c||c|c|c|c|}
    \hline 
   \textbf{Configuration} & \textbf{Total losses} & \textbf{Axis ($s < 10^{-2}$)} & \textbf{Prompt } & \textbf{Transitioning} \\ \hline 
   ARIES-CS & 2470 & 542 & 257 & 1693 \\ \hline 
   NCSX & 2343 & 211  & 1298 & 1745 \\ \hline 
   ITER & 796 & 170 & 20 & 388 \\ \hline 
\end{tabular}
\caption{Number of losses for QA configurations. For reasons discussed in Section \ref{sec:numerics}, axis trajectories which pass through $s< 10^{-2}$ are not included in the classification analysis. Prompt losses leave the plasma boundary in less than $10^{-3}$ seconds. Transitioning losses pass through several trapping classes before leaving the plasma boundary.}
\label{tab:qa_losses}
\end{table}

The raw number of losses in the QA configurations is given in Table \ref{tab:qa_losses}. Here axis losses indicate trajectories that come close to the magnetic axis ($s < 10^{-2}$) and are discarded from the analysis since the magnetic coordinates break down. Prompt losses indicate particles that pass through the plasma boundary in less than $10^{-3}$ seconds. 

\begin{figure}
    \centering
    \begin{subfigure}[b]{\textwidth}
    \centering
    \includegraphics[trim=2cm 11cm 3cm 1cm,clip,width=1.0\textwidth]{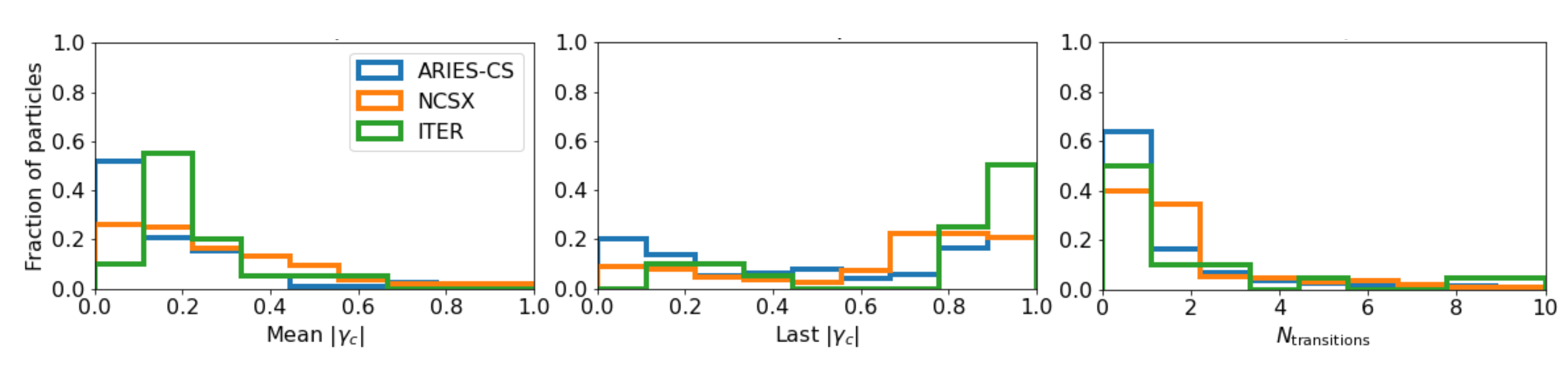}
    \caption{Prompt}
    \end{subfigure}
    \begin{subfigure}[b]{\textwidth}
    \centering
    \includegraphics[trim=2cm 11cm 3cm 1cm,clip,width=1.0\textwidth]{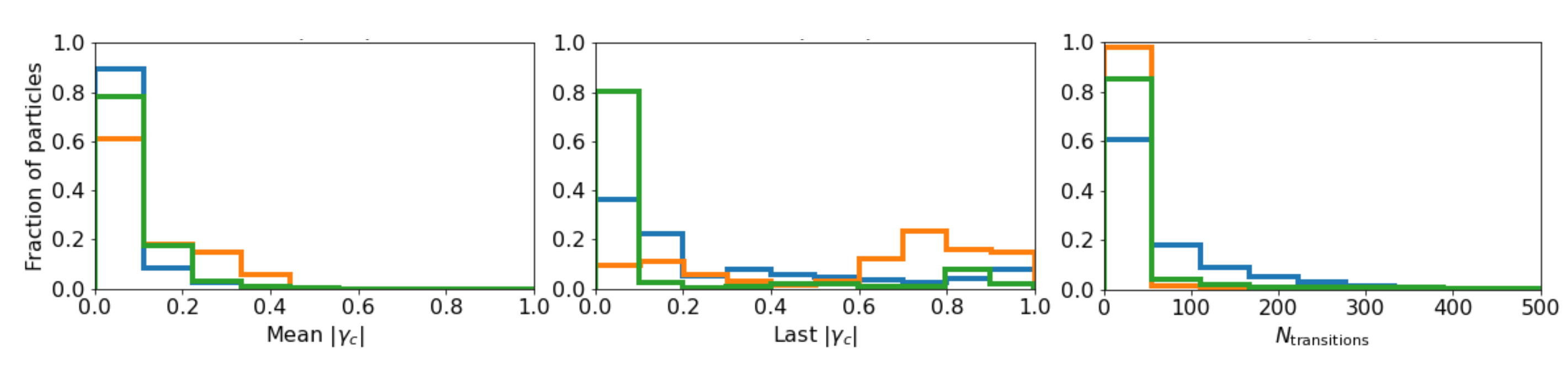} 
    \caption{Non-prompt}
    \end{subfigure}
    \caption{Trajectory statistics for QA configurations. Prompt losses leave the plasma boundary in less than $10^{-3}$ seconds. The left histograms display the distribution of the mean value of $|\gamma_{\mathrm{c}}|$ over bounce segments while the middle histograms display the value of $|\gamma_{\mathrm{c}}|$ for the final bounce segment before being lost (see \eqref{eq:gammac_traj}). The distribution of the number of class transitions along the lost trajectories is shown on the right. Each histogram's counts are scaled by the number of losses so that the distributions can be displayed on the same scale for each configuration.}
    \label{fig:qa_gammac_mean}
\end{figure}

\begin{figure}
    \centering
    \begin{subfigure}[b]{\textwidth}
    \centering
    \includegraphics[trim=1cm 11cm 3cm 1cm,clip,width=1.0\textwidth]{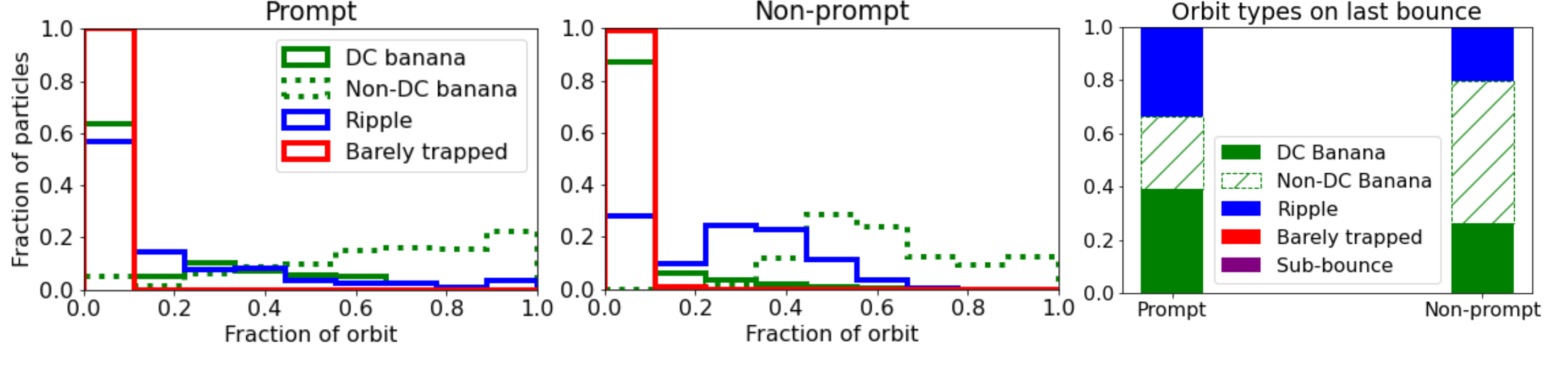}
    \caption{ARIES-CS}
    \label{fig:histograms_ariescs}
    \end{subfigure}
    \begin{subfigure}[b]{\textwidth}
    \centering
    \includegraphics[trim=1cm 11cm 3cm 1cm,clip,width=1.0\textwidth]{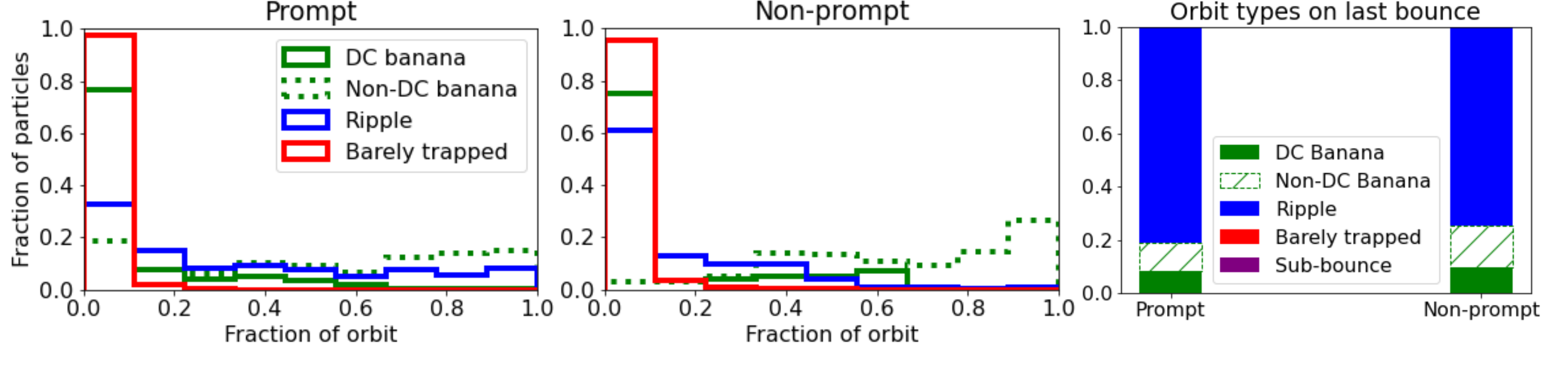}
    \caption{NCSX}
    \label{fig:histograms_ncsx}
    \end{subfigure}
    \begin{subfigure}[b]{\textwidth}
    \centering
    \includegraphics[trim=1cm 11cm 3cm 1cm,clip,width=1.0\textwidth]{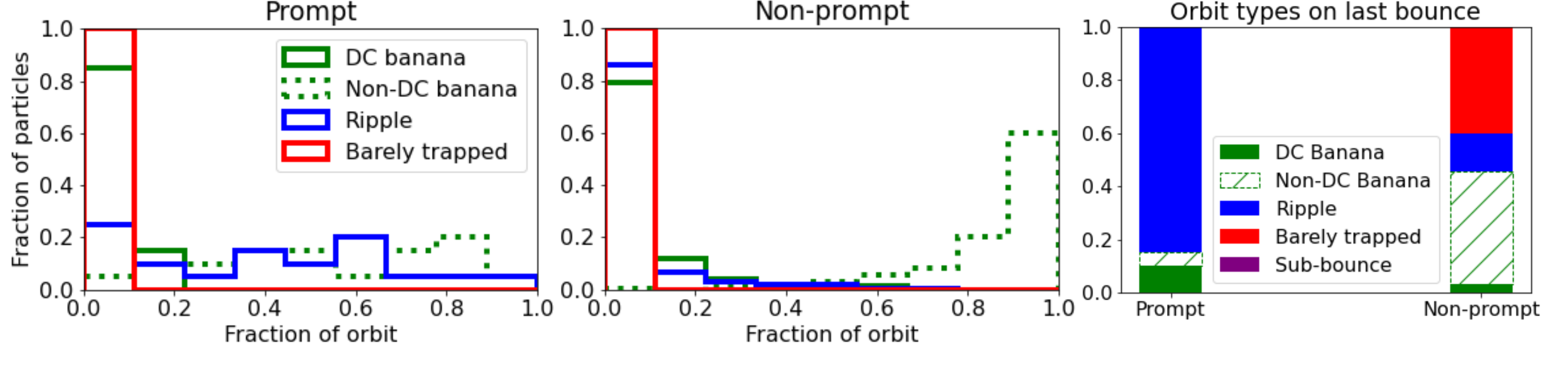}
    \caption{ITER}
    \label{fig:histograms_iter}
    \end{subfigure}
    \caption{Distribution of orbit segment types for QA configurations on prompt ($<10^{-3}$ seconds) and non-prompt timescales. For each half-bounce segment, a trajectory is categorized into the ripple, banana, or barely-trapped categories. Drift-convective (DC) transport is indicated by $|\gamma_{\mathrm{c}}|>0.2$ (see \eqref{eq:gammac_traj}). Each histogram's counts are scaled by the number of losses so that the distributions can be displayed on the same scale for each configuration. The bar charts on the right display the distribution of orbit types on the bounce segment before the trajectory is lost. Here sub-bounce indicates a particle that does not complete a full bounce orbit before being lost.}
    \label{fig:histograms_QA}
\end{figure}

An overview of the statistics for the QA configurations is shown in Figures \ref{fig:qa_gammac_mean} and \ref{fig:histograms_QA}. 
Figure \ref{fig:qa_gammac_mean} displays the distribution of the mean value of $|\gamma_{\mathrm{c}}|$ over bounce segments, the value of $|\gamma_{\mathrm{c}}|$ for the final bounce segment before the particle exits the plasma boundary, and the number of class transitions for prompt and non-prompt losses. The distributions are normalized by the total number of prompt or non-prompt losses in each configuration such that the sum is unity. In Figure \ref{fig:histograms_QA} we display the distribution of orbit segment types for each configuration. 
For a given prompt or non-prompt loss, the fraction of orbit indicates the fraction of bounce segments that fall into each trapping category. For each trapping class in each configuration, the distribution is normalized by the total number of prompt or non-prompt losses such that the sum is unity. For example, in all of the histograms displayed for the QA configurations, we see a distribution of barely-trapped orbit segments centered at an orbit fraction near zero with a fraction of particles close to unity, indicating that almost all losses spend almost none of their orbit in the barely-trapped class. On the other hand, in ARIES-CS, about 20\% of the prompt losses spend close to 100\% of the orbit in the non-DC banana class, indicating that diffusive or resonant banana transport is a substantial prompt loss channel. On the right, we show the distribution of orbit types for the final bounce segment before the trajectory is lost. Here sub-bounce indicates orbits that mirror one or fewer times before exiting such that the segment type cannot be classified. The sub-bounce orbits are not included in the histograms on the left. No passing trajectories are lost from the equilibria. Several example trajectories are shown in the following Sections. Animations corresponding to some of these trajectories are published in an accompanying Zenodo archive \cite{zenodo}.

\noindent \textit{\textbf{Prompt losses}}

\begin{figure}
    \centering
    \begin{subfigure}[b]{0.32\textwidth}
    \includegraphics[width=\textwidth]{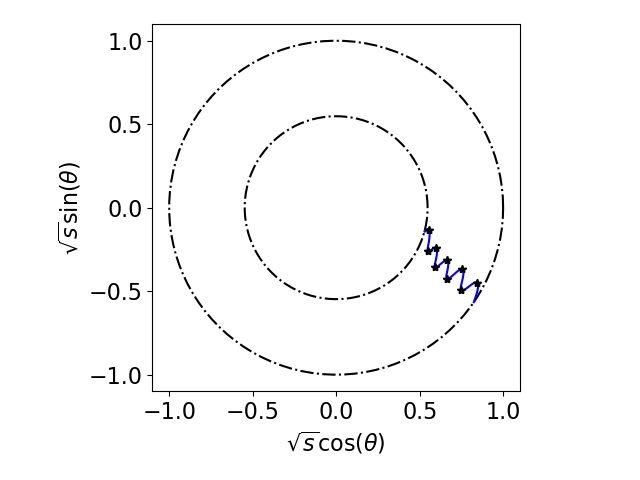}
    \caption{}
    \end{subfigure}
    \begin{subfigure}[b]{0.32\textwidth}
    \includegraphics[width=\textwidth]{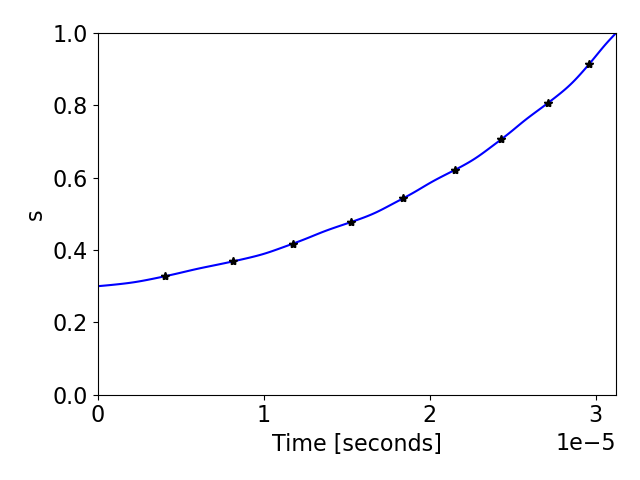}
    \caption{}
    \end{subfigure}
    \begin{subfigure}[b]{0.32\textwidth}
    \includegraphics[width=\textwidth]{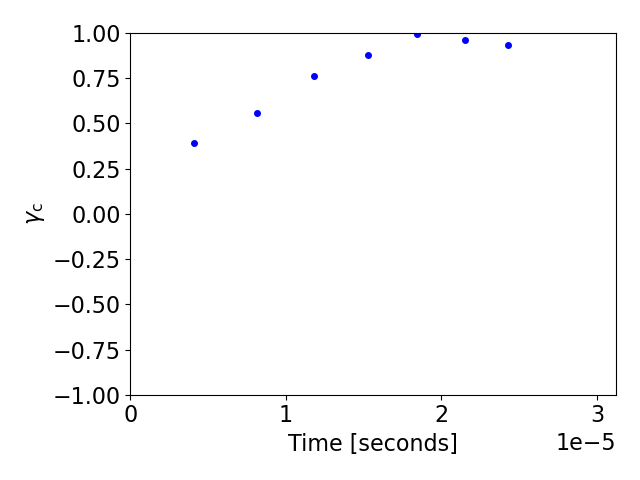}
    \caption{}
    \end{subfigure}
    \begin{subfigure}[b]{0.32\textwidth}
    \includegraphics[width=\textwidth]{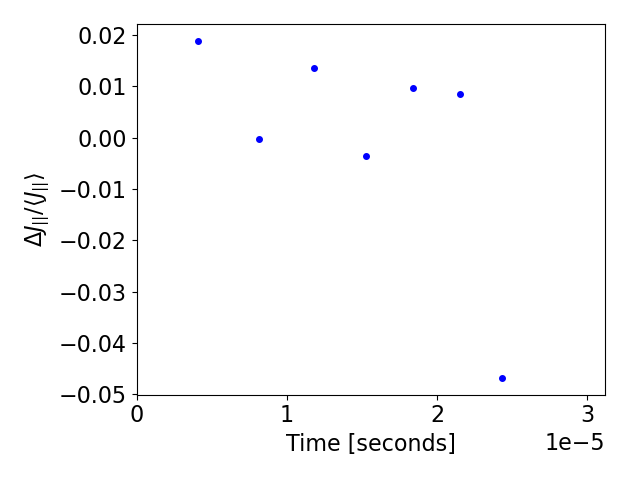}
    \caption{}
    \end{subfigure}
    \begin{subfigure}[b]{0.32\textwidth}
    \includegraphics[width=\textwidth]{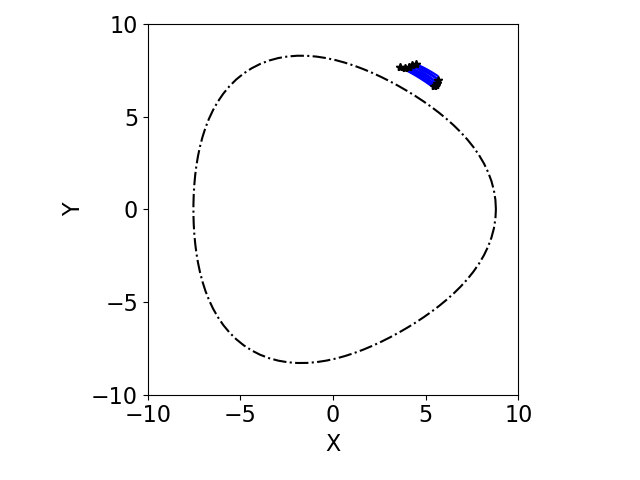}
    \caption{}
    \end{subfigure}
    \caption{ARIES-CS prompt ($3.1\times 10^{-5}$ seconds) loss on a ripple trajectory. Black stars indicate bounce points. (a) Poloidal cross-section. Black dashed lines indicate the initial magnetic surface and plasma boundary. (b) Radial coordinate $s = \psi/\psi_0$ as a function of time. (c) Drift-convection parameter evaluated for each bounce segment indicative of banana drift-convective transport. (d) The normalized change in the parallel adiabatic invariant indicates that $J_{\|}$ is well conserved along this trajectory. (e) Overhead view of orbit indicative of toroidal localization. The magnetic axis is indicated by a black dashed line.}
    \label{fig:ariescs_ripple}
\end{figure}

\begin{figure}
    \centering
    \begin{subfigure}[b]{0.32\textwidth}
    \includegraphics[width=\textwidth]{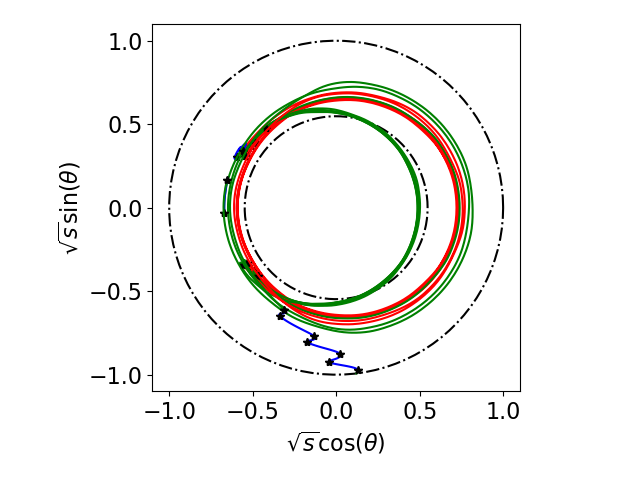}
    \caption{}
    \end{subfigure}
    \begin{subfigure}[b]{0.32\textwidth}
    \includegraphics[width=\textwidth]{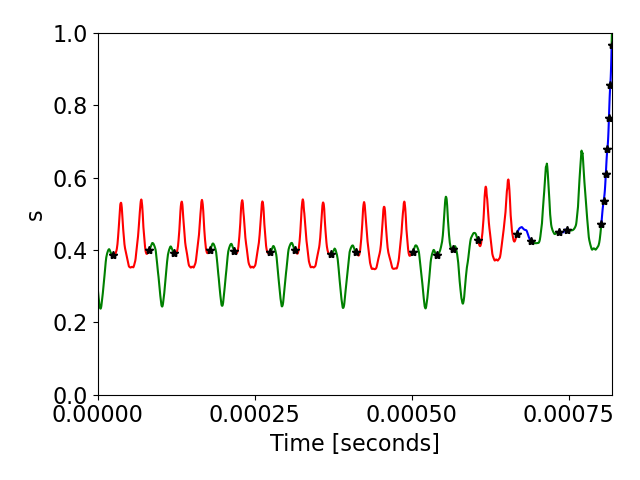}
    \caption{}
    \end{subfigure}
    \caption{NCSX prompt ($8.2 \times 10^{-4}$ seconds) loss featuring several transitions between banana (green) and barely-trapped (red) classes before being lost on a ripple orbit segment (blue). Black stars indicate bounce points. (a) Poloidal cross-section. Black dashed lines indicate the initial magnetic surface and plasma boundary. (b) Radial coordinate $s = \psi/\psi_0$ as a function of time. Irregular transitioning behavior is observed.
    }
    \label{fig:ncsx_prompt_transitioning}
\end{figure}

\begin{figure}
    \centering
    \begin{subfigure}[b]{0.32\textwidth}
    \includegraphics[width=\textwidth]{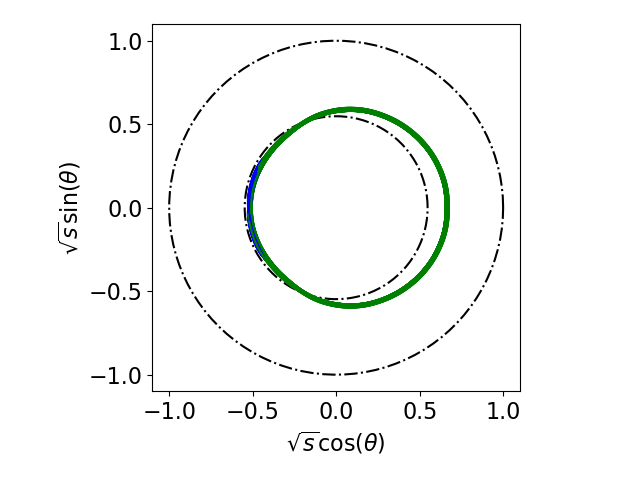}
    \caption{}
    \end{subfigure}
    \begin{subfigure}[b]{0.32\textwidth}
    \includegraphics[width=\textwidth]{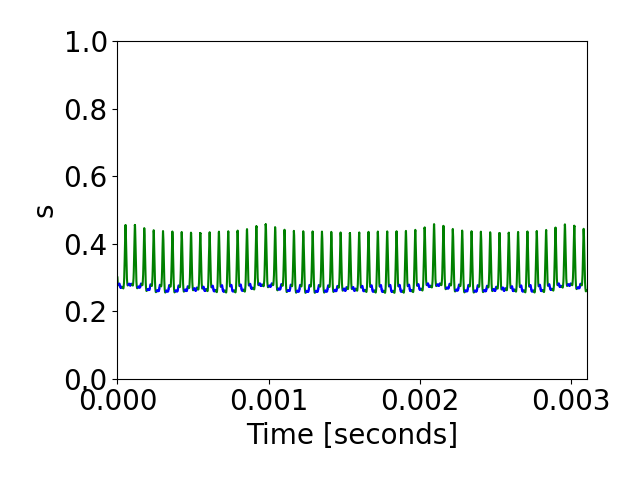}
    \caption{}
    \end{subfigure}
    \begin{subfigure}[b]{0.32\textwidth}
    \includegraphics[width=\textwidth]{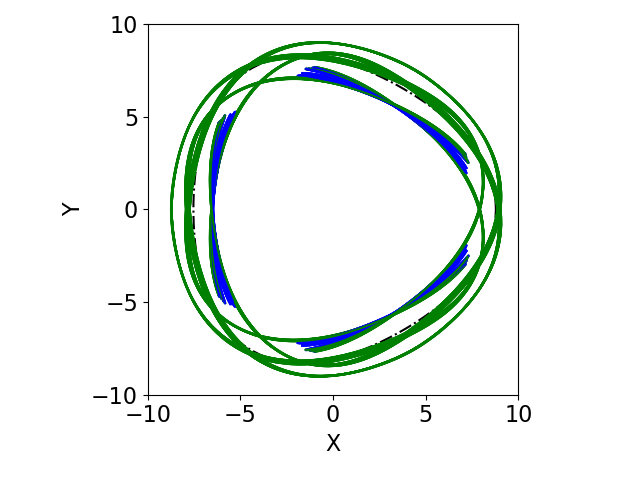}
    \caption{}
    \end{subfigure}
    \caption{ARIES-CS non-prompt ($3.4 \times 10^{-2}$ seconds) loss featuring many transitions between banana (green) and ripple (blue) classes. The particle is eventually lost on a more irregular transitioning orbit (not shown). (a) Poloidal cross-section. Black dashed lines indicate the initial magnetic surface and plasma boundary. (b) Radial coordinate $s = \psi/\psi_0$ as a function of time. 
    (c) Overhead view of orbit. The magnetic axis is indicated by a black dashed line.}
    \label{fig:ariescs_banana_ripple}
\end{figure}

The distribution of the mean and final $|\gamma_{\mathrm{c}}|$ are shifted toward zero for ARIES-CS in comparison with NCSX. This behavior is indicative of the optimization of the ARIES-CS configuration to reduce prompt losses compared to NCSX, as can also be seen from Table \ref{tab:qa_losses}. We can also see from comparing Figures \ref{fig:histograms_ariescs} and \ref{fig:histograms_ncsx} that the fraction of prompt losses arising from ripple and drift-convective transport is diminished in ARIES-CS in comparison with NCSX. However, a few ripple-trapped trajectories remain in ARIES-CS, as seen in Figure \ref{fig:ariescs_ripple}. Ripple trapping is substantial in NCSX, and 10 of the prompt losses from NCSX did not complete a full bounce segment due to rapid ripple losses. As only a small fraction of the initialized particles will be born on ripple trajectories, the trajectories that transition onto ripple segments can greatly impact the transport. This can be seen in Figure \ref{fig:ncsx_prompt_transitioning}, an NCSX orbit that transitions to a ripple-trapped segment near $\theta = -\pi/2$ and is subsequently lost. In contrast, in ARIES-CS the largest ripple wells are localized near $\theta = \pi$ where the radial drift is reduced, resulting in eventual detrapping (Figure \ref{fig:ariescs_banana_ripple}). These differences in ripple trapping partially account for the prompt confinement characteristics of ARIES-CS and NCSX and will be discussed further in Section \ref{sec:ripple_trapping}. 

\begin{figure}
    \centering
    \begin{subfigure}[b]{0.32\textwidth}
    \includegraphics[width=\textwidth]{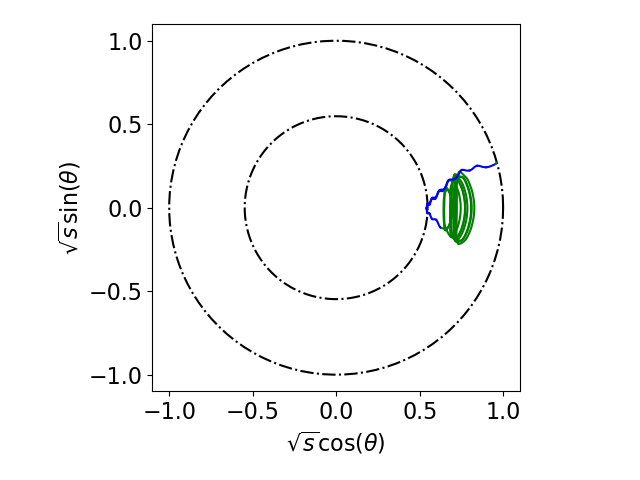}
    \caption{}
    \end{subfigure}
    \begin{subfigure}[b]{0.32\textwidth}
    \includegraphics[width=\textwidth]{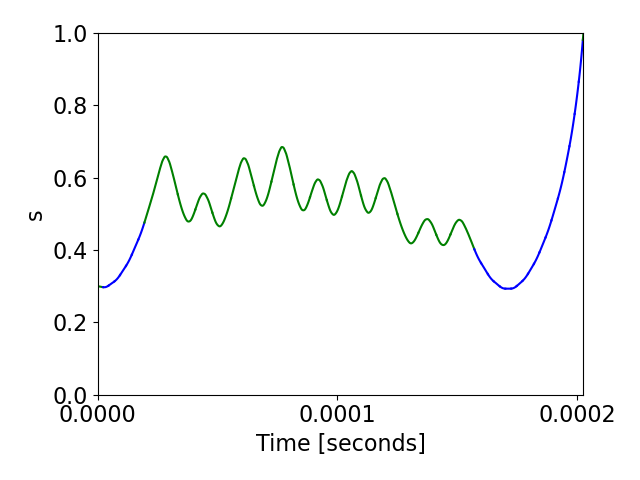}
    \caption{}
    \end{subfigure}
    \begin{subfigure}[b]{0.32\textwidth}
    \includegraphics[width=\textwidth]{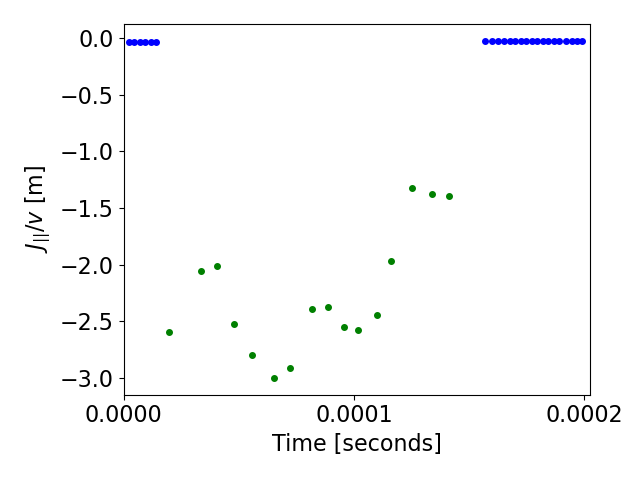}
    \caption{}
    \end{subfigure}
    \caption{ITER prompt ($2.0 \times 10^{-4}$ seconds) loss featuring a deeply-trapped banana orbit (green) and transition to a ripple orbit (blue). (a) Poloidal cross-section. Black dashed lines indicate the initial magnetic surface and plasma boundary. (b) Radial coordinate $s = \psi/\psi_0$ as a function of time. 
    (c) Parallel adiabatic invariant normalized by the velocity magnitude, evaluated for each bounce segment that the trajectory remains in the same trapping class. 
    }
    \label{fig:iter_stagnation}
\end{figure}

Ripple trapping is also quite substantial among prompt losses in ITER. The coil perturbation in ITER is localized on the outboard side, resulting in ripple wells near the outboard midplane. As the radial drift is relatively small near $\theta = 0$, the particles often precess and collisionlessly transition onto deeply-trapped banana orbits. In Figure \ref{fig:iter_stagnation} we show an example of such behavior. 

On prompt timescales, all three configurations have a significant number of transitioning orbits. These often correspond with transitions to ripple loss channels (Figure \ref{fig:iter_stagnation} and \ref{fig:ncsx_prompt_transitioning}). There are also prompt NCSX trajectories featuring many separatrix crossings with somewhat irregular behavior (Figure \ref{fig:ncsx_prompt_transitioning}). In the ARIES-CS configuration, the transitions are often due to periodic ripple detrapping and entrapping, similar to Figure \ref{fig:ariescs_banana_ripple}.

\begin{figure}
    \centering
    \begin{subfigure}[b]{0.32\textwidth}
    \includegraphics[width=\textwidth]{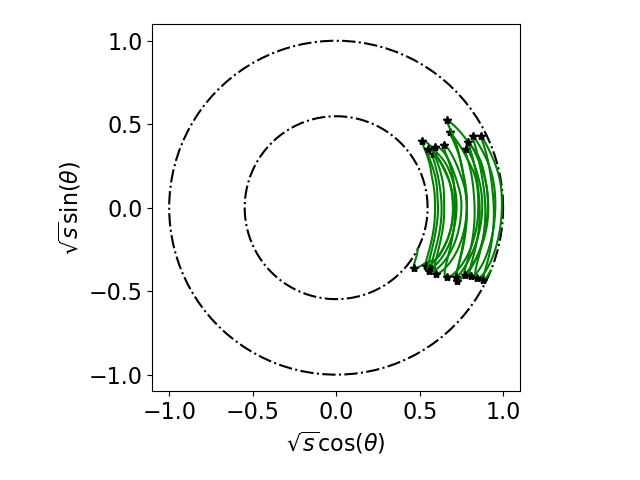}
    \caption{}
    \end{subfigure}
    \begin{subfigure}[b]{0.32\textwidth}
    \includegraphics[width=\textwidth]{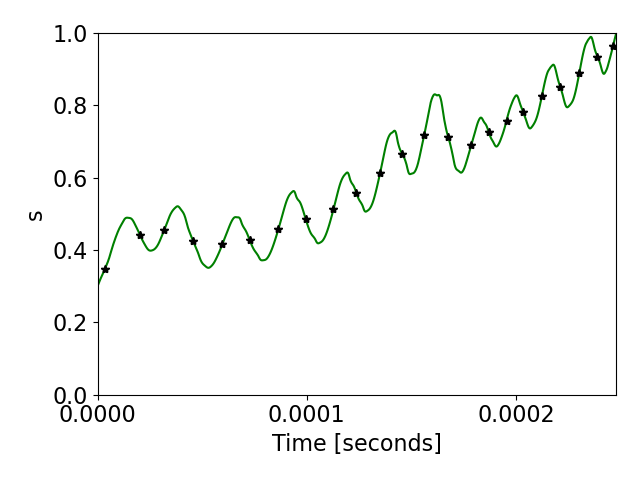}
    \caption{}
    \end{subfigure}
    \begin{subfigure}[b]{0.32\textwidth}
    \includegraphics[width=\textwidth]{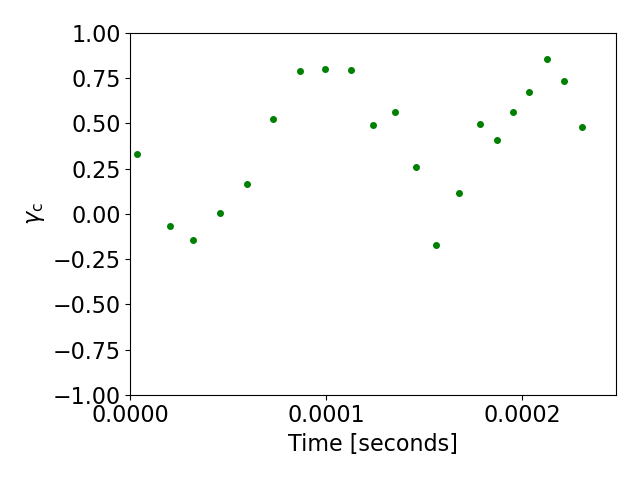}
    \caption{}
    \end{subfigure}
    \begin{subfigure}[b]{0.32\textwidth}
    \includegraphics[width=\textwidth]{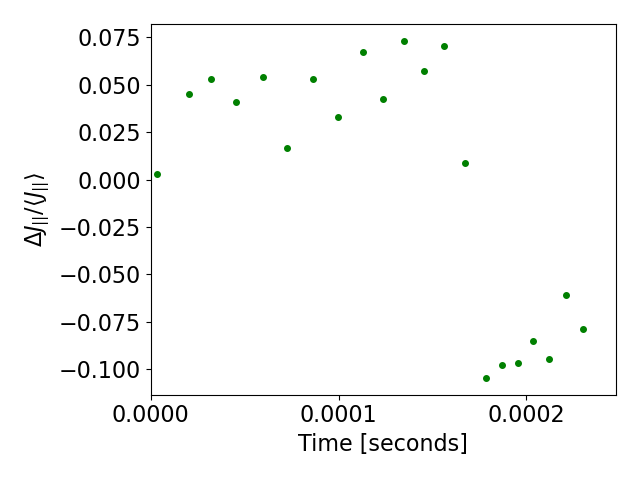}
    \caption{}
    \end{subfigure}
    \begin{subfigure}[b]{0.32\textwidth}
    \includegraphics[width=\textwidth]{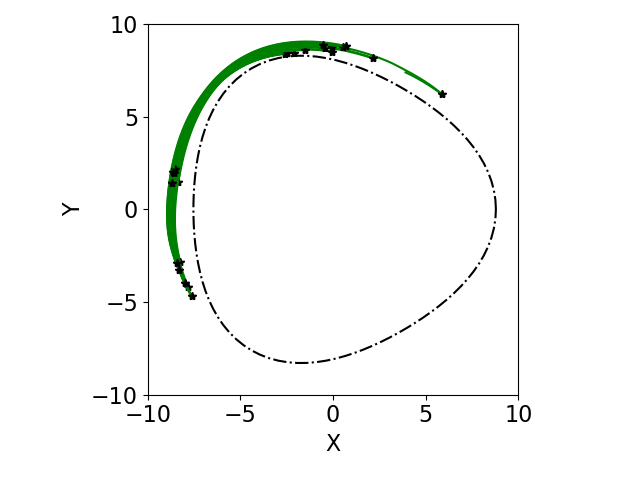}
    \caption{}
    \end{subfigure}
    \caption{ARIES-CS prompt ($2.5\times 10^{-4}$ seconds) loss on a banana drift-convective trajectory. Black stars indicate bounce points. (a) Poloidal cross-section. Black dashed lines indicate the initial magnetic surface and plasma boundary. (b) Radial coordinate $s = \psi/\psi_0$ as a function of time. (c) Drift-convective parameter evaluated for each bounce segment indicative of significant banana drift-convective transport. (d) The normalized change in the parallel adiabatic invariant indicates that $J_{\|}$ is poorly conserved. (e) Overhead view of orbit indicating toroidal localization. The magnetic axis is indicated by a black dashed line.}
    \label{fig:ariescs_sb}
\end{figure}

\begin{figure}
    \centering
    \begin{subfigure}[b]{0.32\textwidth}
    \includegraphics[width=\textwidth]{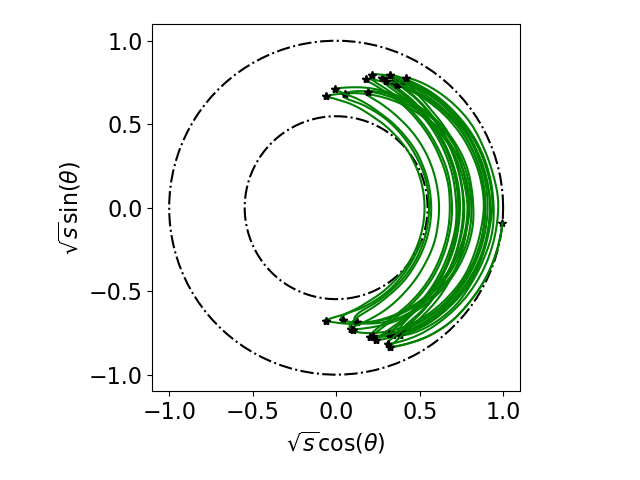}
    \caption{}
    \end{subfigure}
    \begin{subfigure}[b]{0.32\textwidth}
    \includegraphics[width=\textwidth]{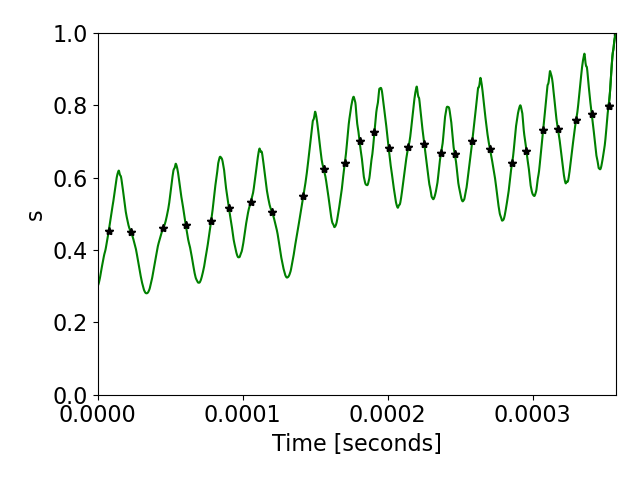}
    \caption{}
    \end{subfigure}
    \begin{subfigure}[b]{0.32\textwidth}
    \includegraphics[width=\textwidth]{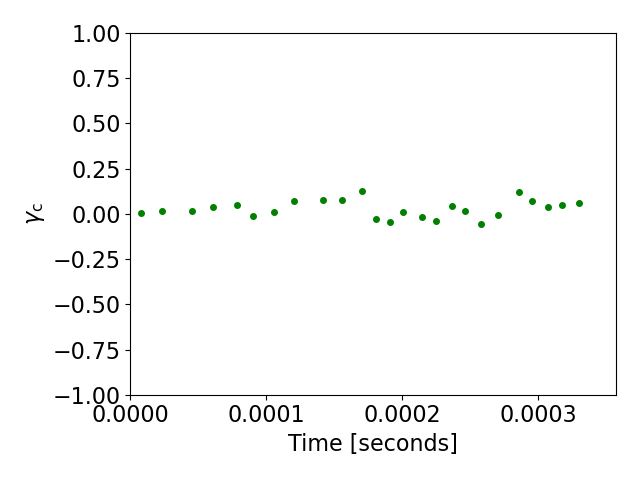}
    \caption{}
    \end{subfigure}
    \begin{subfigure}[b]{0.32\textwidth}
    \includegraphics[width=\textwidth]{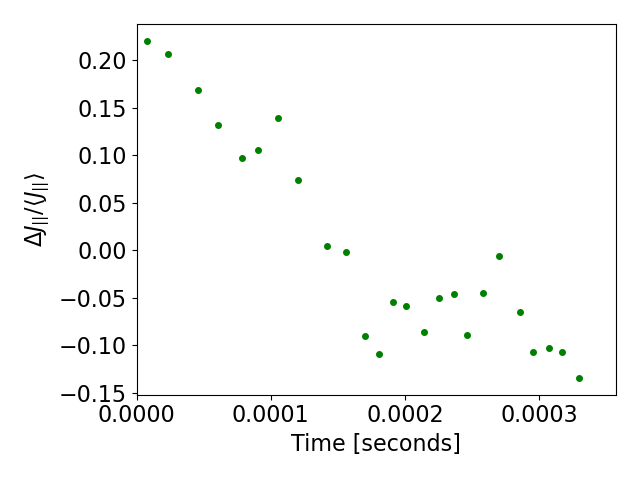}
    \caption{}
    \end{subfigure}
    \begin{subfigure}[b]{0.32\textwidth}
    \includegraphics[width=\textwidth]{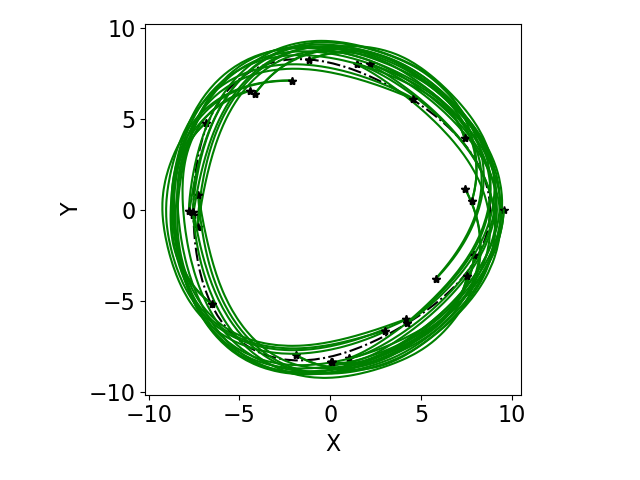}
    \caption{}
    \end{subfigure}
    \caption{ARIES-CS prompt ($3.6 \times 10^{-4}$ seconds) loss featuring diffusive banana tip motion.
    Black stars indicate bounce points. (a) Poloidal cross-section, demonstrating the diffusive motion of banana tips. Black dashed lines indicate the initial magnetic surface and plasma boundary. (b) Radial coordinate $s = \psi/\psi_0$ as a function of time. (c) The drift-convective parameter is evaluated for each bounce segment, indicating that banana drift-convective transport is insignificant. (d) The normalized change in the parallel adiabatic invariant indicates poor conservation along the trajectory. (e) Overhead view of orbit indicating the absence of toroidal localization. The magnetic axis is indicated by a black dashed line.}
    \label{fig:ariescs_diffusive_banana}
\end{figure}

\begin{figure}
    \centering
    \begin{subfigure}[b]{0.32\textwidth}
    \includegraphics[width=\textwidth]{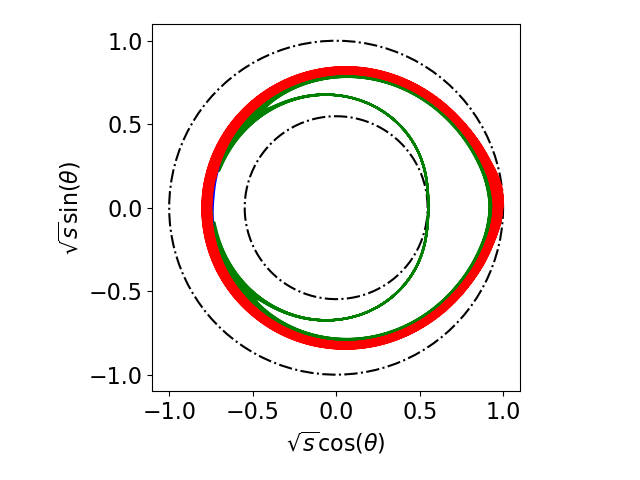}
    \caption{}
    \end{subfigure}
    \begin{subfigure}[b]{0.32\textwidth}
    \includegraphics[width=\textwidth]{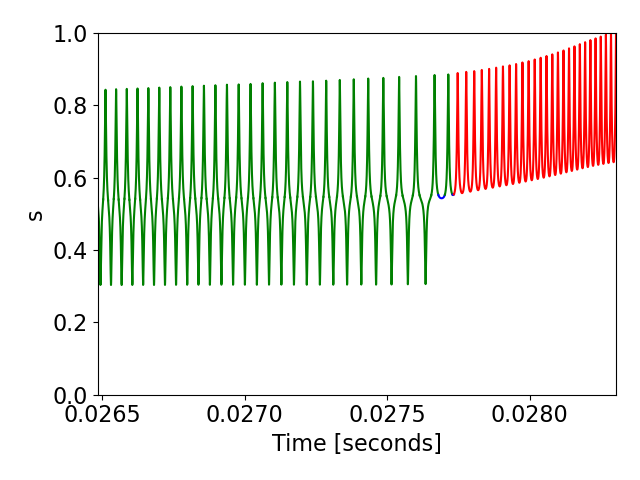}
    \caption{}
    \end{subfigure}
    \begin{subfigure}[b]{0.32\textwidth}
    \includegraphics[width=\textwidth]{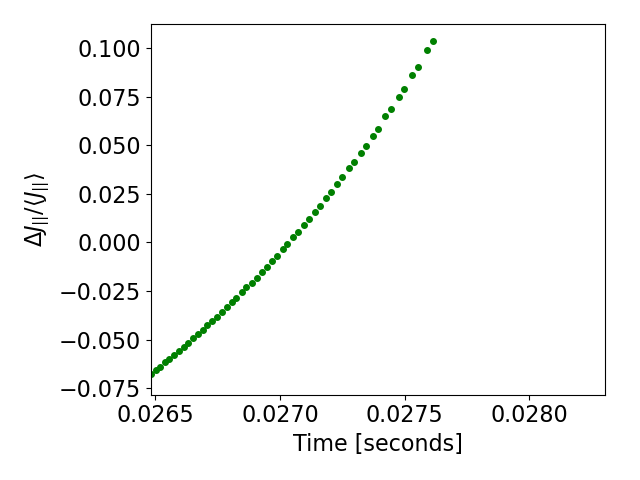}
    \caption{}
    \end{subfigure}
    \begin{subfigure}[b]{0.32\textwidth}
    \includegraphics[width=\textwidth]{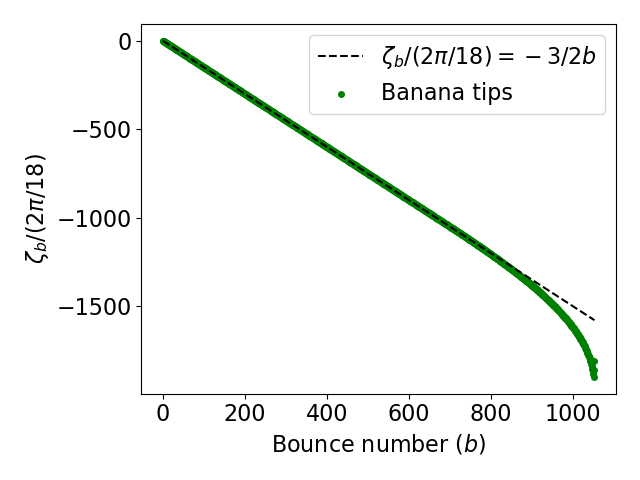}
    \caption{}
    \end{subfigure}
    \begin{subfigure}[b]{0.32\textwidth}
    \includegraphics[width=\textwidth]{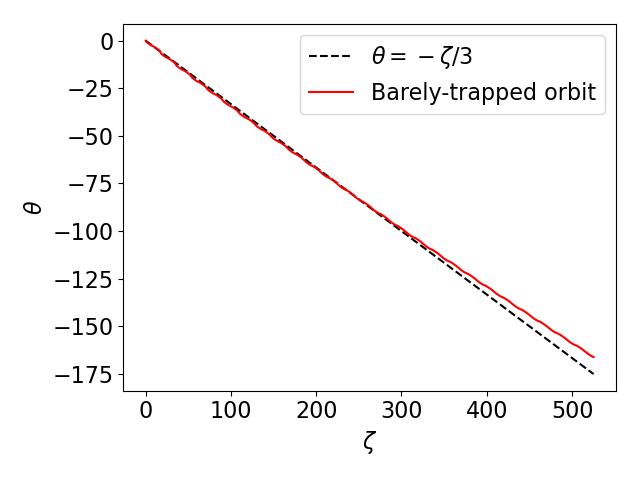}
    \caption{}
    \end{subfigure}
    \caption{ITER non-prompt ($2.8 \times 10^{-2}$ seconds) resonant banana orbit (green) which transitions to a ripple (blue) and resonant barely-trapped orbit (red). (a) Poloidal cross-section. Black dashed lines indicate the initial magnetic surface and plasma boundary. (b) Radial coordinate $s = \psi/\psi_0$ as a function of time. (c) The normalized change in the parallel adiabatic invariant is evaluated for each bounce segment that the trajectory remains in the same trapping class. The secular change in $J_{\|}$ indicates a resonant perturbation. (d) The upper bounce points, $\zeta_b$, are plotted as a function of the number of bounce segments, $b$, to demonstrate the $l = 2$, $k = -3$ periodicity in the notation of \eqref{eq:bounce_transit_resonance}. (e) The final barely-trapped orbit segment is plotted to demonstrate the $l=-1$, $k = 3$ periodicity in the notation of \eqref{eq:passing_resonance_condition}.}
    \label{fig:iter_passing_resonance}
\end{figure}

While ripple and DC orbit segments are present in the ARIES-CS and NCSX prompt losses, the non-DC banana segments dominate the orbit fraction. Evidence for diffusive banana tip motion associated with non-conservation of $J_{\|}$ can be seen in both NCSX and ARIES-CS (Figure \ref{fig:ariescs_diffusive_banana}). Here the normalized change in $J_{\|}$ is computed as $\Delta J_{\|}/\langle J_{\|} \rangle = (J_{\|} - \langle J_{\|} \rangle)/\langle J_{\|} \rangle$ where the average $\langle \dots \dot \rangle$ is performed over the bounce segments. We note that $J_{\|}$ is also relatively poorly conserved for drift-convective orbits in these configurations (Figure \ref{fig:ariescs_sb}), resulting in a small amount of non-secular radial motion of the banana tips. The addition of diffusive motion to convective transport along $J_{\|}$ contours may enhance losses. This behavior is in contrast with DC banana orbits in QH configurations, which better conserve $J_{\|}$ and whose banana tips drift radially in a secular fashion (Figure \ref{fig:aten_sb}). 
We see a similarly significant presence of non-DC banana segments in ITER. In addition to diffusive banana tip motion, some of the non-DC banana transport in ITER is due to resonant orbits. In Figure \ref{fig:iter_passing_resonance} we see an example of such a barely-trapped banana trajectory in ITER which exhibits a bounce-transit resonance with $k = -3$ and $l = 2$ in the notation of \eqref{eq:bounce_transit_resonance}.

Considering the orbit types on the last bounce in Figure \ref{fig:histograms_QA}, most of the prompt losses in the NCSX and ITER equilibria terminate on ripple orbit segments. In contrast, in the ARIES-CS equilibria, the number of DC banana and non-DC banana orbit segments become comparable to ripple segments. Furthermore, in all three equilibria, we see that the distribution of $|\gamma_{\mathrm{c}}|$ on the last bounce is shifted toward larger values compared to the distribution of the mean. This behavior indicates many trajectories that spend most of their orbit time in the non-DC class but are eventually lost due to ripple or DC mechanisms. 

\noindent \textit{\textbf{Non-prompt losses}}

For each configuration, the distribution of the mean $|\gamma_{\mathrm{c}}|$ is shifted toward zero on non-prompt timescales in comparison with prompt timescales (Figure \ref{fig:qa_gammac_mean}), indicative of non-DC transport mechanisms that dominate on longer timescales. Similar behavior is seen in the distribution of the last $|\gamma_{\mathrm{c}}|$, except for in the NCSX equilibrium. The distinct behavior in NCSX reflects orbits which stay confined for a long period of time and are eventually lost on a ripple trajectory, as can be seen from the orbit segment type on the last bounce (Figure \ref{fig:histograms_ncsx}) and final value of $|\gamma_{\mathrm{c}}|$ (Figure \ref{fig:qa_gammac_mean}). A population of trajectories terminates on DC segments in all three configurations, indicating that these loss channels remain important even on longer time scales. 

\begin{figure}
    \centering
    \begin{subfigure}[b]{0.32\textwidth}
    \includegraphics[width=\textwidth]{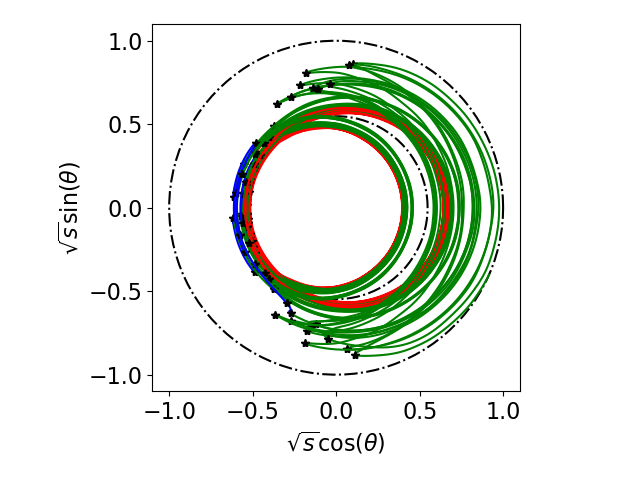}
    \caption{}
    \end{subfigure}
    \begin{subfigure}[b]{0.32\textwidth}
    \includegraphics[width=\textwidth]{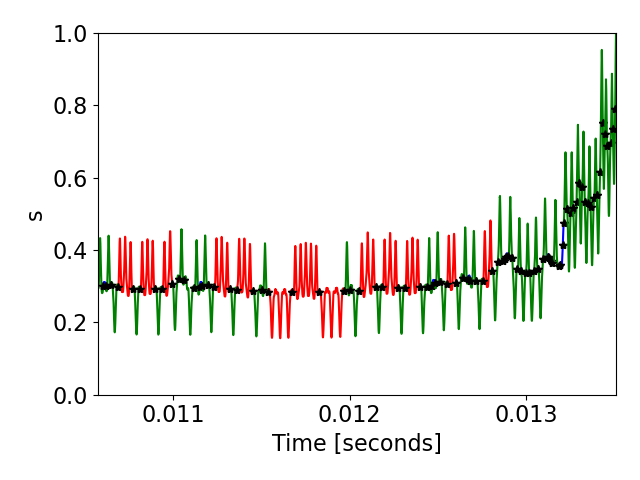}
    \caption{}
    \end{subfigure}
    \caption{ARIES-CS non-prompt ($1.4 \times 10^{-2}$ seconds) loss featuring many transitions between banana (green), barely-trapped (red), and ripple (blue) classes before being lost on a diffusive banana orbit. Black stars indicate bounce points. (a) Poloidal cross-section. Black dashed lines indicate the initial magnetic surface and plasma boundary. (b) Radial coordinate $s = \psi/\psi_0$ as a function of time. Irregular transitioning behavior is observed.
    }
    \label{fig:ariescs_transitions_driven}
\end{figure}

The ARIES-CS equilibrium features many trajectories with hundreds of transitions. For example, Figure \ref{fig:ariescs_transitions_driven} shows a trajectory that irregularly changes classes before being lost on a ripple orbit. Sometimes trajectories featuring many transitions can exhibit more regular behavior. For example, in Figure \ref{fig:ariescs_banana_ripple} we see a trajectory from ARIES-CS which transitions periodically between banana and ripple trapping classes and is eventually lost when the separatrix crossings become more irregular. We see similar periodic transitions between ripple and banana classes in the ITER equilibrium before being lost on a resonant barely-trapped orbit. However, irregular transitioning behavior does not appear among the ITER lost trajectories. 

\begin{figure}
    \centering
    \begin{subfigure}[b]{0.32\textwidth}
    \includegraphics[width=\textwidth]{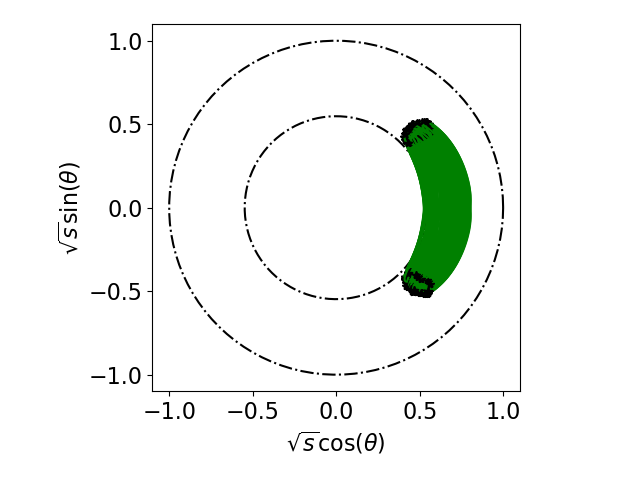}
    \caption{}
    \end{subfigure}
    \begin{subfigure}[b]{0.32\textwidth}
    \includegraphics[width=\textwidth]{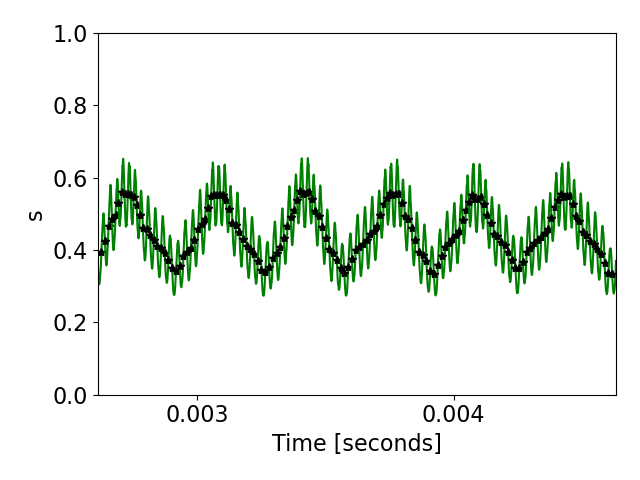}
    \caption{}
    \end{subfigure}
    \begin{subfigure}[b]{0.32\textwidth}
    \includegraphics[width=\textwidth]{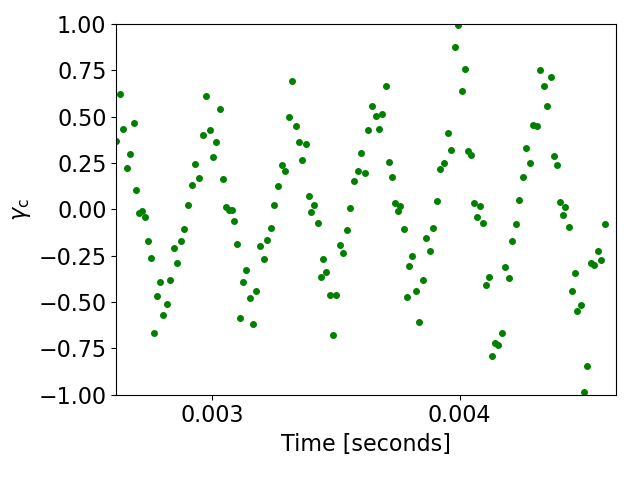}
    \caption{}
    \end{subfigure}
    \begin{subfigure}[b]{0.32\textwidth}
    \includegraphics[width=\textwidth]{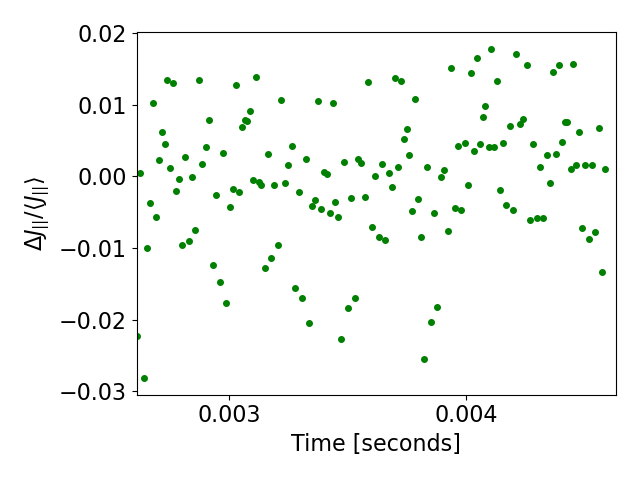}
    \caption{}
    \end{subfigure}
    \begin{subfigure}[b]{0.32\textwidth}
    \includegraphics[width=\textwidth]{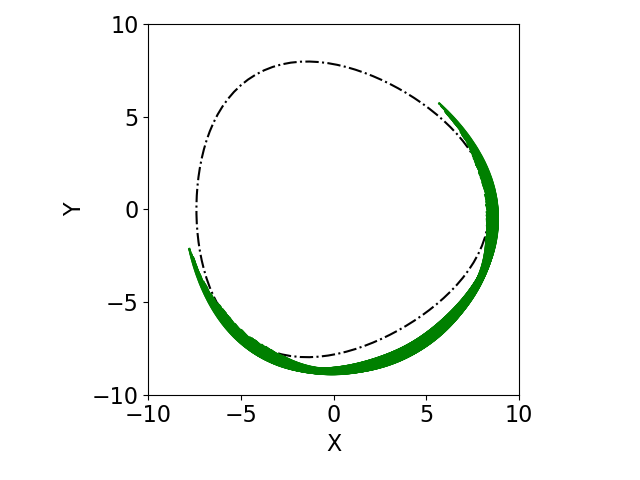}
    \caption{}
    \end{subfigure}
    \caption{NCSX non-prompt ($1.3 \times 10^{-1}$ seconds) loss featuring periodic banana drift-convective motion, eventually lost on a ripple orbit (not shown). Black stars indicate bounce points. (a) Poloidal cross-section. Black dashed lines indicate the initial magnetic surface and plasma boundary. (b) Radial coordinate $s = \psi/\psi_0$ as a function of time. (c) Drift-convective parameter evaluated for each bounce segment. 
    (d) The normalized change in the parallel adiabatic invariant indicates that $J_{\|}$ has deviations of a few percent.
    (e) Overhead view of orbit, indicative of the toroidal localization. The magnetic axis is indicated by a black dashed line.}
    \label{fig:ncsx_periodic_banana}
\end{figure}

Some non-prompt lost trajectories feature other types of periodic behavior. For example, a periodic DC banana orbit in NCSX is shown in Figure \ref{fig:ncsx_periodic_banana} which is eventually lost on a ripple orbit. A similar periodic DC banana orbit is observed in ARIES-CS, eventually lost on a diffusive banana orbit. These can be understood as drift-convective trajectories whose orbits are healed due to global features of the orbit given the closure of $J_{\|}$ contours within the plasma boundary \cite{1983Mynick}. Eventual loss of such an orbit can occur due to slight non-conservation of $J_{\|}$. 

Many of the non-prompt losses in ITER terminate on a barely-trapped orbit (Figure \ref{fig:histograms_iter}). An example of a transition from a banana orbit segment to a resonant barely-trapped orbit segment is shown in Figure \ref{fig:iter_passing_resonance}. The barely-trapped segment is close to the $k = 3$, $l = -1$ resonance in the notation of \eqref{eq:passing_resonance_condition} given the $\iota$ profile which crosses through $1/3$ (Figure \ref{fig:loss_times}). This periodicity gives rise to fluctuations in the radial coordinate concurrent with each toroidal transit and a secular radial drift due to a resonance in the equilibrium. As the loss of a resonant barely-trapped orbit requires many toroidal transits, this mechanism is more prevalent on long timescales. 

\subsection{Quasihelical configurations}

\begin{table}
\centering
\begin{tabular}{|c||c|c|c|c|}
    \hline 
   \textbf{Configuration} & \textbf{Total losses} & \textbf{Axis ($s < 10^{-2}$)} & \textbf{Prompt} & \textbf{Transitioning} \\ \hline 
   HSX & 2825 & 696 & 741 & 1441 \\ \hline 
   Ku5 & 332 & 0 & 0 & 318 \\ \hline 
   Wistell-A & 120 & 14 & 14 & 94 \\ \hline 
\end{tabular}
\caption{Number of losses for QH configurations. Axis trajectories pass through $s< 10^{-2}$ and are not included in the analysis. Prompt losses leave the plasma boundary in less than $10^{-3}$ seconds. Transitioning losses pass through several trapping classes before leaving the plasma boundary. }
\label{tab:qh_losses}
\end{table}

\begin{figure}
    \centering
    \begin{subfigure}[b]{\textwidth}
    \centering
    \includegraphics[trim=2cm 11cm 3cm 1cm,clip,width=1.0\textwidth]{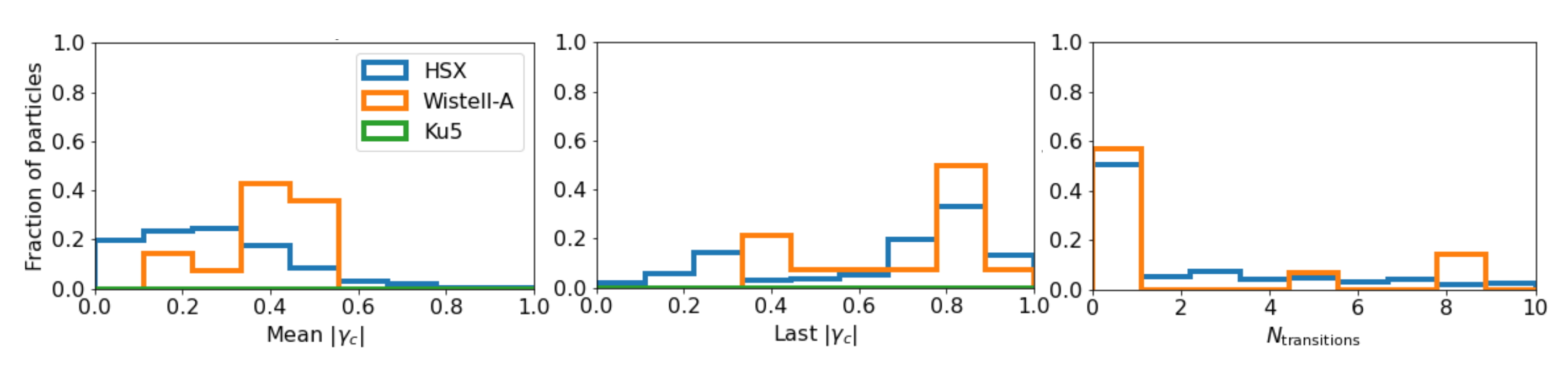}
    \caption{Prompt}
    \end{subfigure}
    \begin{subfigure}[b]{\textwidth}
    \centering
    \includegraphics[trim=2cm 11cm 3cm 1cm,clip,width=1.0\textwidth]{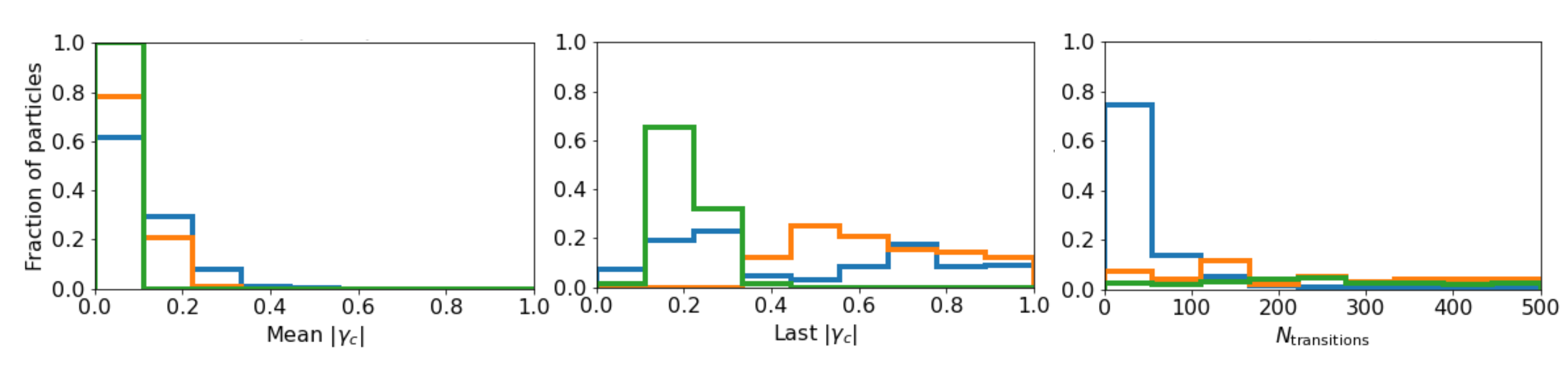} 
    \caption{Non-prompt}
    \end{subfigure}
    \caption{Trajectory statistics for QH configurations. Prompt losses leave the plasma boundary in less than $10^{-3}$ seconds. The left histograms display the distribution of the mean value of $|\gamma_{\mathrm{c}}|$ over bounce segments while the middle histograms display the value of $|\gamma_{\mathrm{c}}|$ for the final bounce segment before being lost (see \eqref{eq:gammac_traj}). The distribution of the number of class transitions along the lost trajectories is shown on the right. Each histogram's counts are scaled by the number of losses so that the distributions can be displayed on the same scale for each configuration.}
    \label{fig:qh_gammac_mean}
\end{figure}

\begin{figure}
    \centering
    \begin{subfigure}[b]{\textwidth}
    \centering
    \includegraphics[trim=1cm 11cm 3cm 1cm,clip,width=1.0\textwidth]{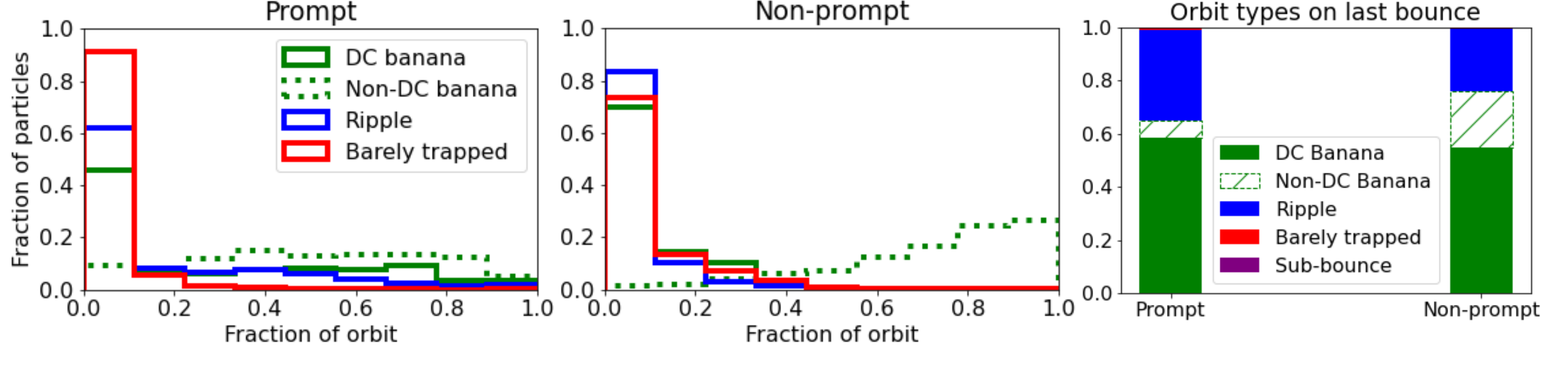}
    \caption{HSX}
    \label{fig:histograms_hsx}
    \end{subfigure}
    \begin{subfigure}[b]{\textwidth}
    \centering
    \includegraphics[trim=1cm 11cm 3cm 1cm,clip,width=1.0\textwidth]{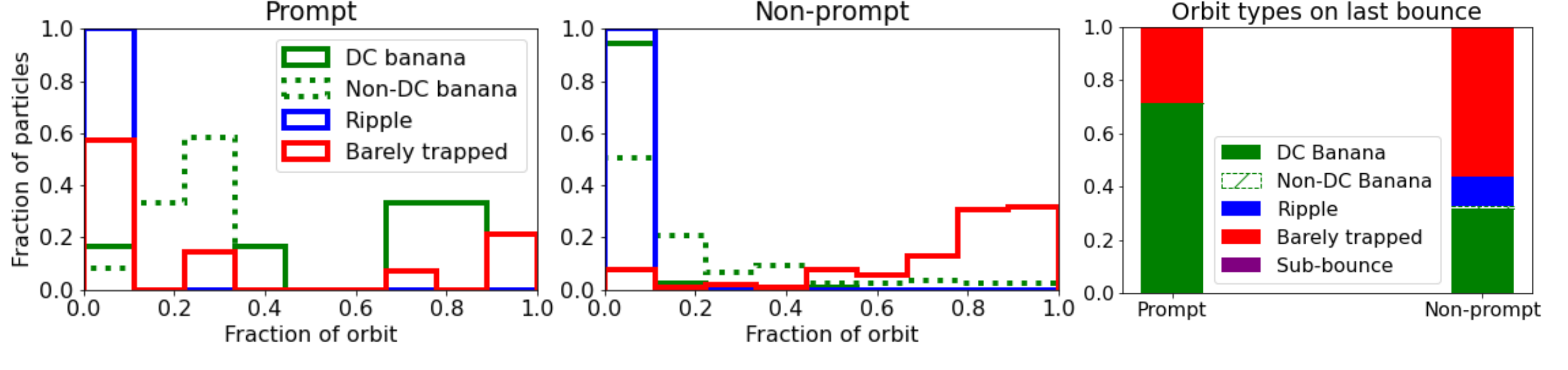}
    \caption{Wistell-A}
    \end{subfigure}
    \begin{subfigure}[b]{\textwidth}
    \centering
    \includegraphics[trim=1cm 11cm 3cm 1cm,clip,width=1.0\textwidth]{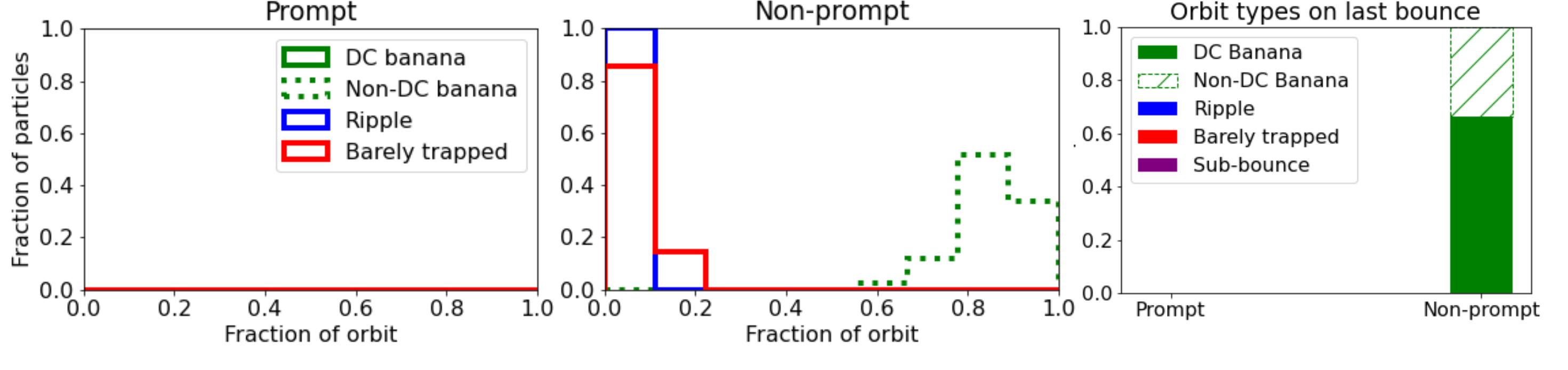}
    \caption{Ku5}
    \end{subfigure}
    \caption{Distribution of orbit segment types for QH configurations on prompt ($<10^{-3}$ seconds) and non-prompt timescales. For each half-bounce segment, a trajectory is categorized into the ripple, banana, or barely-trapped categories. Drift-convective (DC) transport is indicated by $|\gamma_{\mathrm{c}}|>0.2$. Each histogram's counts are scaled by the number of losses so that the distributions can be displayed on the same scale for each configuration. The bar charts on the right display the distribution of orbit types on the bounce segment before the trajectory is lost. Here sub-bounce indicates a particle that does not complete a full bounce orbit before being lost.}
    \label{fig:qh_histograms}
\end{figure}

The raw number of losses in the QH configurations is given in Table \ref{tab:qh_losses}. Compared with the QA equilibria, we note that the total number of prompt losses is relatively small in the Ku5 and Wistell-A configurations. Transitioning orbits remain prominent in all three configurations. An overview of the statistics for the QH configurations are shown in Figures \ref{fig:qh_gammac_mean} and \ref{fig:qh_histograms}. No passing trajectories are lost.

\noindent \textbf{\textit{Prompt losses}}

\begin{figure}
    \centering
    \begin{subfigure}[b]{0.32\textwidth}
    \includegraphics[width=\textwidth]{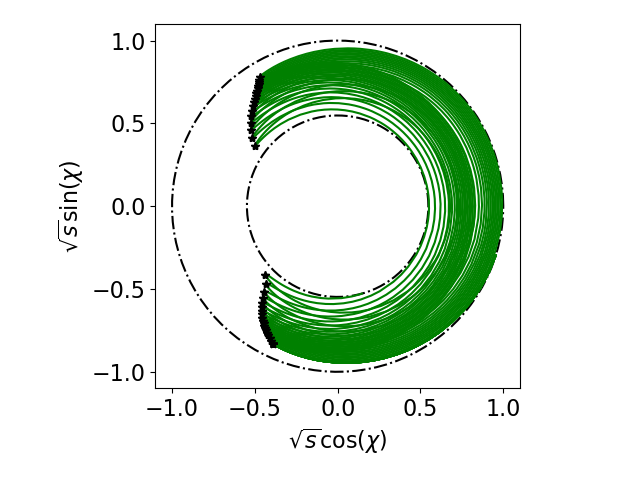}
    \caption{}
    \end{subfigure}
    \begin{subfigure}[b]{0.32\textwidth}
    \includegraphics[width=\textwidth]{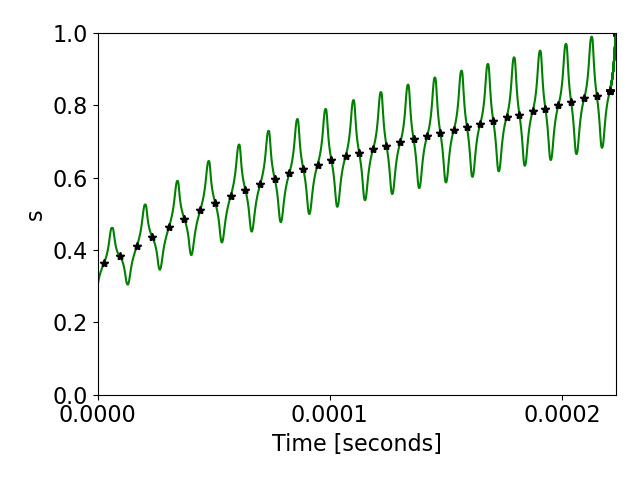}
    \caption{}
    \end{subfigure}
    \begin{subfigure}[b]{0.32\textwidth}
    \includegraphics[width=\textwidth]{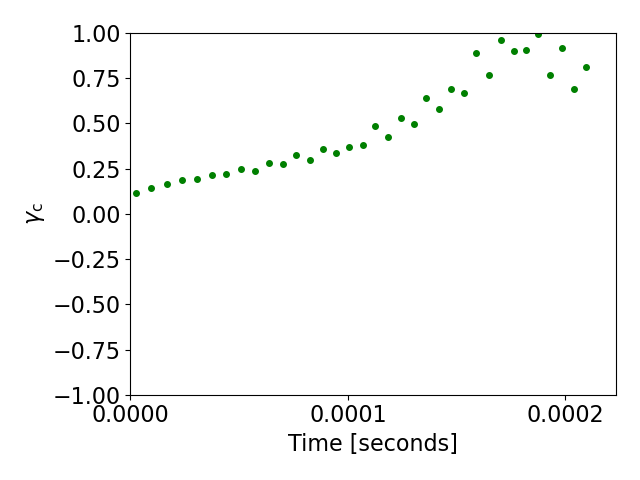}
    \caption{}
    \end{subfigure}
    \begin{subfigure}[b]{0.32\textwidth}
    \includegraphics[width=\textwidth]{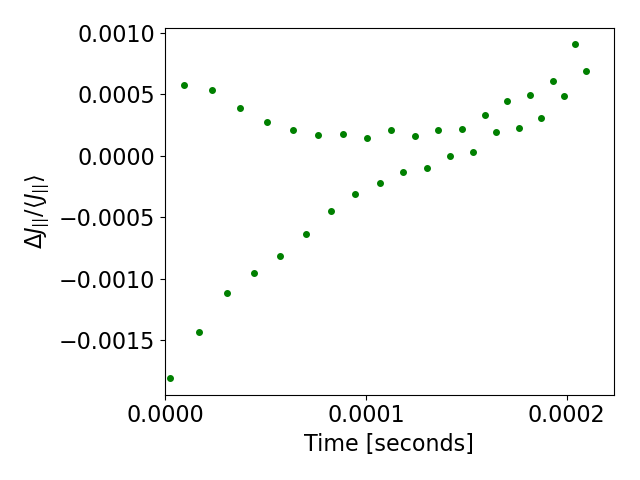}
    \caption{}
    \end{subfigure}
    \begin{subfigure}[b]{0.32\textwidth}
    \includegraphics[width=\textwidth]{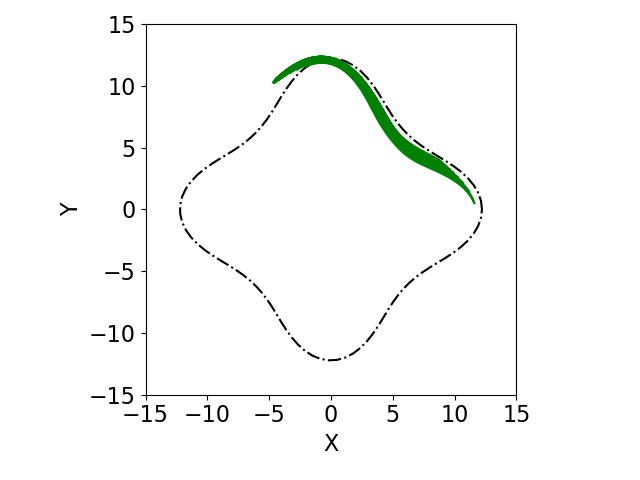}
    \caption{}
    \end{subfigure}
    \caption{Wistell-A prompt ($2.2 \times 10^{-4}$ seconds) loss which features a banana drift-convective orbit. Black stars indicate bounce points. (a) Poloidal cross-section. Black dashed lines indicate the initial magnetic surface and plasma boundary. (b) Radial coordinate $s = \psi/\psi_0$ as a function of time. Secular radial banana tip motion is observed. (c) Drift-convective parameter evaluated for each bounce segment. Drift-convective transport is evident throughout the trajectory. (d) The normalized change in the parallel adiabatic invariant indicates that $J_{\|}$ is well conserved. (e) Overhead view of orbit indicating toroidal localization. The magnetic axis is indicated by a black dashed line. Localization in toroidal angle is evident.}
    \label{fig:aten_sb}
\end{figure}

\begin{figure}
    \centering
    \begin{subfigure}[b]{0.32\textwidth}
    \includegraphics[width=\textwidth]{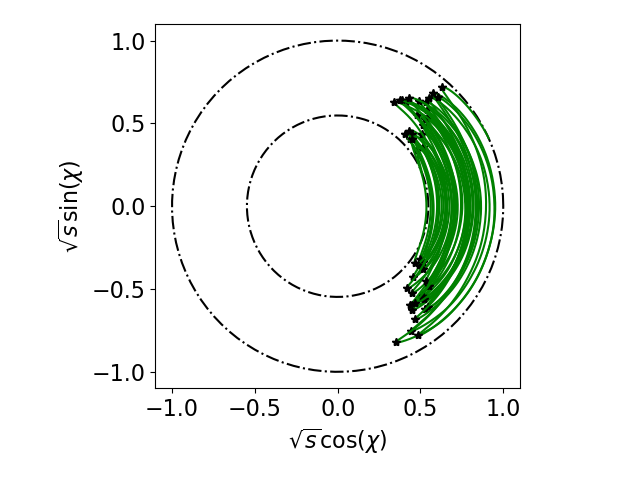}
    \caption{}
    \end{subfigure}
    \begin{subfigure}[b]{0.32\textwidth}
    \includegraphics[width=\textwidth]{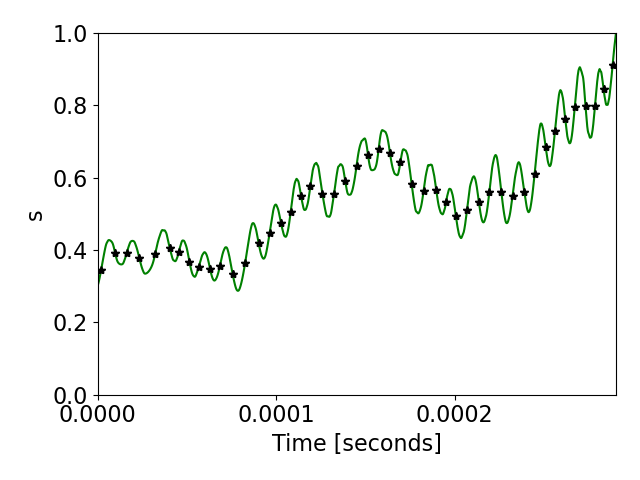}
    \caption{}
    \end{subfigure}
    \begin{subfigure}[b]{0.32\textwidth}
    \includegraphics[width=\textwidth]{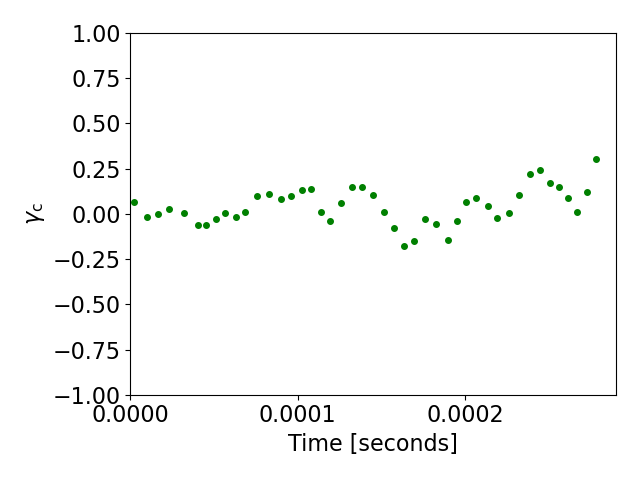}
    \caption{}
    \end{subfigure}
    \begin{subfigure}[b]{0.32\textwidth}
    \includegraphics[width=\textwidth]{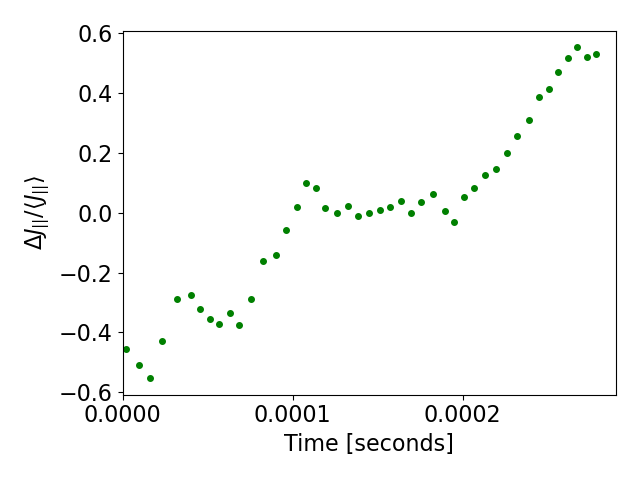}
    \caption{}
    \end{subfigure}
    \caption{HSX prompt ($2.9 \times 10^{-4}$) loss from featuring diffusive banana tip motion. Black stars indicate bounce points. (a) Poloidal cross-section. Black dashed lines indicate the initial magnetic surface and plasma boundary. Diffusive banana tip motion is evident. (b) Radial coordinate $s = \psi/\psi_0$ as a function of time. (c) Drift-convective parameter evaluated for each bounce segment. DC transport is evident as it nears the boundary. (d) The normalized change in the parallel adiabatic invariant demonstrates that $J_{\|}$ is not well conserved, indicative of diffusive banana tip motion.}
    \label{fig:hsx_resonance_banana}
\end{figure}

The small number of prompt losses in Wistell-A have higher values of the mean and final $|\gamma_{\mathrm{c}}|$, indicating the prevalence of drift-convective transport. The histograms confirm this behavior in Figure \ref{fig:qh_histograms}, which suggests that non-DC banana transport is not significant for Wistell-A prompt losses. An example of a DC banana loss in Wistell-A is shown in Figure \ref{fig:aten_sb}. HSX similarly has a significant population of DC banana orbits in addition to losses on non-DC banana orbits. An example of a non-DC banana trajectory can be seen in Figure \ref{fig:hsx_resonance_banana}, which features diffusive banana tip motion. In this example trajectory, we note that $|\gamma_{\mathrm{c}}|$ increases as it is transported outward in radius, indicative of convection along in addition to diffusion across $J_{\|}$ contours. 

\begin{figure}
    \centering
    \begin{subfigure}[b]{0.32\textwidth}
    \includegraphics[width=\textwidth]{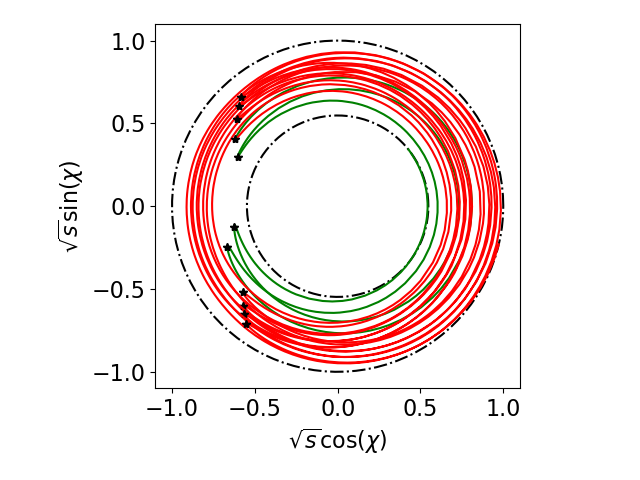}
    \caption{}
    \end{subfigure}
    \begin{subfigure}[b]{0.32\textwidth}
    \includegraphics[width=\textwidth]{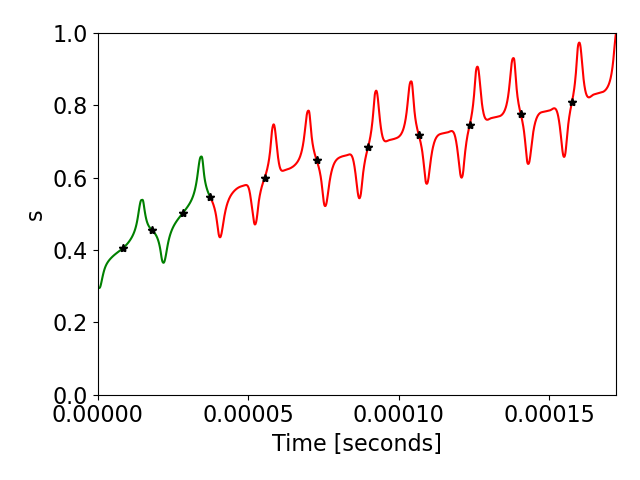}
    \caption{}
    \end{subfigure}
    \begin{subfigure}[b]{0.32\textwidth}
    \includegraphics[width=\textwidth]{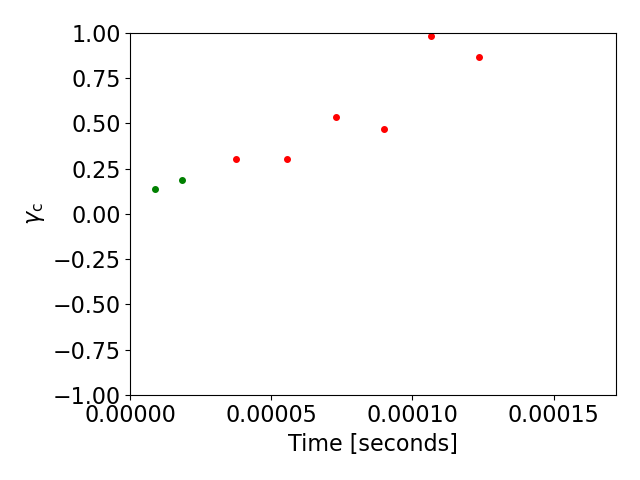}
    \caption{}
    \end{subfigure}
    \caption{Prompt loss ($1.7 \times 10^{-4}$ seconds) from Wistell-A featuring a DC banana orbit (green) which transitions to a DC barely-trapped orbit (red). Black stars indicate bounce points. (a) Poloidal cross-section. Black dashed lines indicate the initial magnetic surface and plasma boundary. (b) Radial coordinate $s = \psi/\psi_0$ as a function of time. (c) Drift-convective parameter evaluated for each bounce segment. DC transport is evident during both the banana and barely-trapped segments.
    }
    \label{fig:aten_sb_passing}
\end{figure}

\begin{figure}
    \centering
    \begin{subfigure}[b]{0.32\textwidth}
    \includegraphics[width=\textwidth]{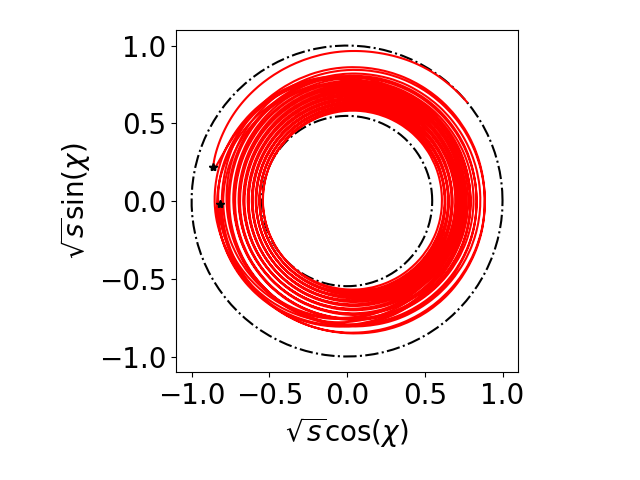}
    \caption{}
    \end{subfigure}
    \begin{subfigure}[b]{0.32\textwidth}
    \includegraphics[width=\textwidth]{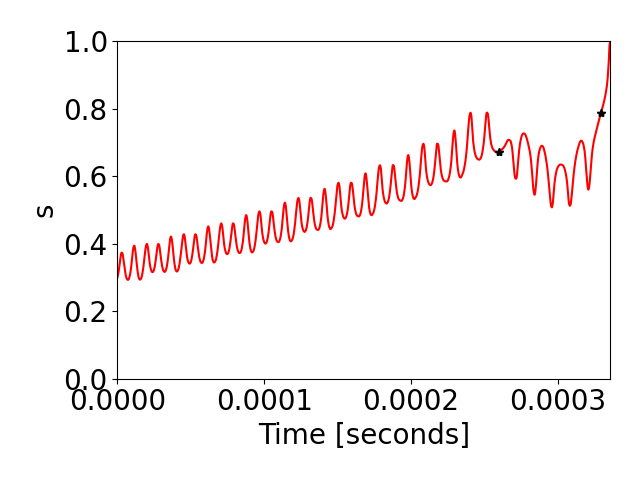}
    \caption{}
    \end{subfigure}
    \begin{subfigure}[b]{0.32\textwidth}
    \includegraphics[width=\textwidth]{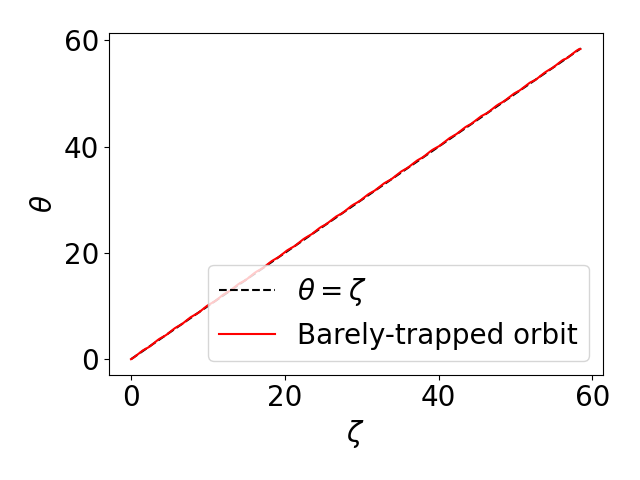}
    \caption{}
    \end{subfigure}
    \caption{Prompt loss ($6.3 \times 10^{-4}$ seconds) from HSX featuring a resonant barely-trapped orbit. Black stars indicate bounce points. (a) Poloidal cross-section. Black dashed lines indicate the initial magnetic surface and plasma boundary. (b) Radial coordinate $s = \psi/\psi_0$ as a function of time. (c) The first barely-trapped trajectory segment is plotted to demonstrate the $l =4$, $k=4$ resonance in the notation of \eqref{eq:passing_resonance_condition}.}
    \label{fig:hsx_resonance}
\end{figure}

In addition to DC banana losses, DC barely-trapped orbits contribute toward the transport in Wistell-A (Figure \ref{fig:aten_sb_passing}). Due to the large variation of the field strength maximum on the surface (Figure \ref{fig:bmin_bmax}), there exists a significant population of trajectories that can traverse through a full period of the field strength variation before bouncing, leading to the presence of barely-trapped orbits among the final orbit types (Figure \ref{fig:qh_histograms}). There are also a small number of resonant barely-trapped orbits In HSX. In Figure \ref{fig:hsx_resonance} we show a barely-trapped orbit close to the $l = 4$, $k=4$ resonance in the notation of \eqref{eq:passing_resonance_condition}, as the rotational transform lies close to $\iota = 1$ with low shear.

\begin{figure}
    \centering
    \begin{subfigure}[b]{0.32\textwidth}
    \includegraphics[width=\textwidth]{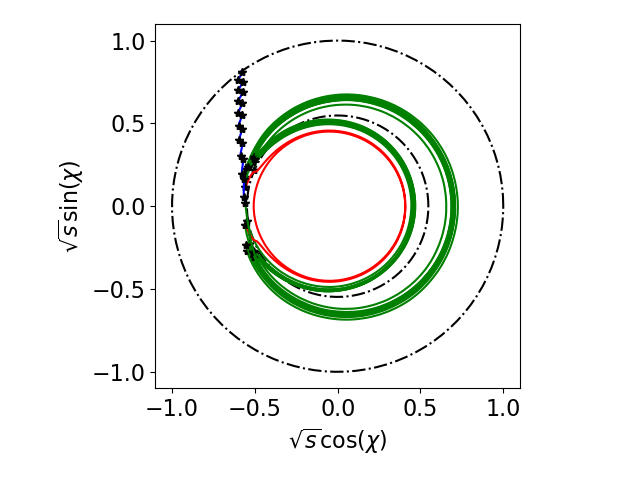}
    \caption{}
    \end{subfigure}
    \begin{subfigure}[b]{0.32\textwidth}
    \includegraphics[width=\textwidth]{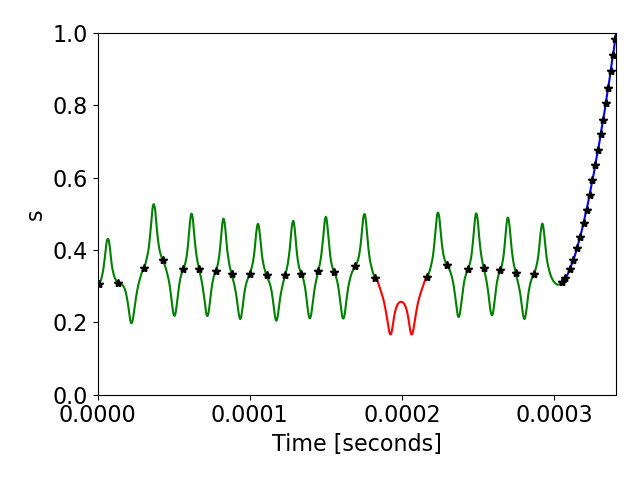}
    \caption{}
    \end{subfigure}
    \caption{HSX prompt ($3.4 \times 10^{-4}$ seconds) loss featuring transitions between the banana (green) and barely-trapped (red) classes before being lost on a ripple orbit (blue). Black stars indicate bounce points. (a) Poloidal cross-section. Black dashed lines indicate the initial magnetic surface and plasma boundary. (b) Radial coordinate $s = \psi/\psi_0$ as a function of time. 
    }
    \label{fig:hsx_final_ripple}
\end{figure}

Non-DC banana and ripple losses are not as prominent among the QH prompt losses as they are among the QA losses (Figures \ref{fig:histograms_QA} and \ref{fig:qh_histograms}). The physical mechanisms behind these differences will be discussed further in Section \ref{sec:optimization_techniques}. Ripple losses are present among the HSX prompt losses (Figure \ref{fig:hsx_final_ripple}): many trajectories transition to ripple-trapped segments and are rapidly lost. In contrast, Wistell-A features minimal ripple trapping. 

As shown in Figure \ref{fig:qh_gammac_mean}, prompt transitions are common in both Wistell-A and HSX as they were in the QA configurations. An example from HSX is shown in Figure \ref{fig:hsx_final_ripple}, which features transitions between the banana and barely-trapped classes before being lost on a ripple orbit. 

\vspace{1cm}

\noindent \textbf{\textit{Non-prompt losses}}

\begin{figure}
    \centering
    \begin{subfigure}[b]{0.32\textwidth}
    \includegraphics[width=\textwidth]{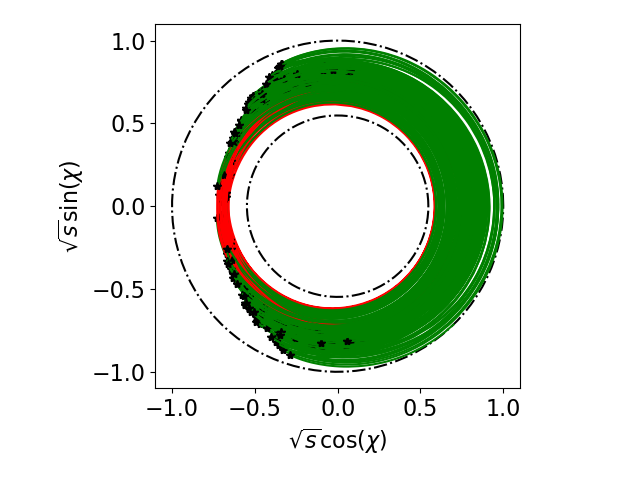}
    \caption{}
    \end{subfigure}
    \begin{subfigure}[b]{0.32\textwidth}
    \includegraphics[width=\textwidth]{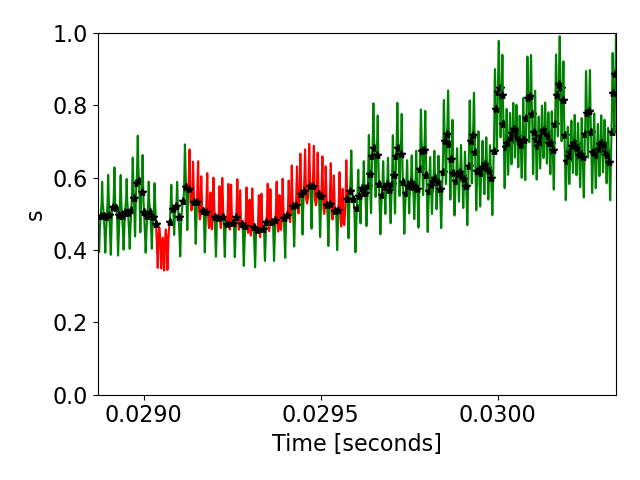}
    \caption{}
    \end{subfigure}
    \caption{Ku5 non-prompt ($3.0 \times 10^{-2}$ seconds) loss featuring transition-driven transport between the banana (green) and barely-trapped (red) classes before being lost on a diffusive banana orbit. Black stars indicate bounce points. (a) Poloidal cross-section. Black dashed lines indicate the initial magnetic surface and plasma boundary. (b) Radial coordinate $s = \psi/\psi_0$ as a function of time. Irregular transitioning behavior is evident.}
    \label{fig:ku_transitions}
\end{figure}

\begin{figure}
    \centering
    \begin{subfigure}[b]{0.32\textwidth}
    \includegraphics[width=\textwidth]{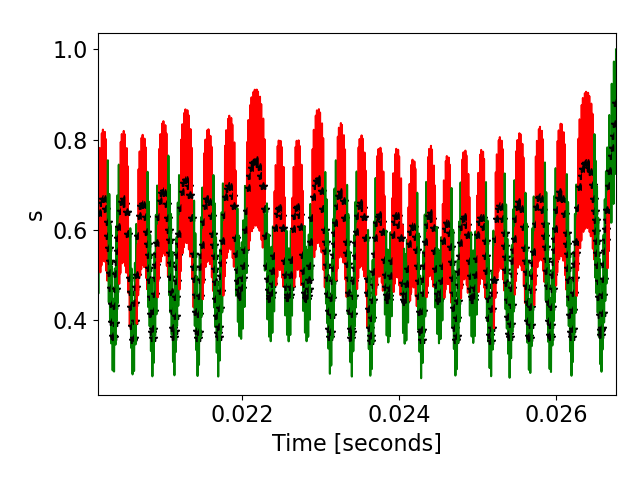}
    \caption{}
    \end{subfigure}
    \begin{subfigure}[b]{0.32\textwidth}
    \includegraphics[width=\textwidth]{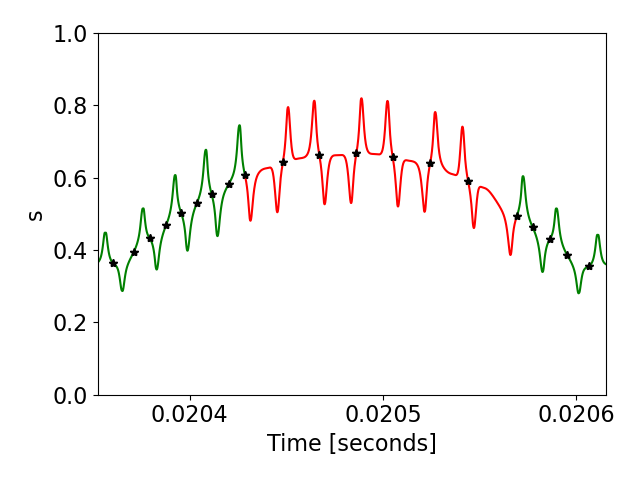}
    \caption{}
    \end{subfigure}
    \begin{subfigure}[b]{0.32\textwidth}
    \includegraphics[width=\textwidth]{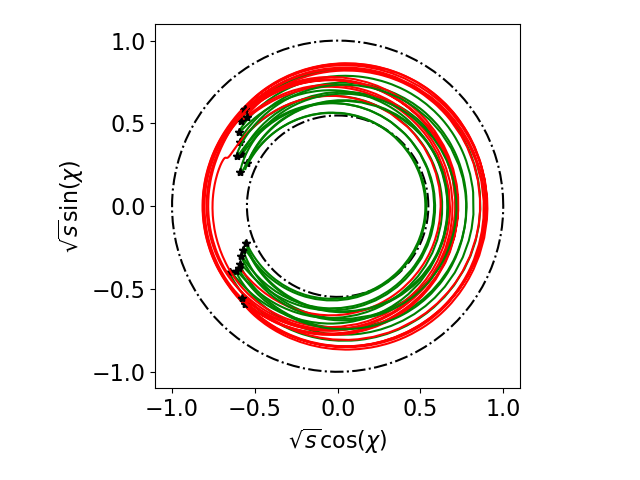}
    \caption{}
    \end{subfigure}
    \begin{subfigure}[b]{0.32\textwidth}
    \includegraphics[width=\textwidth]{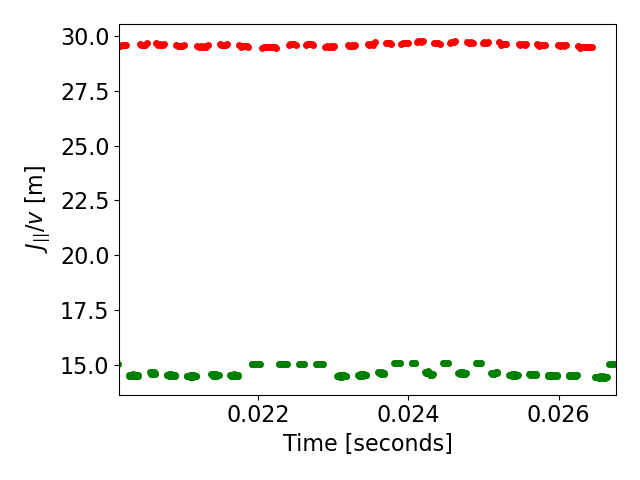}
    \caption{}
    \end{subfigure}
    \caption{Wistell-A non-prompt ($2.7 \times 10^{-2}$ seconds) loss featuring periodic transitions between the banana (green) and barely-trapped (red) classes. The particle is eventually lost on a DC banana segment. Black stars indicate bounce points. (a) Radial coordinate $s = \psi/\psi_0$ as a function of time for a long segment where the periodic behavior is observed.
    (b) Radial coordinate $s = \psi/\psi_0$ as a function of time for one period. (c) The poloidal cross-section for one period. Black dashed lines indicate the initial magnetic surface and plasma boundary. (d) Parallel adiabatic invariant normalized by the velocity magnitude, evaluated for each bounce segment that the trajectory remains in the same trapping class for a long segment where the periodic behavior is observed. Jumps in $J_{\|}$ associated with separatrix crossings are evident.}
    \label{fig:aten_periodic}
\end{figure}

Transitions become much more prominent on non-prompt timescales for all three configurations, especially for Wistell-A and Ku5 (Figure \ref{fig:qh_gammac_mean}). An example of irregular transition-driven transport in Ku5 is shown in Figure \ref{fig:ku_transitions}. Other losses feature regular periodic transitions, as can be seen in Wistell-A in Figure \ref{fig:aten_periodic}. Due to the variation of the field strength maximum on the surface, transitions between banana and barely-trapped segments are prevalent. In the Ku5 equilibrium, only one lost orbit does not undergo transitions but exhibits periodic banana tip motion.

Similar to the QA non-prompt losses, we note that the QH non-prompt losses feature a distribution of the mean $|\gamma_{\mathrm{c}}|$ that is shifted toward zero in comparison with the prompt losses. This behavior indicates that non-DC mechanisms are prominent along the lost trajectories on long timescales. However, the final value of $|\gamma_{\mathrm{c}}|$ remains substantial, indicating that a trajectory may be confined for an extended period before it is lost on a DC orbit segment. An example Wistell-A trajectory is shown in Figure \ref{fig:aten_periodic}, which exhibits periodic transitioning behavior before being lost on a DC banana orbit segment. 

\begin{figure}
    \centering
    \begin{subfigure}[b]{0.32\textwidth}
    \centering
    \includegraphics[width=\textwidth]{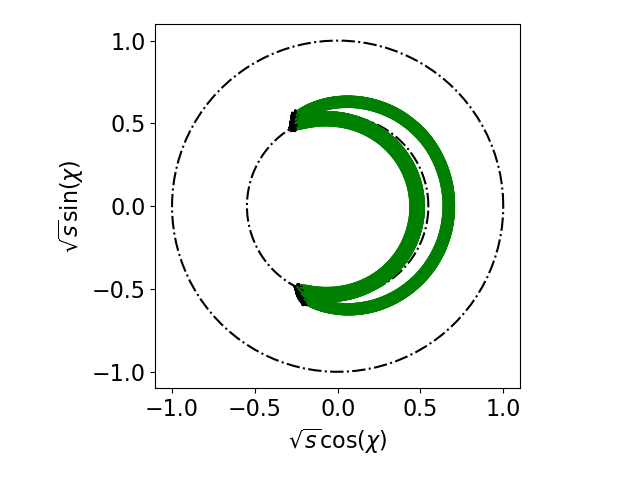}
    \caption{}
    \end{subfigure}
    \begin{subfigure}[b]{0.32\textwidth}
    \centering
    \includegraphics[width=\textwidth]{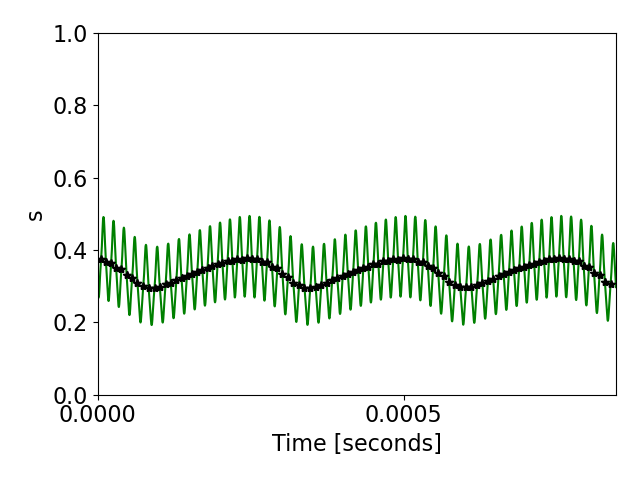}
    \caption{}
    \end{subfigure}
    \begin{subfigure}[b]{0.32\textwidth}
    \centering
    \includegraphics[width=\textwidth]{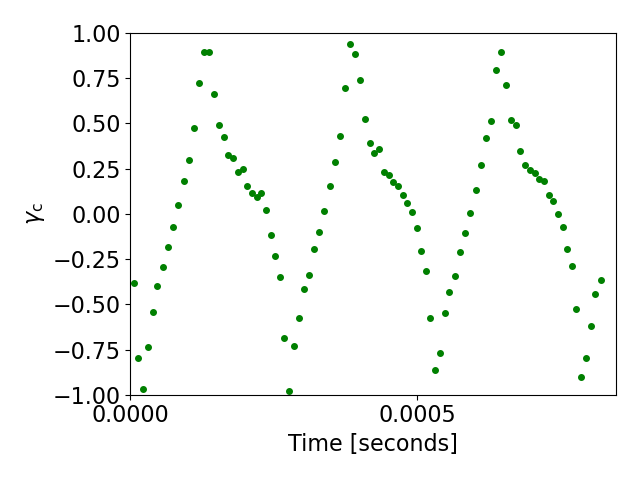}
    \caption{}
    \end{subfigure}
    \begin{subfigure}[b]{0.32\textwidth}
    \centering
    \includegraphics[width=\textwidth]{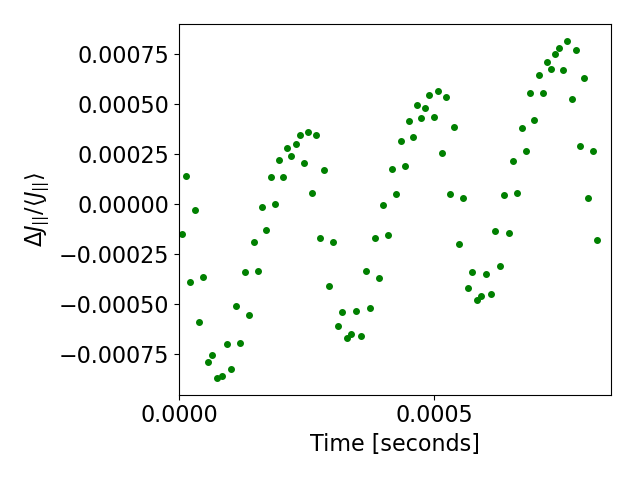}
    \caption{}
    \end{subfigure}
    \begin{subfigure}[b]{0.32\textwidth}
    \centering
    \includegraphics[width=\textwidth]{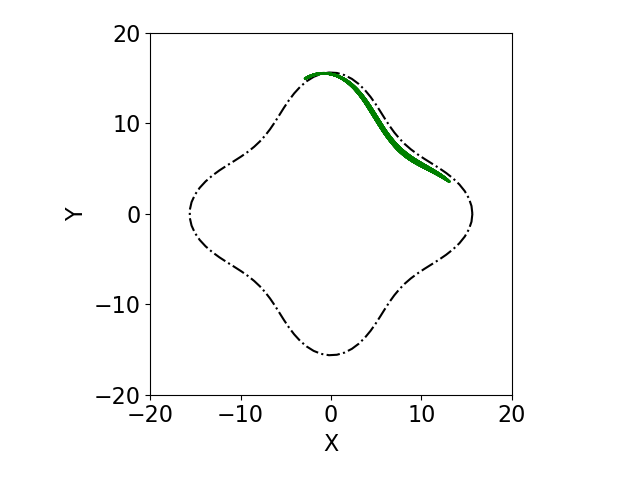}
    \caption{}
    \end{subfigure}
    \caption{HSX non-prompt ($1.6 \times 10^{-1}$ seconds) loss featuring a periodic DC banana, eventually lost on a ripple-trapped segment (not shown). Black stars indicate bounce points. (a) Poloidal cross-section. Black dashed lines indicate the initial magnetic surface and plasma boundary. (b) Radial coordinate $s = \psi/\psi_0$ as a function of time. Periodic behavior is evident. (c) Drift-convective parameter evaluated for each bounce segment. DC transport is evident. (d) The normalized change in the parallel adiabatic invariant indicates that $J_{\|}$ is well conserved. (e) Overhead view of orbit, indicative of toroidal localization. The magnetic axis is indicated by a black dashed line.}
    \label{fig:hsx_periodic_banana}
\end{figure}

Similar to the QA configurations (Figure \ref{fig:ncsx_periodic_banana}), there remain DC orbits that exhibit periodic motion and are confined for long time scales. In the example HSX trajectory shown in Figure \ref{fig:hsx_periodic_banana}, the trajectory is eventually lost on a ripple orbit. 

\section{Optimization techniques}
\label{sec:optimization_techniques}

Having identified the relevant transport channels for the given equilibria, we discuss the impact of the equilibrium magnetic field and possible optimization techniques. Here we consider a magnetic field which is close to a quasisymmetric magnetic field, $B = B(s,\chi)$ with $\chi = M \theta - N \zeta$
and a model for the perturbation from quasisymmetry with mode numbers $m_{\epsilon}$ and $n_{\epsilon}$,
\begin{equation}
    \delta B(s,\theta,\zeta) = \overline{B} \epsilon(s) \cos(m_{\epsilon} \theta - n_{\epsilon} \zeta),
    \label{eq:model_perturbation}
\end{equation}
where $\overline{B}$ is a characteristic field strength of the equilibrium.
We will use the notation $\chi_{\epsilon} = m_{\epsilon} \theta - n _{\epsilon} \zeta$.

\subsection{Orbit-width effects}
\label{sec:orbit_width_effects}

The conserved canonical momentum of guiding center motion in a quasisymmetric magnetic field is,
\begin{equation}
    p_{\zeta}(s,\chi) =  \frac{m v_{||}\left(MG(s) + N I(s) \right)}{ B(s,\chi)}  + q \left(N s\psi_0  - M \psi_P(s)\right),
\end{equation}
where $2\pi\psi_P(s)$ is the poloidal flux and $q$ is the charge. At the inflection points where $d\psi/d\chi = 0$, the field strength is minimized along the orbit, corresponding to $\chi = 0$. Labeling the surface where $s$ is maximized/minimized along the banana orbit as $s_{\pm}$, we Taylor expand $p_{\zeta}$ about $s_-$ and use conservation of $p_{\zeta}$ to obtain the expression,
\begin{equation}
    \Delta s = \frac{2 m v_{\|}(s_-,0)\left(MG(s_-) + N I(s_-) \right)}{ B(s_-,0) q\psi_0 (N - \iota(s_-) M)},
\end{equation}
dropping higher order terms in $\rho_{*}$.
We can define a metric for the orbit width,
\begin{equation}
    f_{\mathrm{orbit width}} = \left \vert \frac{M G + NI}{M \iota -N} \right \vert.
    \label{eq:orbit_width}
\end{equation}
From this expression we note the impact of the symmetry helicity on orbit width. For typical stellarator configurations $|I(s)| \ll |G(s)|$ such that quasi-poloidal configurations ($M = 0$) have the smallest orbit width. Quasihelical configurations ($M \ne 0$, $N\ne 0$) also enjoy reduced orbit width since typically $|N - \iota M| > 1$. Quasiaxisymmetric configurations ($N = 0$) typically have the widest orbits, although the width can be reduced by increasing $\iota$. Finally, in both QA and QH configurations, the orbit width decreases with decreasing $G$.
 The expression for $G$ on the magnetic axis is $G_0 = L B_0/(2\pi)$, where $B_0$ is the field strength on the axis and $L$ is the length of the magnetic axis \cite{2018Landreman}. As we compare equilibria scaled to the same volume and field strength, smaller aspect ratio configurations enjoy reduced orbit widths.
 
\begin{figure}
    \centering    
    \begin{subfigure}[b]{0.48\textwidth}
    \centering
    \includegraphics[width=1.0\textwidth]{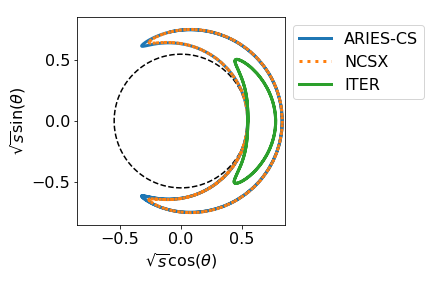}
    \caption{}
    \end{subfigure}
    \begin{subfigure}[b]{0.48\textwidth}
    \centering 
    \includegraphics[width=1.0\textwidth]{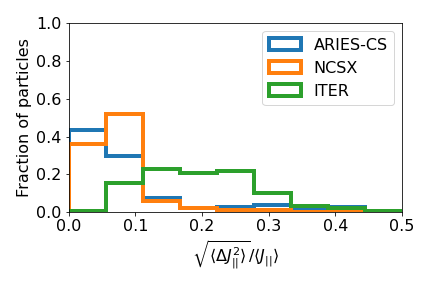}
    \caption{}
    \end{subfigure}
    \begin{subfigure}[b]{0.48\textwidth}
    \centering
    \includegraphics[width=1.0\textwidth]{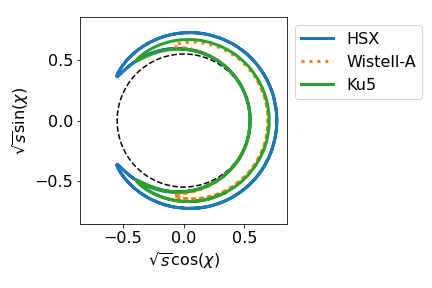}
    \caption{}
    \end{subfigure}
    \begin{subfigure}[b]{0.48\textwidth}
    \centering 
    \includegraphics[width=1.0\textwidth]{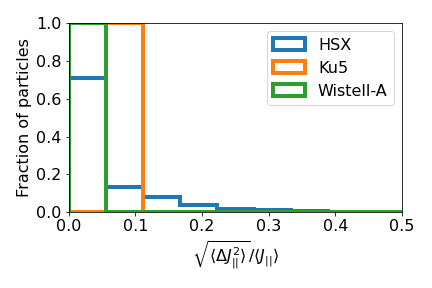}
    \caption{}
    \end{subfigure}
    \caption{(a) and (c) A banana trapped particle is followed in each configuration with the same initial condition, and its position is shown in the poloidal plane to compare the orbit widths across configurations. The orbits in ARIES-CS and NCSX and in Wistell-A and Ku5 nearly overlap. (b) and (d) Distribution of the normalized variation in $J_{\|}$ over non-transitioning lost guiding center orbits. 
    The QA configurations show more non-adiabatic behavior than the QH configurations.}
    \label{fig:jpar_conservation}
\end{figure}

In Figure \ref{fig:jpar_conservation} we compare the banana orbit widths for the same initial condition in Boozer coordinates and pitch angle, corresponding to moderate trapping in the main well, across the six QA and QH configurations. The non-symmetric modes are artificially removed such that the orbits close in the poloidal plane. The Wistell-A and Ku5 configurations have the smallest orbit widths given their helical symmetry and small aspect ratio in comparison to HSX; see Table \ref{tab:configurations}. The widest orbit widths are found in the ARIES-CS and NCSX configurations, which feature a larger aspect ratio than ITER. 
 
 Reducing the orbit width can prevent the prompt loss of wide banana orbits that intersect the boundary. These orbits can be present, for example, in QA equilibria with relatively small values of the rotational transform \cite{2022Landreman}. Furthermore, smaller orbit width reduces the transport due to diffusive banana tip motion. Since the transition to stochasticity occurs due to the resonant overlap experienced in an orbit width, wider banana orbits can contribute more strongly to banana tip diffusion \cite{1981Goldston}, as discussed in the following Section.
 
\subsection{Diffusive banana transport}
\label{sec:diffusive_transport}

In analyzing transport mechanisms in QA and QH configurations, we have found that non-DC banana mechanisms are dominant among the lost trajectories. Furthermore, a significant number of lost orbits terminate on a non-DC banana orbit, especially in QA configurations (Figure \ref{fig:histograms_QA} and \ref{fig:qh_histograms}). One such non-DC banana mechanism that emerges is diffusive banana tip motion (Figure \ref{fig:ariescs_diffusive_banana}). 

The diffusive motion of banana tips arises due to the non-conservation of $J_{\|}$. If $J_{\|}$ is conserved in addition to the magnetic moment and energy, the guiding center motion in 3D magnetic fields is (nearly) integrable. Deviations of $J_{\|}$ arise due to the change in the shape of the magnetic field strength well during one bounce period. If the shape of the effective potential well changes on a timescale comparable to the periodic bounce motion, the assumption of adiabaticity breaks down. Thus, this effect is enhanced for particles with increased in-surface magnetic drifts and more significant deviations from quasisymmetry. While drift-convective transport arises due to the misalignment of $J_{\|}$ contours with magnetic surfaces and is present in the collisionless dynamics as long as $J_{\|}$ is roughly conserved, diffusive banana transport becomes more substantial as energy increases due to the increase in the precessional magnetic drift.

For a given quasisymmetric magnetic field $B(s,\chi)$, the maximum bounce time is,
\begin{equation}
    \tau(s) = \int_0^{2\pi} \frac{d \chi}{v_{\|}\hat{\textbf{b}} \cdot \nabla \chi} = \frac{G(s)+ \iota(s) I(s)}{M \iota(s) - N} \int_0^{2\pi} \frac{d\chi}{B(s,\chi) v_{\|}(s,\chi)}.
\end{equation}
The time to precess through one period of the perturbation is,
\begin{equation}
    \tau_{\epsilon} = \int_0^{2\pi} \frac{d \chi_{\epsilon}}{\textbf{v}_{\mathrm{m}} \cdot \nabla \chi_{\epsilon}},
\end{equation}
where $\chi_{\epsilon} = m_{\epsilon} \theta - n_{\epsilon} \zeta$ (see \eqref{eq:model_perturbation}). 
For smaller values of,
\begin{equation}
    f_{\mathrm{diffusive}} = \frac{\epsilon \tau}{\tau_{\epsilon}},
    \label{eq:f_diffusion}
\end{equation}
a given perturbation with amplitude $\epsilon$ (see \eqref{eq:model_perturbation}) will not break adiabaticity as easily, thus minimizing the diffusive banana response. In the vacuum limit, the above expression scales as $f_{\mathrm{diffusive}} \sim \epsilon m_{\epsilon} \left|\partial B/\partial s\right| /(M\iota - N)$.
The inverse scaling with $(M\iota - N)$ explains the reduction of banana-diffusive transport in QH (in comparison with QA) configurations seen in Section \ref{sec:summary_data}. However, some non-adiabaticity is present in HSX due to the higher mode number ripple perturbations (Figure \ref{fig:modB}). The scaling with $|\partial B/\partial s|$ implies that less compact configurations may be advantageous \cite{2018Landreman}.

In addition to the in-surface precession, rapid radial drift can change the well shape and cause the breaking of adiabaticity. This behavior can be seen in a DC trajectory in ARIES-CS (Figure \ref{fig:ariescs_sb}), which features $\approx  15\%$ deviations in $J_{\|}$ due to the significant radial drift although the orbit remains toroidally localized. As a result, the banana tips wander slightly in this example. 

In Figure \ref{fig:jpar_conservation} we present the distribution of the relative change in $J_{\|}$ along guiding center trajectories for the configurations. The distribution is only presented for trajectories that do not transition between classes. The Ku5 configuration has only one lost trajectory that does not transition, having $\sqrt{\left\langle \Delta J_{\|}^2\right\rangle}/\langle J_{\|} \rangle = 0.08$ and the Wistell-A configuration only has 12 non-transitioning orbits with a maximum value of $\sqrt{\left\langle \Delta J_{\|}^2\right\rangle}/\langle J_{\|} \rangle = 0.03$. 
There are significant deviations from adiabaticity for each configuration (except for Wistell-A), as had been anticipated for stellarator configurations \cite{1998Spong}. The spread in deviation can also be significant, which is accounted for by the difference in precession velocities among trajectories. For example, this can be seen in comparing the evolution of $J_{||}$ between a DC trajectory (Figure \ref{fig:ariescs_ripple}) and non-DC trajectory (Figure \ref{fig:ariescs_diffusive_banana}) in the same configuration. The adiabaticity is generally improved in the QH configurations compared to the QA configurations, consistent with \eqref{eq:f_diffusion}. We note that the non-adiabaticity is especially noticeable in the rippled ITER configuration. In this case, the resonant banana-drift mechanism (Figure \ref{fig:iter_passing_resonance}) is present in addition to banana diffusion.

\begin{figure}
    \centering
    \includegraphics[width=0.4\textwidth]{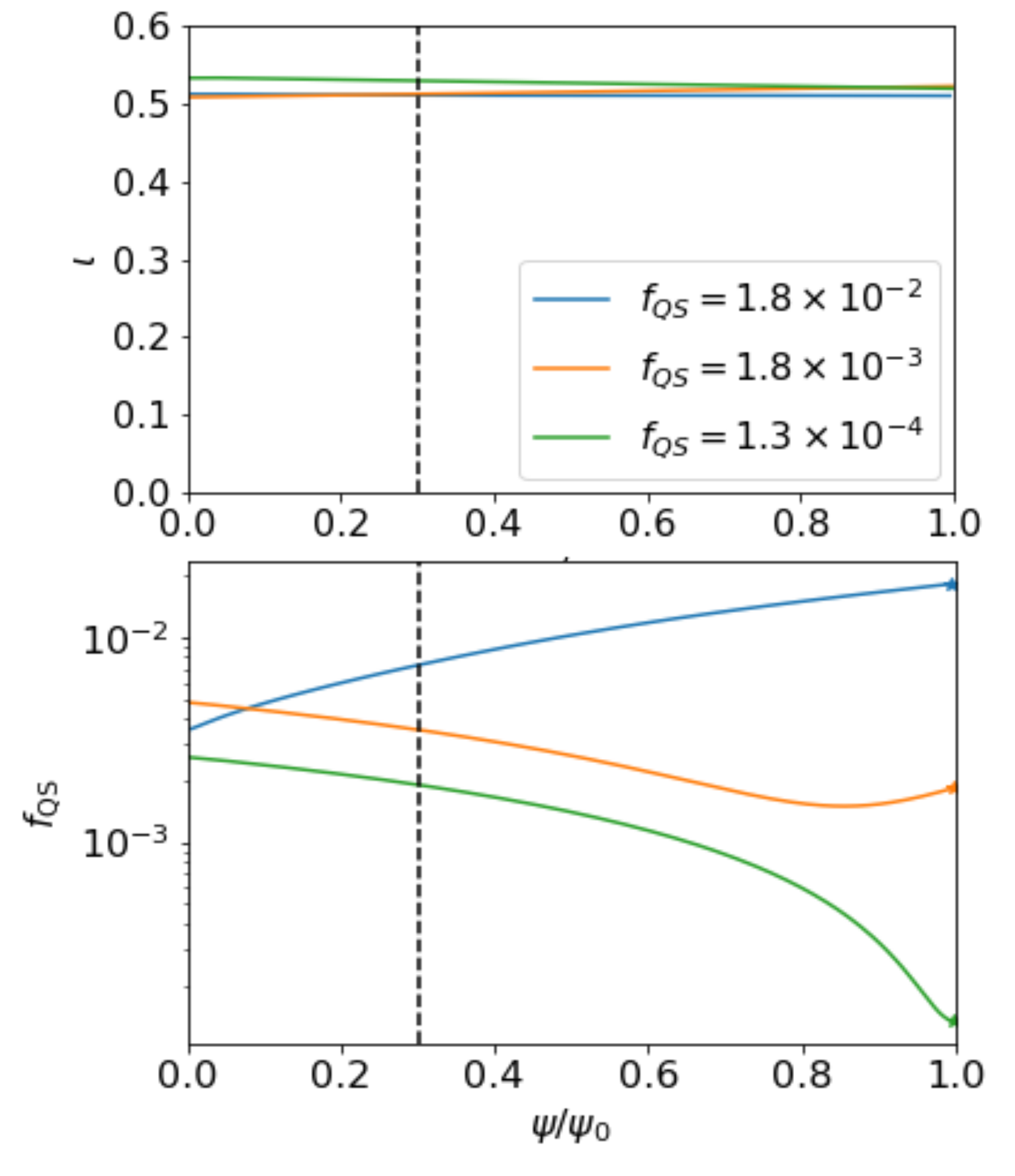}
    \caption{Rotational transform and quasisymmetry error \eqref{eq:qs_error} as a function of normalized flux for the set of configurations optimized for quasisymmetry on the boundary. The black dashed line represents the surface on which trajectories are initialized for the $J_{\|}$ conservation studies.}
    \label{fig:des_iota_QS}
\end{figure}

\begin{figure}
    \centering
    \begin{subfigure}[b]{0.32\textwidth}
    \centering
    \includegraphics[width=\textwidth]{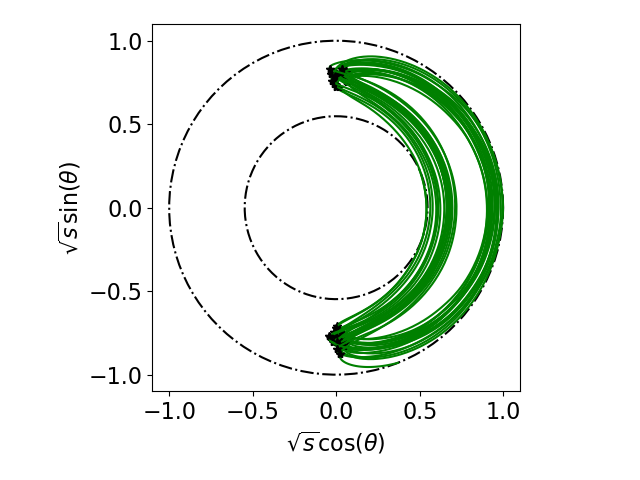}
    \caption{}
    \end{subfigure}
    \begin{subfigure}[b]{0.32\textwidth}
    \centering
    \includegraphics[width=\textwidth]{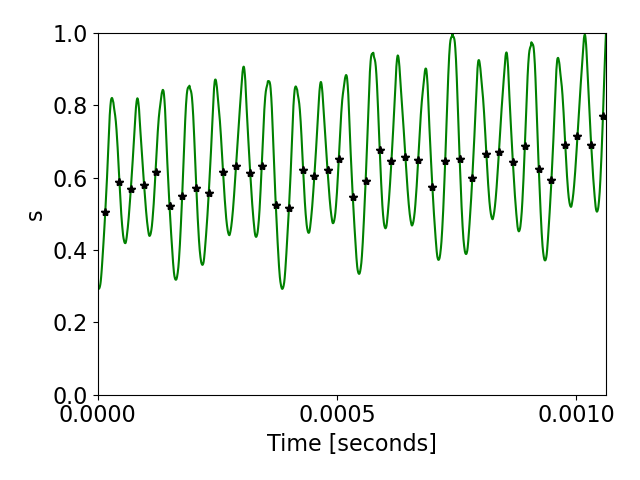}
    \caption{}
    \end{subfigure}
    \begin{subfigure}[b]{0.32\textwidth}
    \centering
    \includegraphics[width=\textwidth]{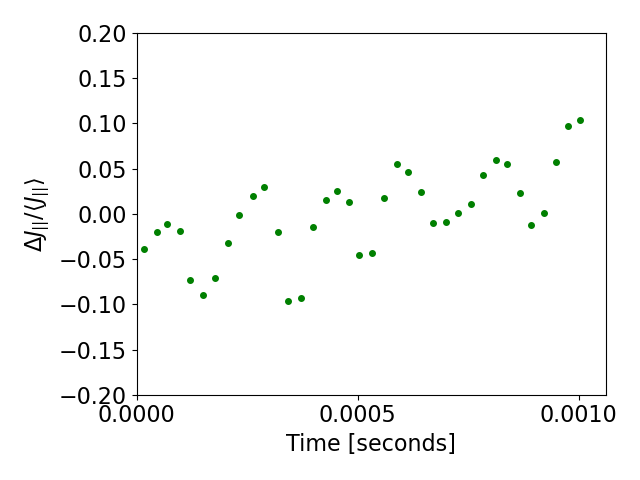}
    \caption{}
    \end{subfigure}
    \caption{Prompt ($1.0 \times 10^{-3}$ seconds) loss featuring diffusive banana tip motion. Black stars indicate bounce points. (a) Poloidal cross-section. Black dashed lines indicate the initial magnetic surface and plasma boundary. (b) Radial coordinate $s = \psi/\psi_0$ as a function of time. Diffusive banana tip motion is evident. 
    (c) The normalized change in the parallel adiabatic invariant indicates that $J_{\|}$ is not conserved well, featuring some secular increase in addition to periodic oscillations. 
    }
    \label{fig:example_diffusive}
\end{figure}

\begin{figure}
    \centering
    \includegraphics[width=1.0\textwidth]{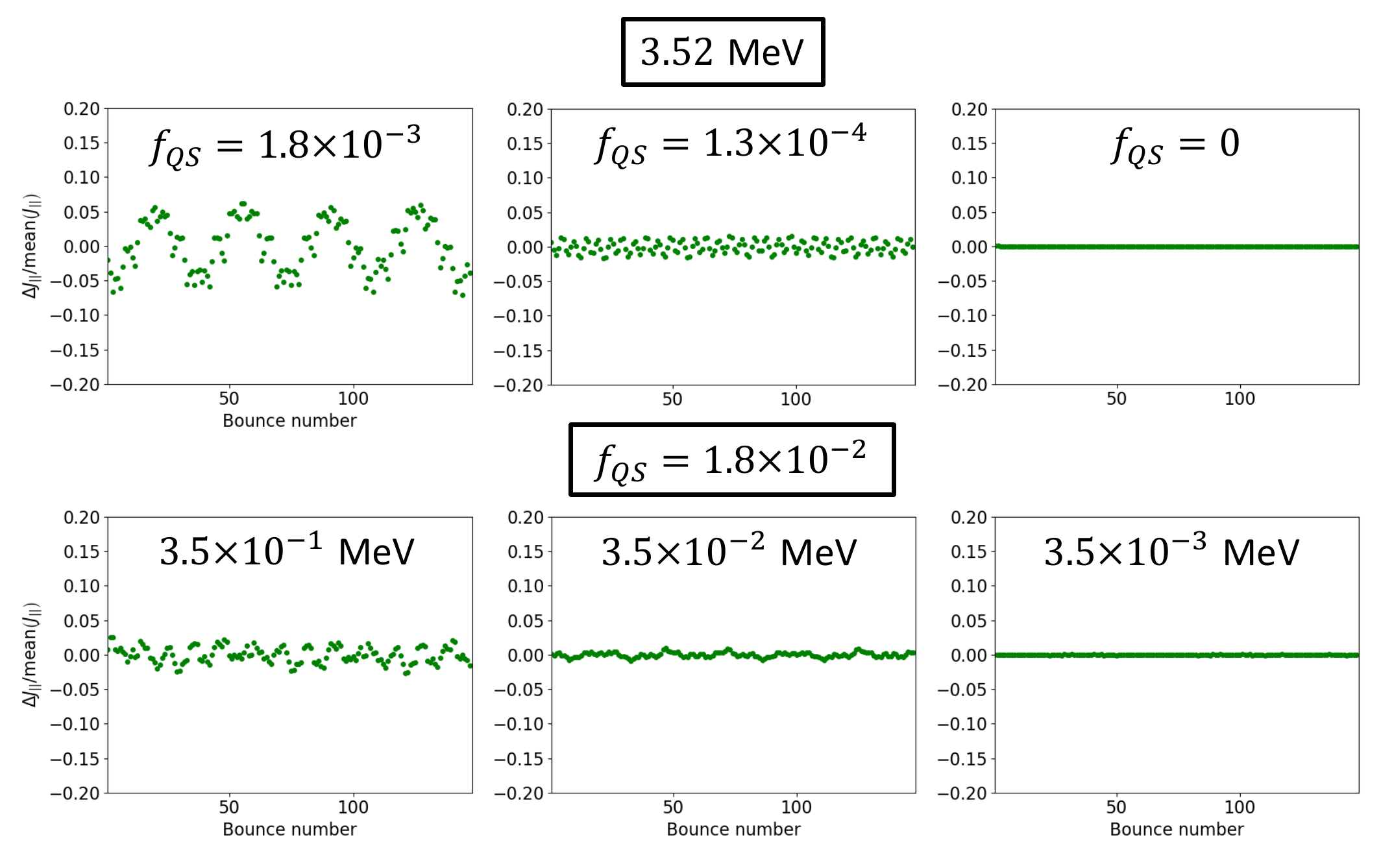}
    \caption{The variation in the parallel adiabatic invariant is shown for a diffusive banana orbit in the three configurations optimized for quasiaxisymmetry on the boundary \eqref{fig:des_iota_QS}. The same initial conditions are used with varying energy and configuration. At a small perturbation amplitude, periodic oscillations in $J_{\|}$ emerge. In equilibria closer to quasiaxisymmetry and at lower energy, $J_{\|}$ is better conserved.}
    \label{fig:Jpar_conservation}
\end{figure}

To further study the mechanisms for the deviations in $J_{\|}$, we perform GC integration in several vacuum equilibria that are optimized for quasiaxisymmetry on the boundary \cite{2022Nies} in addition to having a value of $\iota = 0.52$ on the boundary and aspect ratio 6. For the base equilibrium, the optimization is terminated when the quasisymmetry on the boundary is $f_{QS} = 1.8 \times 10^{-2}$, see \eqref{eq:qs_error}. We compare with two other equilibria, for which the optimization is terminated at $f_{QS} = 1.8 \times 10^{-3}$ and $f_{QS} = 1.3 \times 10^{-4}$. The quasisymmetry error and rotational transform profiles for these equilibria are shown in Figure \ref{fig:des_iota_QS}. 

We consider initial conditions for a 3.52 MeV fusion-born alpha particle on the $s = 0.3$ surface corresponding to a trajectory that is lost due to diffusive banana transport in the base equilibrium, see Figure \ref{fig:example_diffusive}. 
In the base equilibrium we note that $J_{\|}$ has periodic oscillations on top of a secular increase up to $15\%$ above the mean value. In Figure \ref{fig:Jpar_conservation} we compare the conservation of $J_{\|}$ in the two other equilibria with reduced $f_{QS}$
and for an equilibrium where the non-symmetric modes of the field strength are artificially removed, $f_{QS} = 0$. As the quasisymmetry error is reduced, the secular increase is eliminated, and periodic oscillations in $J_{\|}$ remain. The periodicity hints at the existence of a higher-order adiabatic invariant \cite{1967Hastie,1982Hastie}. As the symmetry-breaking is further reduced, $J_{\|}$ is well conserved. Similar behavior is seen as the particle energy is reduced in the base equilibrium. 

The periodic oscillations of $J_{\|}$ about a conserved mean value at low perturbation amplitudes are reminiscent of oscillations in $\mu$, which has been termed super-adiabaticity in the spherical tokamak \cite{2001Carlsson,2021Escande} and mirror \cite{1972Rosenbluth} literature. The periodic oscillations of $J_{\|}$ are consistent with transport due to a single island chain in the banana tip motion. At larger perturbation amplitudes, islands may begin to overlap, leading to stochastic transport.

\subsection{Ripple trapping}
\label{sec:ripple_trapping}

In all of the QA configurations (Figure \ref{fig:histograms_QA}) in addition to HSX (Figure \ref{fig:histograms_hsx}), ripple trapping makes up a significant fraction of the losses, especially on prompt timescales. Considering the model for a magnetic field close to quasisymmetry \eqref{eq:model_perturbation}, the criterion for the perturbation to be large enough to form a local minimum at ($s_r,\theta_r, \zeta_r$) along field lines is, 
\begin{equation}
    \epsilon(s_r) \gtrsim  \Bigg|\frac{M\iota -N }{(m_{\epsilon} \iota - n_{\epsilon})\sin(m_{\epsilon} \theta_r - n_{\epsilon} \zeta_r)} \partder{B(s,M \theta_r - N \zeta_r)}{\chi} \Bigg|.
    \label{eq:critical_ripple}
\end{equation}
An equilibrium will be less susceptible to ripple wells for smaller values of,
\begin{equation}
    f_{\mathrm{ripple}} = \left \vert \frac{m_{\epsilon} \iota - n_{\epsilon}}{M \iota - N} \left(\partder{B(s,\chi)}{\chi} \right)^{-1}\right \vert.
    \label{eq:ripple}
\end{equation}
Given the factor $(M \iota - N)^{-1}$, this presents a potential advantage for QH configurations in avoiding ripple formation for a given perturbation. Increasing $\iota$ may also reduce the ripple formation for small values of $m_{\epsilon}/M$. Given the factor $\left(\partial B/\partial \chi \right)^{-1}$, more compact configurations are advantageous. Furthermore, we can note from \eqref{eq:critical_ripple} that higher mode number, $m_{\epsilon}$ and $n_{\epsilon}$, perturbations form ripple more easily. Similar to the diffusive mechanisms \eqref{eq:f_diffusion}, perturbations with smaller mode number $m_{\epsilon}$ are less deleterious. We see that the factor $1/(M \iota -N)$ is in common with features that reduce diffusive \eqref{eq:f_diffusion} and orbit width \eqref{eq:orbit_width} effects. However, the factor $\left(\partial B/\partial \chi \right)^{-1}$ is in conflict with features that reduce non-adiabaticity. From the GC statistics, ripple-trapping is less prevalent among the lost orbits in QH configurations \eqref{fig:qh_histograms}, except for in the HSX equilibrium, which features higher mode number ripple perturbations (Figure \ref{fig:modB}). 

 Finally, we note that ripples form more easily near the minima and maxima of the field strength. Indeed, although the ripple perturbation amplitude is relatively small on the inboard side of ITER, ripple trapping can occur near the maximum of the field strength (Figure \ref{fig:iter_passing_resonance}). However, ripple wells localized near the field maxima or minima may not be as deleterious, as the radial drift scales as $|\partial B/\partial \chi|$. Particles trapped in these wells mainly drift parallel to surfaces and may eventually detrap as the ripple perturbation amplitude diminishes. This behavior can be seen in the ARIES-CS trajectories (Figure \ref{fig:ariescs_banana_ripple}), which periodically detrap and entrap in ripple wells. On the other hand, the quasisymmetry deviations in NCSX lead to ripple wells away from the maxima and minima \cite{2006Mynickb}, which causes transitions from banana trapping to prompt ripple losses (Figure \ref{fig:ncsx_prompt_transitioning}). 

\subsection{Transitioning particles}

While the GC motion in perfectly quasisymmetric magnetic fields is free from class transitions, even small perturbations can introduce chaotic regions in phase space for trajectories trapped near the field maximum \cite{2013Lichtenberg}. We have found that a significant fraction of lost trajectories in both the QA and QH configurations undergo transitions between classes (Figures \ref{fig:qa_gammac_mean} and \ref{fig:qh_gammac_mean}). In some cases, these class transitions are associated with irregular behavior, characterized by many separatrix crossings without periodic structure (Figure \ref{fig:ariescs_transitions_driven} or \ref{fig:ku_transitions}). While the separatrix crossings appear to be accompanied by small radial displacements, the radial transport is often not rapid compared to the diffusive banana tip or drift-convective mechanisms. In both trajectory examples, relatively little radial transport occurs during the irregular transition behavior, but the particle is lost due to diffusive banana tip motion. In other cases, irregular transitioning appears responsible for the eventual loss (Figure \ref{fig:ariescs_banana_ripple}).

Furthermore, transitions are not always irregular but display periodic behavior (Figures \ref{fig:ariescs_banana_ripple} and \ref{fig:aten_periodic}).
In the case of the trajectory shown in Figure \ref{fig:ariescs_banana_ripple}, we see a ripple-trapped particle that drifts vertically, allowing it to detrap from the ripple localized on the inboard side. The trajectory then transitions to a banana segment. The trajectory can entrap and repeat the cycle as the banana-trapped segment nears the inboard side. While $J_{\|}$ is not conserved along such trajectories, the periodicity indicates the presence of a higher-order adiabatic invariant. In analogy with the context of banana diffusion, if the periodic orbit is associated with a large radial excursion, such as in Figure \ref{fig:aten_periodic}, slight wandering of the super-orbit may occur due to the breaking of higher-order adiabaticity. Thus transitions associated with wide super-orbits should especially be avoided.

\begin{figure}
    \centering
    \begin{subfigure}[b]{0.48\textwidth}
    \centering 
    \includegraphics[width=1.0
    \textwidth]{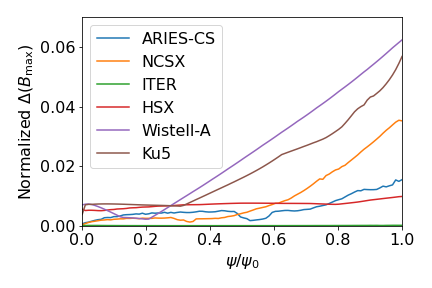}
    \end{subfigure}
    \caption{Normalized variation across field lines of the maximum field strength along the field line. On each surface, $\Delta B_{\mathrm{max}}$ is the maximum deviation of $B_{\max}$ from the average over field line label. This value is normalized by the average $B_{\max}$ over the field line label.}
    \label{fig:bmin_bmax}
\end{figure}

Several equilibrium characteristics prevent class transitions. 
\begin{enumerate}
    \item Ensuring that the maxima of the field strength along a field line are the same for all field lines on a magnetic surface prevents transitions due to in-surface precession \cite{1998Spong,2014Drevlak,2022Sanchez}. In Figure \ref{fig:bmin_bmax} we compare the variation of $B_{\mathrm{max}}$ on surfaces for the set of QA and QH equilibria. We see that the two equilibria with the largest fraction of transitioning lost trajectories (Tables \ref{tab:qa_losses} and \ref{tab:qh_losses}), Ku5 and Wistell-A, feature the largest variation of $B_{\mathrm{max}}$. As this metric does not account for transitions due to ripple wells away from the field maxima, it does not explain the behavior in ARIES-CS and NCSX, the two equilibria with the largest total number of transitioning lost trajectories. 
    \item Pseudosymmetry \citep{1999Isaev} is defined as the quantity,
\begin{align}
    f = \frac{\textbf{B} \times \nabla \psi \cdot \nabla B}{\textbf{B} \cdot \nabla B},
\end{align}
being non-zero everywhere such that the radial magnetic drift vanishes at field maxima and minima. Pseudosymmetry implies that particles trapped near the field maxima cannot transition by drifting radially and ensures the confinement of deeply-trapped particles.
\item Isodrasticity \cite{2021Mackay} is formulated as a condition on the surface $\Sigma^{-}$, defined by the local maxima of the field strength along a field line ($\hat{\textbf{b}}\cdot \nabla B = 0$ and $\hat{\textbf{b}} \cdot \nabla  \left(\hat{\textbf{b}} \cdot \nabla B  \right) < 0$). On $\Sigma^{-}$, the isocurves of $J_{\|}$ and $B$ must coincide. Isodrasticity prevents transitions due to the bounce-averaged radial or in-surface drifts as the transition probability is proportional to \cite{1990Kovrizhnykh,2001Beidler},
\begin{equation}
    D \propto \partder{J_{\|}(s,\alpha,\lambda)}{\alpha} \partder{B(s,\alpha)}{s} -\partder{J_{\|}(s,\alpha,\lambda)}{s} \partder{B(s,\alpha)}{\alpha},
\end{equation}
where $\alpha$ is a field line label and $D$ is evaluated on $\Sigma^-$.
If condition (i) is satisfied, then $\partial B/\partial \alpha = 0$ on the intersection of $\Sigma^{-}$ with a magnetic surface.
If condition (ii) is satisfied, then $\partial J_{\|}/\partial \alpha = 0$ on $\Sigma^{-}$. If both (i) and (ii) are satisfied, then the transition probability vanishes. However, the condition of isodrasticity is more general than enforcing (i) and (ii).
\end{enumerate} 

\subsection{Drift-convective transport}
\label{sec:superbanana_transport}

A significant fraction of trajectories is lost on drift-convective orbits, especially in the QH configurations (Figure \ref{fig:qh_histograms}). Drift-convective transport arises for a class of particles that do not precess quickly in comparison to their radial drift, quantified by large values of $|\gamma_{\mathrm{c}}|$, see \eqref{eq:gamma_c}. While the bounce-averaged radial drift will depend on the specific perturbation applied to a given equilibrium, the bounce-averaged poloidal drift will be largely driven by the equilibrium field \cite{2017Calvo}. Under the vacuum field assumption, the poloidal magnetic drift, $\textbf{v}_{\mathrm{m}} \cdot \nabla \theta$, is proportional to $\partial B/\partial s$ while $\textbf{v}_{\mathrm{m}} \cdot \nabla \zeta = 0$. As $|\partial B/\partial s|$ scales inversely with the aspect ratio, more compact configurations may be advantageous \cite{2018Landreman} for avoiding DC transport. 
Increasing $|\partial B/\partial s|$ can also be accomplished by peaking the pressure profile in some configurations. The resulting increased precession leads to improved fast ion confinement with increasing pressure in some W7-X configurations \cite{2014Drevlak}, while fast ion confinement decreases in high-beta extrapolated LHD configurations \cite{2014Miyazawa}. As the QA configurations under consideration are more compact, we observe that proportionately more losses terminate on a DC orbit in the QH than the QA configurations (Figures \ref{fig:histograms_QA} and \ref{fig:qh_histograms}). As the precession drift decreases with increasing normalized flux, more DC trajectories are likely to be found near the plasma boundary. This can be seen in the DC trajectories (Figures \ref{fig:ariescs_ripple}, \ref{fig:aten_sb}, and \ref{fig:aten_sb_passing}) which demonstrate increasing values of $|\gamma_{\mathrm{c}}|$ as the particle drifts radially. 

Many techniques have been discussed in the literature to prevent DC transport \cite{1999Wobig}. Alignment of $J^*$, an approximation of $J_{\|}$, with flux surfaces has improved the collisionless confinement in compact configurations \cite{1998Spong}. Other techniques based on the alignment of $J_{\|}$ with surfaces improved confinement of marginally-trapped particles in the W7-X configuration space \cite{2014Drevlak}. The $\Gamma_{\mathrm{c}}$ metric averages $|\gamma_{\mathrm{c}}|^2$ over velocity space \cite{2005Nemov,2008Nemov} in order to quantify the misalignment of flux surfaces and $J_{\|}$ surfaces. Optimization including $\Gamma_{\mathrm{c}}$ has successfully improved confinement in QH \cite{2019Bader} and QA \cite{2022LeViness} configurations. 
However, the existence of DC orbits does not always indicate prompt losses. For example, we find DC orbits that exhibit periodic motion that are confined for long time scales (Figure \ref{fig:ncsx_periodic_banana} and \ref{fig:hsx_periodic_banana}). To this end, the metric $\Gamma_{\alpha}$ improves over the $\Gamma_{\mathrm{c}}$ metric. Some degree of misalignment between $J_{\|}$ surfaces and flux surfaces is not penalized, and the possible closure of $J_{\|}$ contours is considered. The $\Gamma_{\alpha}$ model also accounts for the in-surface precession drift onto unconfined orbits \cite{2021Velasco}.

\begin{figure}
    \centering
    \includegraphics[width=1.0\textwidth]{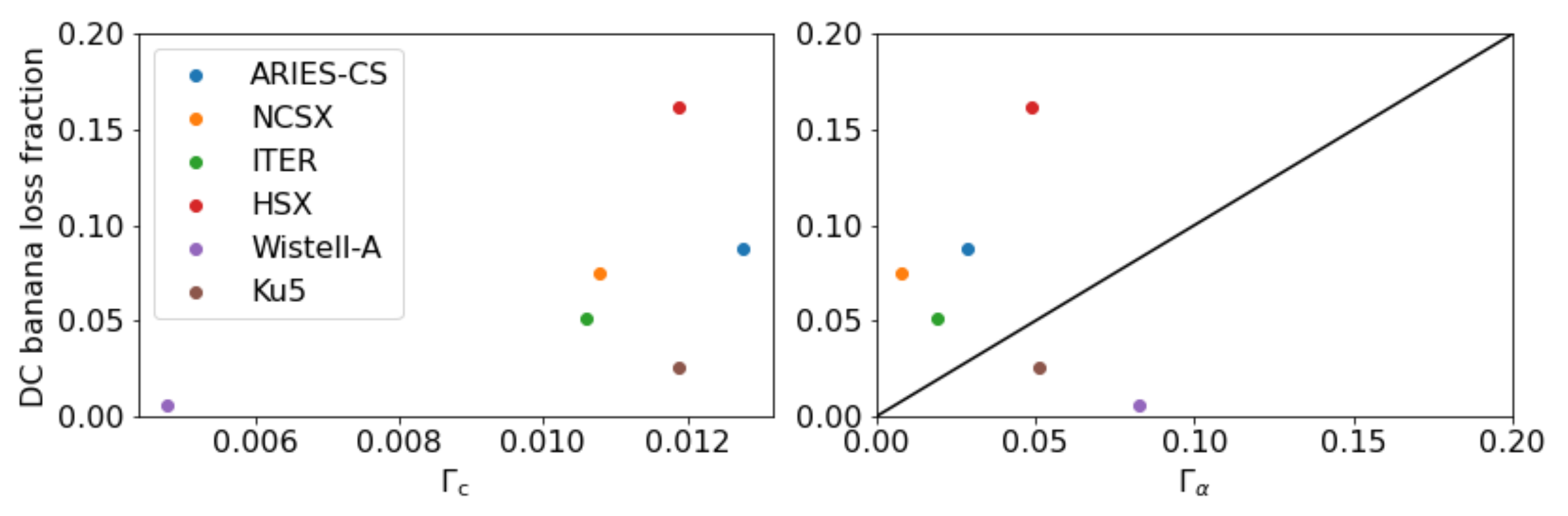}
    \caption{Comparison of banana drift-convection (DC) loss fraction with the (left) $\Gamma_{\mathrm{c}}$ and (right) $\Gamma_{\alpha}$ metrics.
    }
    \label{fig:knosos_comparison}
\end{figure}

In Figure \ref{fig:knosos_comparison}, we compare our classification of DC orbits with the prediction of DC losses based on $\Gamma_{\mathrm{c}}$ and $\Gamma_{\alpha}$ evaluated with the KNOSOS code \cite{2020Velasco} on the $s = 0.3$ surface. Here we define DC banana losses as trajectories which are classified as banana trapped with $|\gamma_{\mathrm{c}}| > 0.2$ for at least one bounce segment. In the computation of $\Gamma_{\alpha}$, the threshold $|\gamma_{\mathrm{c}}| > 0.2$ is also used to identify drift-convective losses. 
While both metrics should be correlated with the DC banana loss fraction, only $\Gamma_{\alpha}$ provides a quantitative estimate of the loss fraction. For this reason in Figure \ref{fig:knosos_comparison} we include the line $y = x$ in the right plot in order to compare the magnitude of the DC banana losses with $\Gamma_{\alpha}$. We see that $\Gamma_{\alpha}$ is indeed a better predictor of the magnitude of DC banana losses than $\Gamma_{\mathrm{c}}$. However, we see little correlation between either metric and the DC loss fraction. 
Both the $\Gamma_{\mathrm{c}}$ and $\Gamma_{\alpha}$ models are limited as they are radially local. Furthermore, they do not account for class transitions and possible unconfined trajectories due to the variation of the minimum of the field strength along a field line. 

If considering configurations within the same class, $\Gamma_{\alpha}$ and $\Gamma_{\mathrm{c}}$ appear to have better correlations with GC confinement. Good correlation has been observed for a set of QA equilibria close to \texttt{li383} \cite{2022LeViness} as well as for W7-X configurations \cite{2021Velasco}. These results suggest that these metrics may perform well as optimization criteria, although the correlation is not as good when comparing across configuration types. 

\section{Summary}
\label{sec:summary}

We have developed classification methods to investigate the relevant collisionless guiding center loss mechanisms of fusion-born alpha particles in equilibria close to quasiaxisymmetry (QA) and quasihelical (QH) symmetry.

\subsection{Prompt loss mechanisms}

In each configuration, the dominant fraction of the orbit is spent in the banana trapping class without drift convection (DC) characteristics. These characteristics indicate that $J_{\|}$ contours align well with flux surfaces and convective radial transport is small for much of the orbit. DC banana transport is also significant throughout the trajectories in each configuration except Ku5. For the QA configurations and HSX, ripple trapping additionally makes up a substantial fraction of the orbits. Many barely-trapped orbit segments, which can pass through at least one field period of the field strength variation, are present in the QH configurations. Most trajectories transition among these classes on short and long timescales. Promptly lost orbits often terminate on a DC orbit, indicating that proxies that align the contours of $J_{\|}$ with flux surfaces may reduce prompt losses. 

In the ITER and NCSX configurations, most promptly lost trajectories terminate on a ripple-trapped segment. On the other hand, the prominence of ripple transport is reduced in ARIES-CS due to the localization of ripple wells on the inboard side \cite{2006Mynickb}. Instead, many ARIES-CS losses terminate on a banana orbit with DC or non-DC characteristics. 
Some non-DC orbits appear to exhibit diffusive banana tip motion (Figure \ref{fig:ariescs_diffusive_banana}).
A fraction of HSX losses also terminates on non-DC banana orbit segments, which appear consistent with diffusive banana tip motion (Figure \ref{fig:hsx_resonance_banana}). This mechanism is associated with non-conservation of the parallel adiabatic invariant, $J_{\|}$. 

Ripple trapping is reduced in the QH configurations due to their reduced connection lengths. The exception is HSX, where higher mode number perturbations due to coil ripple are present (Figure \ref{fig:modB}), increasing the propensity for local well formation. Overall, the number of prompt losses is reduced in the QH configurations compared to the QA configurations, given the reduction of diffusive banana tip motion and ripple trapping. A similar observation of the reduction of losses in QH in comparison to QA configurations has been observed in other studies \cite{2021Bader,2022Landremanb}.

Orbits lost due to barely-trapped trajectory segments are observed in Wistell-A and ITER. In the case of Wistell-A, the significant variation of the field strength maximum on the surface allows for an increased number of transitions between banana and barely-trapped orbit types. These barely-trapped orbits can be in the DC class (Figure \ref{fig:aten_sb_passing}), leading to prompt losses. In the case of ITER, periodic barely-trapped orbits can resonate with the $N = 18$ coil ripple. A small number of resonant barely-trapped orbits are also observed in HSX (Figure \ref{fig:hsx_resonance}). No losses of passing particles are observed.

\subsection{Non-prompt losses}

On non-prompt timescales, we find that many trajectories transition between trapping classes, as has been anticipated by theory \cite{2001Beidler} and numerical studies \cite{2016Faustin,2019Cole}. However, transitioning trajectories do not always imply rapid radial diffusion. Some configurations feature irregular transitioning behavior (Figures \ref{fig:ariescs_transitions_driven} and \ref{fig:ku_transitions}) accompanied by relatively small radial displacements. Transitioning orbits may be confined for a long time but are subsequently lost due to other mechanisms, such as diffusive banana tip motion (Figure \ref{fig:ariescs_transitions_driven}).

There exist a class of particles that perform complex periodic orbits (Figures \ref{fig:ariescs_banana_ripple} and \ref{fig:aten_periodic}). Although $J_{\|}$ is not conserved along transitioning orbits, the presence of complex periodic orbits hints toward the existence of a higher-order adiabatic invariant \cite{1967Hastie,1982Hastie}. This observation indicates that optimization metrics that minimize class transitions may be overly restrictive. 

On non-prompt timescales we also observe DC trajectories whose orbits are healed due to closure of the $J_{\|}$ contours within the plasma boundary (Figures \ref{fig:ncsx_periodic_banana} and \ref{fig:hsx_periodic_banana}). The orbits feature periodic radial displacements and toroidal localization. The effective width of the DC orbit may imply additional diffusive motion when collisions are included.

\subsection{Optimization considerations}

While precise quasisymmetry can yield excellent GC confinement \cite{2022Landreman}, relaxing the requirements on the symmetry level may accommodate the optimization for other metrics of interest, such as microstability. To this end, we consider equilibrium properties that can reduce alpha losses. 
The geometric dependence of the transport mechanisms is summarized in Table \ref{tab:mechanisms}. 
The banana tip diffusion due to non-conservation of $J_{\|}$ is associated with short timescales of precession through one period of a given perturbation compared to the bounce period. In contrast with DC losses, which are present in the collisionless dynamics even at low energy, the prevalence of this loss channel increases with particle energy. This transport mechanism is reduced in configurations with large values of the effective transform, $M\iota -N$, increased aspect ratio, and reduced field strength variation on a magnetic surface (see \eqref{eq:orbit_width}).

The propensity for forming ripple wells along field lines is reduced in configurations with significant effective transform and decreased aspect ratio. Perturbations localized near the field maxima and minima form wells more easily. On the other hand, since the radial drift is minimal near the field maxima and minima, particles trapped in these regions often transition to deeply-trapped or barely-trapped banana trajectories (Figures \ref{fig:iter_stagnation} and \ref{fig:ariescs_banana_ripple}) rather than being promptly lost. Higher mode number perturbations, such as those arising from coil ripple, are more deleterious for both ripple formation and breakdown of $J_{\|}$ conservation. Such perturbations have also been found to increase flow damping substantially \cite{2014Calvo}. 

In addition to the considerations above, banana or barely-trapped orbits that resonate with a perturbation in the equilibrium should be avoided. Resonant barely-trapped orbits have been observed in the HSX (Figure \ref{fig:hsx_resonance}) and rippled ITER (Figure \ref{fig:iter_passing_resonance}) configurations, which arise in the neighborhood of surfaces with low-order rational values of the rotational transform. If a configuration has low magnetic shear throughout the volume, as in the case of HSX (Figure \ref{fig:loss_times}), this can lead to the persistence of the orbit periodicity through the volume \cite{2021White}.

As DC transport is present in all six equilibria, we compare the number of DC losses with the  $\Gamma_{\mathrm{c}}$ and $\Gamma_{\alpha}$ metrics based on the radially-local bounce-averaged motion (Figure \ref{fig:knosos_comparison}). Due to the prevalence of class transitions and finite orbit width effects, the number of DC losses can deviate significantly from the $\Gamma_{\alpha}$ metric. However, it captures the order of magnitude and approximate scaling. The main limitation of $\Gamma_{\mathrm{c}}$ and $\Gamma_{\alpha}$ is the assumption of radial locality, which cannot be justified in a naturally radially-global problem. Implementation of a radially-global drift kinetic equation in the KNOSOS code is underway to overcome this limitation. 

\begin{table}
\centering 
\begin{tabular}{|c||c|c|c|c|}
\hline 
  & \textbf{Wide banana} & \textbf{Diffusive banana} & \textbf{Ripple} & \textbf{Drift convection} \\
  \hline 
  Aspect ratio & Increases & Decreases & Increases & Increases \\
  $|M \iota - N|$ & Decreases & Decreases & Decreases & \\
  $m_{\epsilon}$, $n_{\epsilon}$ & & Increases & Increases & \\
\hline 
\end{tabular}
\caption{Summary of magnetic configuration impact on transport mechanisms. ``Increases''/``decreases'' indicates that increasing the given parameter increases/decreases the corresponding transport mechanisms. No entry implies no apparent dependence. The vacuum assumption is assumed for a quasisymmetric field with helicity $M$, $N$ subject to a perturbation with mode numbers $m_{\epsilon}$, $n_{\epsilon}$ as described in \eqref{eq:model_perturbation}.}
\label{tab:mechanisms}
\end{table}

\section{Discussion and conclusions}
\label{sec:conclusions}

\noindent We highlight the following conclusions from our analysis of loss mechanisms. 
\begin{itemize}
    \item While the quasisymmetry levels in the six equilibria are of comparable magnitudes (except for ITER), the GC confinement properties are markedly distinct. This behavior is partially accounted for by differences in the equilibria that provide natural protection against ripple wells and diffusive transport. As has been observed in several studies \cite{2021Bader,2022Landremanb}, QH configurations enjoy reduced losses due to their short connection lengths. In addition, the spatial localization of quasisymmetry errors is an important consideration. Ripple localized near the field maximum and minima tends to promote transitions between ripple and banana classes, while in other regions it can more easily yield prompt ripple losses. This feature could be accounted for in coil optimization problems or the definition of a spatially-localized quasisymmetry metric. 
    \item The transport mechanisms responsible for both prompt and long-timescale losses differ widely between equilibria. Therefore, analyzing the relevant loss mechanisms should inform further optimization for improved confinement.
    \item Non-conservation of $J_{\|}$ is prevalent among both prompt and long-timescale losses due to class transitions, resonances, or significant deviations of the field strength well shape along a trajectory. This can lead to additional diffusive effects on DC orbits (Figure \ref{fig:ariescs_sb}) and additional prompt losses due to rapid banana tip diffusion (Figure \ref{fig:ariescs_diffusive_banana}). Diffusion across $J_{\|}$ surfaces may contribute toward convective losses by enabling transport onto lost orbits. 
    Quantifying the impact of non-conservation of $J_{\|}$ will require a comparison of GC with orbit-averaged calculations, which will be explored in future work. Radially local metrics for fast ions based on the bounce-averaged motion may not always be applicable. As the radially-global bounce-averaged guiding center equations are valid in a $\rho_* \ll 1$ expansion, the breakdown of $J_{\|}$ adiabaticity may indicate the eventual breakdown of the GC approximation. Therefore, full-orbit simulations may eventually be required to model the dynamics faithfully.
    \item Many prompt losses are not lost on DC banana orbits but on diffusive banana, ripple, or resonant orbits. Some of the non-DC banana losses can be mitigated by improving the alignment of $J_{\|}$ surfaces with flux surfaces, while other losses require different avenues for improving prompt loss metrics in 3D configurations. One approach is integration over the full guiding center motion to identify diffusive or DC mechanisms rather than relying solely on bounce-averaged metrics. Similar objectives involving the full guiding center motion have been employed \cite{2006Ku}, for example, in the design of ARIES-CS \cite{2008Ku}. Another approach is directly targeting integrability in the banana tip map \cite{1981Goldston}. Metrics similar to those used to eliminate magnetic islands in stellarators, such as Greene's residue \cite{1979Greene}, may be applicable. 
    \item It is commonly assumed that the diffusive transport associated with many separatrix crossings is responsible for long-timescale losses \cite{2022Alonso,2019Cole}. While this mechanism is present, it is often the same mechanisms that lead to prompt losses (ripple trapping, DC, or diffusive banana-drift motion) that result in the eventual loss of the trajectory. Thus reducing these prompt loss channels may consequently improve the long-timescale behavior. Furthermore, transitions often appear in a periodic rather than irregular fashion. 
    \item Several features of GC confinement are common to stellarators close to QS and rippled tokamaks, including the prevalence of ripple trapping, diffusive banana tip motion, and collisionless detrapping processes. On the other hand, resonant barely-trapped orbits are more prevalent in ITER and are responsible for many non-prompt losses. 
    In the stellarator equilibria, such resonant lost orbits are not as common, although this may be a potential loss mechanism in other equilibria \cite{2021White}.
    The stellarator equilibria also uniquely feature irregular transitioning behavior, while the rippled ITER equilibrium features only periodic transitioning. Furthermore, the ITER configuration features enhanced losses over several QH stellarators on long timescales. This behavior is due to the high mode-number perturbations arising from coil ripple that enables enhanced non-adiabatic losses, ripple trapping, and resonant barely-trapped losses. To compare tokamaks and stellarators on equal footing would require free-boundary stellarator equilibria close to quasisymmetry.
\end{itemize}

\noindent Several physical effects may introduce additional subtleties. 
\begin{itemize}
    \item We have not included collisions and instead consider the dynamics on timescales shorter than a collisional slowing-down time ignoring the pitch-angle scattering. As collisions are more likely to modify the dynamics on long timescales, additional diffusive mechanisms may arise due to wide periodic transitioning orbits (Figure \ref{fig:aten_periodic}) or periodic DC orbits (Figure \ref{fig:hsx_periodic_banana}).
    \item Throughout, we have assumed that continuously nested magnetic surfaces exist in each equilibrium. However, fast particle losses can be enhanced when flux surfaces are destroyed \cite{2014Nemov}. Additional transport mechanisms will likely arise in this case due to the interaction of particle trajectories with a magnetic island or chaotic magnetic field lines \cite{2007Gunter,2008Strumberger}.
    \item We have assumed that a particle is considered lost at the plasma boundary. This assumption may lead to overestimating the losses due to re-entering trajectories. A more realistic quantification of the losses in a reactor requires modeling of plasma-facing components \cite{2013Seki,2008Sagara} as orbits can generally re-enter the plasma volume. Such an analysis may provide insight into the localization of losses on the first wall and possible damage to material structures.
    \item We have neglected the possible interaction of fast particles with fluctuations, such as Alfv\'{e}nic instabilities. It has been suggested that because stellarators can operate at higher density, the relative alpha particle pressure will be small enough that the instability drive will be sufficiently small \cite{2012Helander}. On the other hand, additional gaps in the continuum spectrum arise in the presence of 3D fields \cite{2011Spong}. Furthermore, there may be different loss mechanisms not considered in this paper, caused by the breakdown of adiabatic invariants due to wave-particle resonances. Further analysis is needed to determine how these interactions will impact stellarator reactors. 
\end{itemize}  

\section*{Acknowledgements}

The authors acknowledge fruitful conversations with R. White, H. Mynick, A. Boozer, R. Mackay, and F. Parra. A. Bader provided the Wistell-A, Ku5, NCSX, and ARIES-CS equilibria and assistance with benchmarking of the SIMSOPT code. 

EJP was supported by the Presidential Postdoctoral Research Fellowship at Princeton University. This research was also supported by the Simons Foundation/SFARI (560651, AB and ML). Some calculations were performed using computing resources at Princeton Plasma Physics Laboratory under DOE contract DE-AC02-09CH11466.

\newcommand{\newblock}{}
\bibliographystyle{unsrt}
\bibliography{main}

\end{document}